\documentclass[review]{elsarticle}

\usepackage{lineno,hyperref}
\modulolinenumbers[5]
\usepackage{amssymb} 
\usepackage{amsmath} 
\usepackage{stmaryrd} 
\usepackage{geometry} 
\usepackage{graphicx}
\usepackage[T1]{fontenc}
\usepackage{subfig}
\usepackage{sectsty}
\sectionfont{\fontsize{12}{8}\selectfont}
\subsectionfont{\fontsize{12}{8}\selectfont}
\subsubsectionfont{\fontsize{12}{8}\selectfont}

\journal{Computer Methods in Applied Mechanics and Engineering}

\begin{document}

\begin{frontmatter}

\title{Concurrent multiscale analysis without meshing: microscale representation with CutFEM and micro/macro model blending}

\author[1]{E. Mikaeili\corref{aaa}}
\ead{mikaeilie@cardiff.ac.uk}
\author[2]{S. Claus\corref{aaa}}
\ead{susanne.claus@onera.fr}
\author[1,3]{P. Kerfriden\corref{aaa}}
\ead{pierre.kerfriden@mines-paristech.fr}

\cortext[aaa]{Corresponding Authors.}

\address[1]{Cardiff University, School of Engineering, \\ The Parade, CF243AA Cardiff, United Kingdom }

\address[2]{ONERA, Université Paris-Saclay, 8 Chemin de la Huni\`{e}re, 91120 Palaiseau, France }

\address[3]{MINES ParisTech, PSL Research University, MAT--Centre des Matériaux, \\ CNRS UMR 7633, BP 87 91003 Evry, France}

\begin{abstract}
In this paper, we develop a novel unfitted multiscale framework that combines two separate scales represented by only one single computational mesh. Our framework relies on a mixed zooming technique where we zoom at regions of interest to capture microscale properties and then mix the micro and macroscale properties in a transition region. Furthermore, we use homogenization techniques to derive macro model material properties. The microscale features are discretized using CutFEM. The transition region between the micro and macroscale is represented by a smooth blending function. To address the issues with ill-conditioning of the multiscale system matrix due to the arbitrary intersections in cut elements and the transition region, we add stabilization terms acting on the jumps of the normal gradient (ghost-penalty stabilization). We show that our multiscale framework is stable and is capable to reproduce mechanical responses for heterogeneous structures in a mesh-independent manner. The efficiency of our methodology is exemplified by 2D and 3D numerical simulations of linear elasticity problems.\\
\end{abstract}

\end{frontmatter}

\linenumbers

\section{Introduction}

Many porous materials in biology and engineering, such as bones and composites, have inherently complex inner structures. These complex inner structures yield highly oscillatory irregular numerical solution fields. And hence an accurate finite element solution requires a discretization that is able to capture the heterogeneities (i.e., pores) within acceptable precision. While resolving the entire structure with extremely high-resolution meshes may result in accurate simulations, the computational cost is also adversely affected and may not be affordable for large structures. One feasible approach for a numerically tractable description of these media is to locally refine the mesh inside the region of interest and coarsen the mesh outside. However, this approach is more applicable when the errors corresponding to the discretization and mechanical behavior in the coarse mesh region remain negligible \cite{ODEN1997, Akbari15}.

To improve the approximation of the coarse scale region a multiscale system can be constructed to incorporate microscale features (such as micro-pores and inclusions) in macroscale solutions. This can be carried out either over the entire structure or only inside regions of interest. These two approaches can be classified as hierarchical and partitioned-domain concurrent multiscale methods \cite{tadmor_miller_2011}, respectively, at which both scale solutions are computed simultaneously. In the hierarchical approach, micro and macroscale solutions are addressed at the same time and positions, while for the partitioned-domain approach, a particular part of the domain is resolved with microscale governing equations and the rest of the medium with macroscale governing equations. The hierarchical approach, also known as computational homogenization, was proposed in a simplified version based on the effective medium by \cite{Eshelby57,Mori1973} to homogenize heterogeneities in terms of volume fractions, and then developed with fewer restrictions and for a broad range of problems; such as first and second order computational homogenization methods (see for instance \cite{Sanchez2013,Nguyen2012, Akbari.kerfriden.ea.18}). These approaches rely on the assumption of the existence of scale separation. When this assumption is violated, mainly in critical regions with local defects such as crack tips, damages, and holes, the partitioned domain approaches can be employed instead to alleviate this issue by directly modeling the critical regions with microscale governing equations. Partitioned domain approaches can be used to link either same or different mathematical models (in terms of physics and/or scales), such as continuum-to-continuum problems \cite{FISH1992539, Dhia06}, continuum models coupled with molecular dynamic simulations \cite{Xiao2004, Belytschko03,Silani.Talebi.ea.15, Talebi.Silani.ea.13} and coupled atomic-to-continuum models \cite{Tadmor1996,rokovs2016variational,rokovs2017adaptive,magliulo2020contact,beex2015mechanical}. Belytschko et al. \cite{belytschko2008multiscale} proposed another class of partitioned-domain approaches for aggregation of discontinuities across different scales, which mainly treats material instabilities at critical regions. See \cite{krokos2021bayesian} and the references therein for recent advances of partitioned-domain approaches based on machine learning.
We can use both hierarchical and partitioned-domain methods in a problem simultaneously \cite{Lloberas-Valls2012, Talebi.Silani.ea.14}.

There are numerous coupling techniques available in the literature extending the FEM-based mono-models for the partitioned-domain multiscale approaches. The examples of these methods are overlapping domain decomposition methods \cite{Guidault.07, MASSING2013SIAM} such as the Arlequin method \cite{Dhia.98, Dhia.Rateau.05} and non-overlapping domain decomposition methods \cite{ZOHDI2001,Gendre09} like the s-method \cite{FISH1992539} and the mortar method \cite{Lamichhane.Wohlmuth.04}, which all rely on superimposing local models with a fine mesh to the lowermost coarse mesh global model. These methods aim at reconstructing a multi-mesh framework using gluing conditions for the common interface between meshes. The interface conditions are implemented in a framework of coupling operators, such as the Lagrange multiplier approach \cite{BURMAN20102680} that is extensively used in the Arlequin method \cite{Dhia.08, Sun.17} and the Nitsche approach \cite{ BURMAN2012328, MASSING2018262} which imposes interface conditions weakly. In contrast to the mentioned coupling techniques that are all intrusive, Gendre et al. \cite{Gendre09, Gendre11} introduced a non-intrusive strategy for the coupling of global and local models; where the local model does not modify the global model, and all computations are performed with standard FEM. This approach is computationally efficient for large-scale problems with nonlinear phenomena that occur in small portions of the total domain.

From a computational standpoint, classical FEM is not sufficiently flexible for complex and time-dependent geometries, which are prevalent features in microscale phenomena. Mesh refinement and regeneration are obvious remedies to preserve the accuracy but these are very costly for large scale problems. 
%In an alternative approach, the CutFEM technique \cite{HANSBO20025537,Burman15, Hansbo2004} as a generalization of FEM, aims to facilitate the computations of complex and evolving geometries.(version before reviewing)
In an alternative approach, the CutFEM technique \cite{HANSBO20025537,Burman15, Hansbo2004} as a variation of XFEM \cite{Moes99,bordas2022}, aims to facilitate the computations of complex and evolving geometries.
In this method, the geometry is decoupled from the finite element mesh and the boundary of the computational domain is represented by a level set function or a given surface mesh over a fixed background mesh. The computation and update of the geometry are done in the discretized formulation, reducing the pre-processing computational cost of meshing.
Aside from the robust geometry description, the method ensures the stability of the discretization by introducing ghost penalty regularization terms in cut elements  \cite{Burman10,mikaeili2022novel}. This leads to improved conditioning of the resulting system matrix, which is a challenge in unfitted FEM approaches (see for instance \cite{agathos2016well,agathos2019improving}). The CutFEM technique has been applied for a range of single-scale problems, such as unilateral contact \cite{Claus18.2}, multiphase phenomena \cite{Claus19.2,Kerfriden.20} and fiber-reinforced composites \cite{Claus21b}.
It has also been recently developed for modeling multi-component structures using different meshes for each component. In this multi-mesh framework proposed by \cite{JOHANSSON2019672, DOKKEN2020113129}, multiple meshes can overlap in an arbitrary manner and intersected elements are regularized using ghost penalty regularization.

In this paper, we use the fact that CutFEM allows us to decouple the geometry from the finite element mesh (background mesh). We use CutFEM to capture the fine-scale geometry, which is expressed either in terms of an analytical distance function or in terms of a given surface mesh. Then, we project this fine-scale geometry description onto a background mesh which is fine in areas of interest and coarse elsewhere. \\
We demonstrate that a straightforward projection of the geometry description onto an adapted background mesh is insufficient and consequently, we develop a multiscale CutFEM framework. The straightforward projection is insufficient because the piecewise linear signed distance function approximation, in a mesh with the combination of fine resolution in areas of interest and very coarse mesh elements elsewhere, gives rise to the random appearance of geometrical artifacts in the coarse mesh region, yielding stress singularities. In order to alleviate this issue, we replace the signed distance function description in the coarse mesh domain with a homogenized domain. To couple the fine scale region or "\textit{zoom region}" with the coarse scale region, we develop a smooth mixing approach of the homogenized material and the fine scale "CutFEM" region. This approach belongs to the class of smoothened domain coupling methods, such as Arlequin or the Bridging Domain Method referenced previsouly. As such, we avoid the difficulty of meshing the coupling interface in a way that is compatible with the microscale description, which is known to create numerical issues \cite{Akbari15}. Moreover, we do not have to choose coupling conditions within the usual dictionary of possible interface gluing strategies. Instead, the gluing is performed "naturally", using a decaying weighted average of the two models within a smooth interface.
Then, we demonstrate the efficiency, robustness and accuracy of the smooth mixing approach between homogenized macroscale and CutFEM microscale region.

In our multiscale framework, the smooth mixing technique is inspired by the Arlequin method.
However, in contrary to the Arlequin mixing strategy, we do not cross and glue a high-resolution mesh to the underlying mesh but use a level set function over a single background mesh to define a transition region. We then mix the scales in the elements inside the transition region.
This implicit level-set-based description of geometrical and mixing properties (i.e. macro and microscale domains, zooming location and transition region) yields a highly versatile mesh-independent multiscale framework. In this framework, we can modify the geometrical and mixing properties by only changing the level set functions, leading to less computational pre-processing cost in comparison to the previous methods. 

The outline of the paper is as follows. In Section 2, we present the continuous formulation of the multiscale framework, in strong and weak forms. Then, in Section 3, we discretize the formulations using CutFEM and introduce the transition area for mixing purposes. In Section 4, we first test the idea of taking the functional description of microscale and projecting it onto a adaptive mesh background. Then we corroborate the efficiency of the proposed smooth mixed multiscale framework with 2D and 3D elasticity problems. In the 2D examples, we study a heterogeneous structure where the micro-pores are either distributed locally or uniformly. In the 3D case, we simulate a trabecular bone with complex micro-structure derived directly from micro-CT image data (also available in \cite{perilli07.2, perilli07}).

\section{Governing equations of the mixed multiscale problem}

In this Section, the formulation for the mixed elasticity problem is presented. First, we will introduce definitions and notations related to the domain partitioning of the method and then present the strong and weak formulation for the concurrent multiscale elasticity problem. Finally, we will present the discretization of the governing equations.\\

\subsection{Domain partitioning}
\begin{figure}[!ht]
   \centering
    \subfloat[ \label{Domain1}]{%
     \includegraphics[scale=0.4]{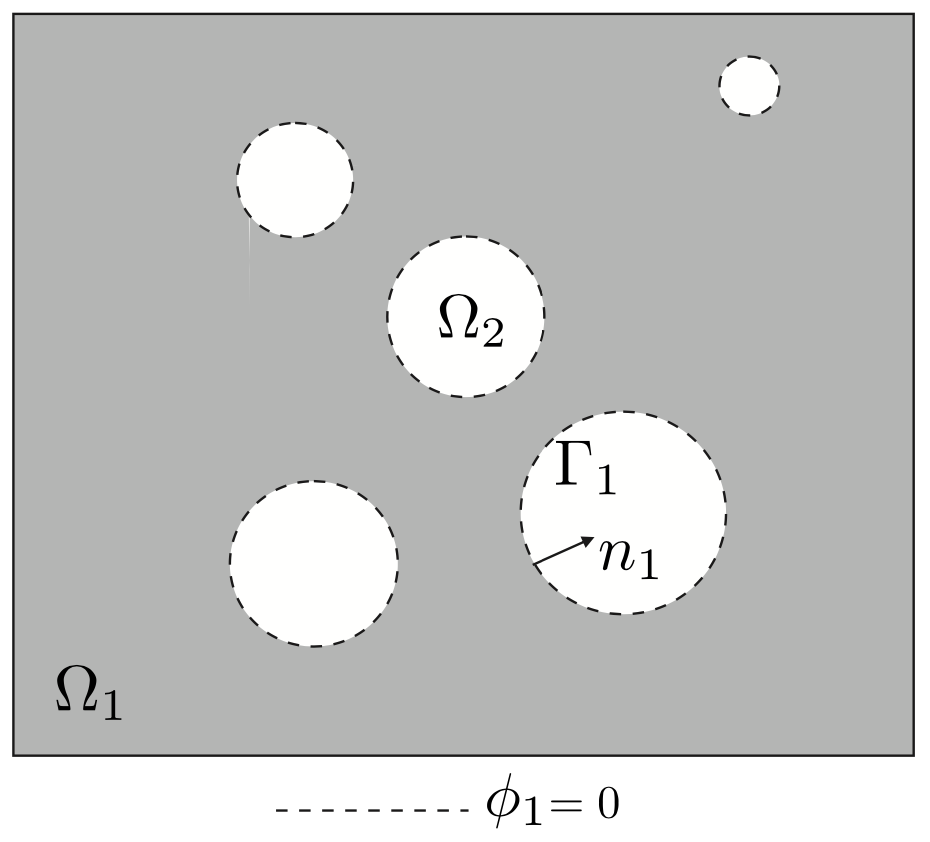}
     }
    \hfill
     \subfloat[ \label{Domain2}]{%
       \includegraphics[scale=0.4]{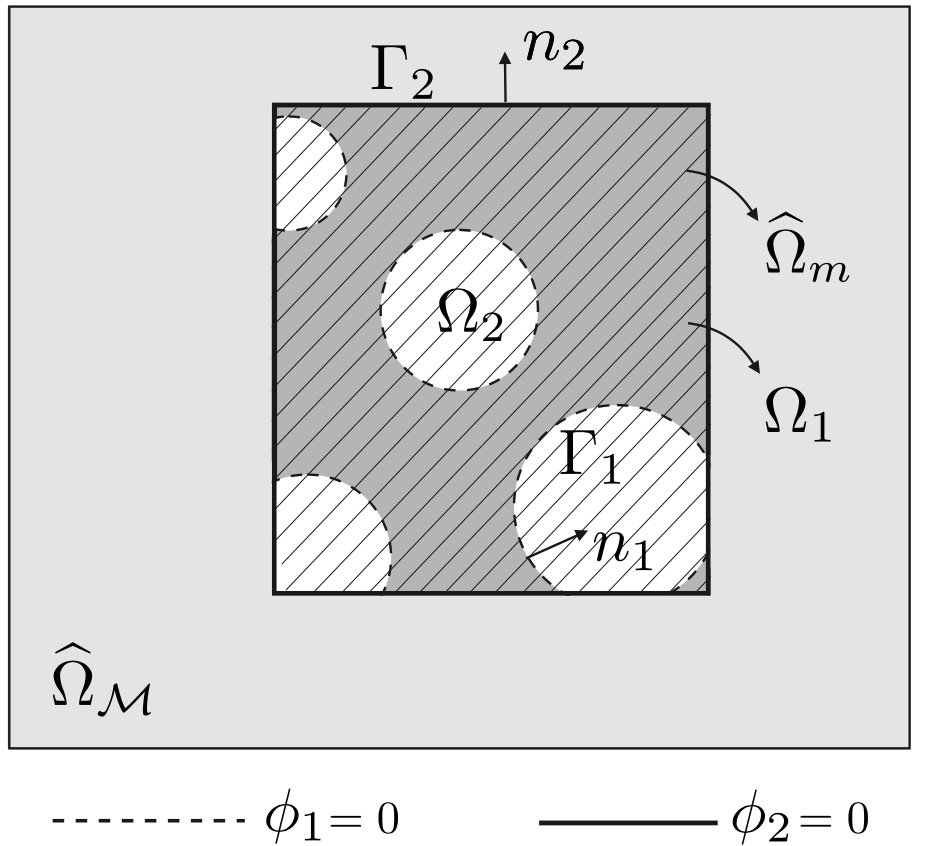}
     }
    \hfill
     \caption{Domain partitioning for the mixed multiscale method, a) micro-porous domain $\Omega$ partitioned  into matrix subdomain $\Omega_1$ and pore subdomain $\Omega_2$ ($\Omega = \Omega_1 \cup \Omega_2$) with interface $\Gamma_1$ and b) partition of the computational domain $\Omega$ into macro subdomain $\widehat{\Omega}_{\mathcal{M}}$ and micro subdomain $\widehat{\Omega}_{z}$ ($\Omega = \widehat{\Omega}_{\mathcal{M}} \cup \widehat{\Omega}_{z}$). Here, $\widehat{\Omega}_{m}=\widehat{\Omega}_{z} \setminus \Omega_2$ denotes the porous micro-domain.}
     \label{fig:Domains1&2}
\end{figure}

Let $\Omega$ be the computational domain of a micro-porous heterogeneous medium comprised of a matrix subdomain ${\Omega}_1$ and a pore subdomain ${\Omega}_2$, as illustrated in Figure~\ref{Domain1} and  
\begin{equation}
   {\Omega}_i\subset \mathbb{R}^d  , \ \ \ i=1,2, \ \ d=2,3,
\label{Eq:1}
\end{equation}
where the interface between ${\Omega}_1$ and ${\Omega}_2$ is determined by a continuous level set function $\phi_1$ defined as follows
\begin{equation}
    \phi_1(x) = \begin{cases}
         \textgreater 0 \qquad x \in {\Omega}_2 ,  \\
         =0 \qquad x \in \Gamma_1 ,\\
         \textless 0 \qquad x \in {\Omega}_1.
    \end{cases}  
\end{equation}
%For an illustration see Figure \ref{fig:Domains1&2}a.
The normal vector in $x \in \Gamma_1$, pointing from $\Omega_1$ to $\Omega_2$, is given by
\begin{equation}
    n_1 = \frac{\nabla \phi_1(x)}{\left\Vert\nabla \phi_1(x)\right\Vert}.
\end{equation}
In the previous definition, $\left\Vert x \right\Vert$ denotes the Euclidean norm $\left\Vert x \right\Vert = \sqrt{x \cdot x}$. 
Next, we define the microscale zoom region $\widehat{\Omega}_z$ for our multiscale analysis by a continuous level set function $\phi_2$ given by

\begin{equation}
    \phi_2(x) = \Bigg\{ \begin{array}{ll}
         \textgreater 0 \qquad x \in \widehat{\Omega}_{\mathcal{M}} ,  \\
         =0 \qquad x \in \Gamma_2, \\
         \textless 0 \qquad x \in \widehat{\Omega}_z,
    \end{array}  
\end{equation}
whose zero iso-line defines the boundary of the zoom. The macro-domain, denoted by $\widehat{\Omega}_{\mathcal{M}}$, is the domain outside of the zoom. For an illustration see Figure \ref{fig:Domains1&2}b, with $\widehat{\Omega}_z$ shown as the shaded area. Furthermore, the normal vector on the interface $\Gamma_2$ pointing from $\widehat{\Omega}_z$ to $\widehat{\Omega}_{\mathcal{M}}$ is given by
\begin{equation}
    n_2 = \frac{\nabla \phi_2(x)}{\left\Vert\nabla \phi_2(x)\right\Vert}.
\end{equation}
Furthermore, let $\widehat{\Omega}_m=\widehat{\Omega}_z \setminus \Omega_2 = \widehat{\Omega}_z \cap \Omega_1$, denote the microscale region without the pores. The porous microscale domain can be expressed by a combination of the two level set functions
\begin{equation}
\widehat{\Omega}_m = \{ x \in \Omega: \phi_1(x)< 0 \mbox{ and } \phi_2(x) < 0 \}. 
\end{equation}
%\subsection{Strong form of governing equations}
%%%%%%%%%%%%%%%%%%
\subsection{Field equations of the multiscale elasticity problem}
Let us consider linear elastic behavior for the macro-domain $\widehat{\Omega}_\mathcal{M}$ and the porous micro-domain $ \widehat{\Omega}_m$. In our multi-model, we are seeking the deformation field $ u : \widehat{\Omega}_\mathcal{M} \times \widehat{\Omega}_m \rightarrow \mathbb{R}^d \times \mathbb{R}^d $ which satisfies

\begin{subequations}
    \begin{align}
    \text{div} \, \sigma_\mathcal{M}+ f_\mathcal{M} = 0 \qquad  \text{in} \  \widehat{\Omega}_\mathcal{M}, \\
        \text{div} \, \sigma_m + f_m = 0 \qquad  \text{in} \  \widehat{\Omega}_m,
    \end{align}
    \label{Eq: StrongF}
\end{subequations}
where
\begin{subequations}
    \begin{align}
    \sigma_\mathcal{M} (u):=D_\mathcal{M} :\nabla_s u   \\
        \sigma_m (u):=D_m :\nabla_s u 
    \end{align}
\end{subequations}

The boundary of the background domain $\Omega$ is partitioned into $\partial \Omega_u$ and $\partial \Omega_t$ ($\partial \Omega = \partial \Omega_t  \cup \partial \Omega_u$), where $\partial \Omega_u$ is the part where the body is clamped and $\partial \Omega_t$ is the part where traction $t$ is applied with $ \partial \Omega_t  \cap \partial \Omega_u = \emptyset$.

In the expressions above, $f_\mathcal{M}$ and $f_m$ are volume source terms, $\nabla_s . =\frac{1}{2}(\nabla.+(\nabla.)^T)$ is the symmetric gradient operator, and $D \in (\mathbb{R}^d)^4$ is the fourth order Hooke tensor of isotropic linear elastic material given by 
\begin{equation}
     D: \nabla_s. = \lambda \textrm{Tr}(\nabla_s.)\mathbb{I}+2\mu\nabla_s.
\end{equation}
where $\textrm{Tr}$ is the tensor trace operator, $\lambda=\frac{E\nu}{(1+\nu)(1-2\nu)}$, $\mu = \frac{E}{2(1+\nu)}$ are the Lamé parameters expressed by the Young's module $E$ and the Poisson's ratio $\nu$.

On the zooming interface, $\Gamma_2$, between micro and macro model, the traction is required to satisfy the following coupling condition

\begin{equation}
    \sigma_m \cdot {n}_{2} = - \sigma_\mathcal{M} \cdot {n}_{2} \qquad \text{on} \qquad \Gamma_2.
\end{equation}

Integrating governing equations (\ref{Eq: StrongF}a)-(\ref{Eq: StrongF}b) over the given domains, i.e. macro-domain $\widehat{\Omega}_\mathcal{M}$ and micro-domain $\widehat{\Omega}_{m}$, the weak form of the multiscale elasticity problem is given as follows. We seek a displacement field $ u: \widehat{\Omega}_\mathcal{M} \times \widehat{\Omega}_m \rightarrow \mathbb{R}^d \times \mathbb{R}^d $, $u \in H^1(\Omega_\mathcal{M}) \times H^1(\Omega_m)$,  satisfying
 \begin{equation}
\int_{\widehat{\Omega}_{\mathcal{M}}} \sigma_\mathcal{M}( \nabla_s u) : \nabla_s \delta u \, dx + \int_{\widehat{\Omega}_m} \sigma_m( \nabla_s u) : \nabla_s \delta u \, dx
=   \int_{\widehat{\Omega}_{\mathcal{M}}}  f_{\mathcal{M}} \cdot \delta u \, dx + \int_{\widehat{\Omega}_{m}} f_m \cdot \delta u \, dx +  \int_{\partial \Omega_t} t \cdot \delta u  \, dx \, , 
\label{Eq:Weak}
\end{equation}
for all test functions $\delta u: \widehat{\Omega}_\mathcal{M} \times \widehat{\Omega}_m \rightarrow \mathbb{R}^d \times \mathbb{R}^d$, $\delta u \in H^1_0(\Omega_\mathcal{M}) \times H^1_0(\Omega_m)$ which satisfy the homogeneous Dirichlet boundary condition
\begin{equation}
\delta u = 0 \qquad \textrm{on} \ \ \ \partial \Omega_u.
\end{equation}

\section{Discretization for mixed multiscale problems}

In this Section, we introduce a CutFEM-based approximation scheme of the multiscale elasticity problem proposed in Section 2 using a novel mixed cut finite element approach. The arbitrary intersection of the porous domain by the sharp zooming interface, $\Gamma_2$, can result in bad conditioning for the assembled system matrix. To alleviate this problem, we introduce a mixing strategy between the macroscale and microscale regions. In this mixed approach, we create an overlap between the two models. We refer to the overlapping domain as "\textit{transition domain}", as highlighted in Figure \ref{{fig:Domain3}}a in yellow.

\begin{figure}[!ht]
   \centering
    \subfloat[ \label{domain3}]{%
     \includegraphics[scale=0.4]{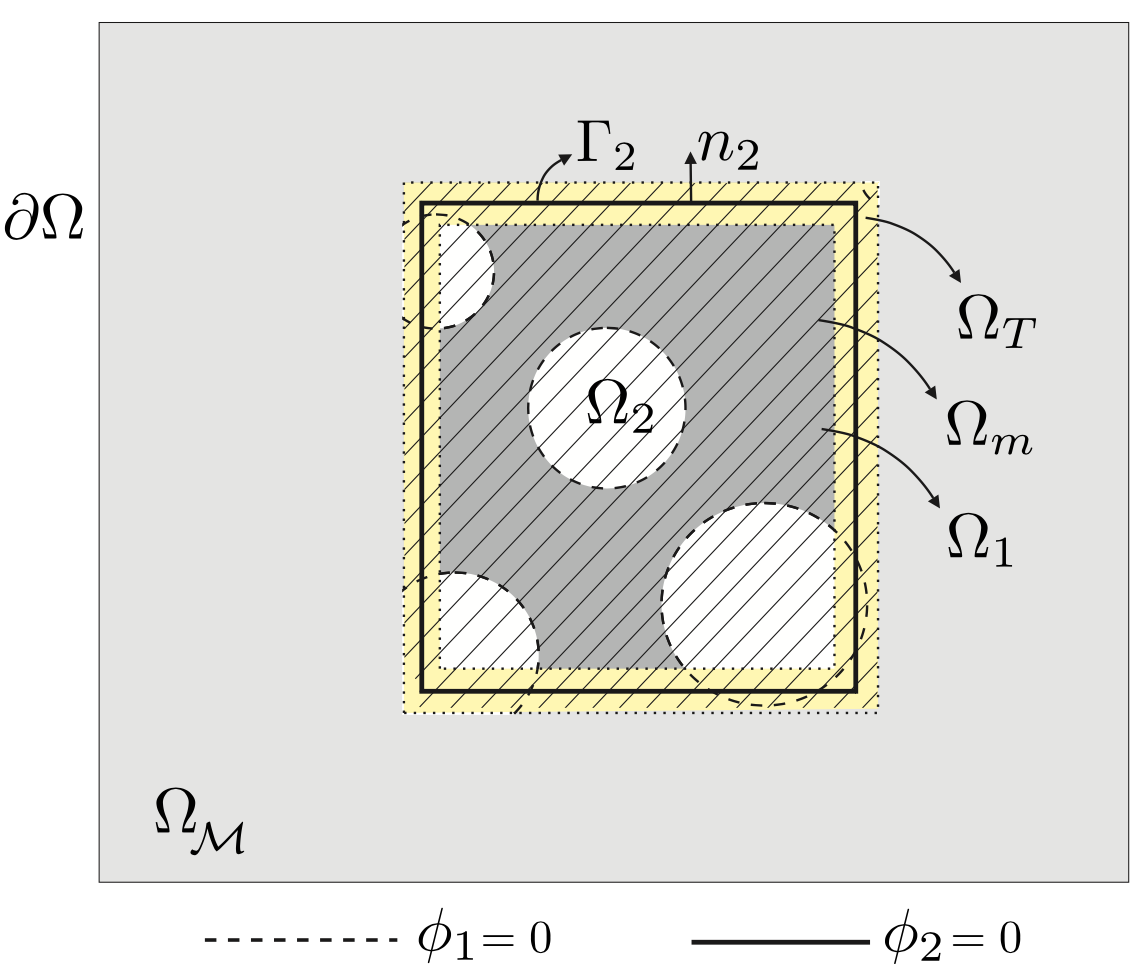}
     }
    \hfill
     \subfloat[ \label{domain2}]{%
       \includegraphics[scale=0.4]{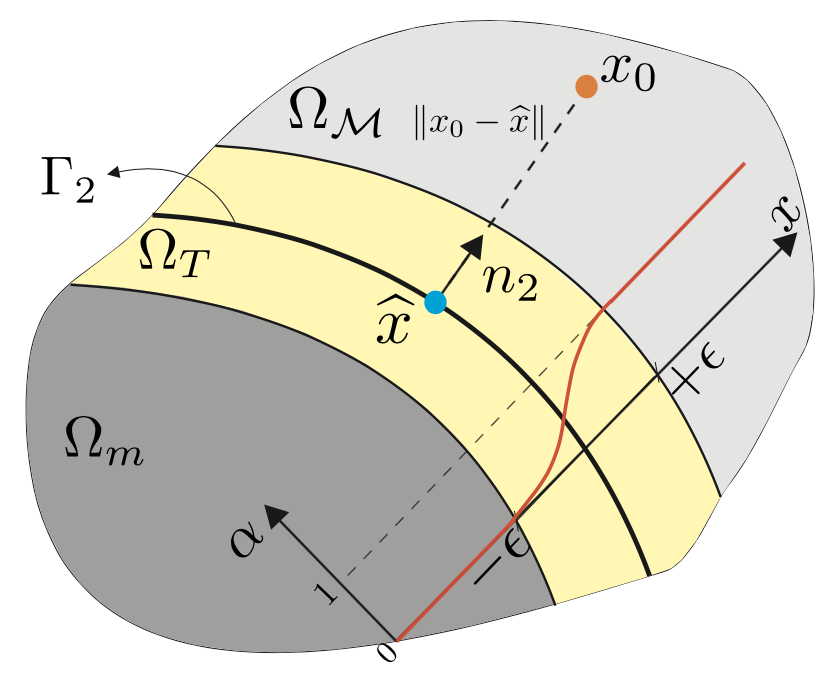}
     }
    \hfill
     \caption{Transition domain in the mixed multiscale method (a) and the distance dependent weight function, $\alpha$, in the transition domain (b).}
     \label{{fig:Domain3}}
\end{figure}

For mixing purposes, we extend the macro and micro domains defined in the previous Section into the transition region. First we extend the macro domain by $\Omega_{\mathcal{M}} := \widehat{\Omega}_{\mathcal{M}} \cup \Omega_T$. Then we extend the micro domain inside zoom to $\Omega_m := \widehat{\Omega}_m \cup \Omega_T$, where $\Omega_T$ is the transition domain. In this framework, $\Omega_{\mathcal{M}}$ and $\Omega_m$ are overlapping in the transition domain. The transition domain is determined by the level set function ${\phi}_2$, which is the signed distance function to ${\Gamma}_2$. We set the width of the transition region to $2\epsilon$, which is given by the signed distance from $-\epsilon$ to $+\epsilon$.

We define a smooth weight function $\alpha$ for the mixing, as shown in Figure \ref{{fig:Domain3}}b, and express it in terms of $\phi_2$ by

\begin{equation}
    \alpha(x) = \Bigg\{ \begin{array}{ll}
         0 \qquad \text{if} \ \ \phi_2(x)\leq-\epsilon,   \\
         \mathcal{S} \qquad \text{if} \ \ -\epsilon<\phi_2(x)<\epsilon, \\
         1 \qquad \text{if} \ \ \phi_2(x)\geq\epsilon.
    \end{array}  
\end{equation}

In the previous expression, $\mathcal{S}$ is a smooth function varying from 0 to 1 inside the transition zone, and as shown in Figure \ref{{fig:Domain3}}b, $-\epsilon$ and $+\epsilon$ are the lower and upper bounds of $\Omega_T$ inside micro and macro domains, respectively. We will blend and mix the macro and micro-model using the weights $\alpha$ and $1-\alpha$, i.e. 
\begin{equation}
  \alpha \,  {a}_{\mathcal{M}}({u},v) + (1-\alpha) \, {a}_{m}({u}_h,v)
= \alpha \, {l}_{\mathcal{M}}(v)+ (1-\alpha) \, {l}_{m}(v).
\end{equation}

\subsection{multiscale finite element space}
Here we discretize the weak form (\ref{Eq:Weak}) of the multiscale model, which locally modifies a global problem by using only one mesh, unlike other similar methods such as the Arlequin method \cite{Dhia.08} and multi-mesh CutFEM \cite{DOKKEN2020113129} that superimpose a high resolution mesh onto a coarse background mesh. In our framework, we introduce triangulation $\mathcal{T}$ for the background domain $\Omega$ and then define the corresponding finite element space of continuous linear function as 
\begin{equation}
{\mathcal{Q}}_h : = \{ w \in  \mathcal{C}^0(\Omega): w |_{K} \in \mathcal{P}^1(K) \, , \forall K \in \mathcal{T} \},
\end{equation}
where the corresponding mixed multiscale model physical domains, $\Omega^h _\mathcal{M}$ and ${\Omega}^h _{{m}}$ are approximated as
\begin{equation}
    \Omega^h _\mathcal{M} = \{ x \in \Omega | {\phi}^h _2 (x) \geq -\epsilon \},
    \label{Eq:physDfem1}
\end{equation}
\begin{equation}
    \Omega^h _m = \{ x \in \Omega | {\phi}^h _1 (x)\leq 0 \ \ \text{and} \ \ \phi^h _2 (x)\leq 
     \epsilon \}.
    \label{Eq:physDfem2}
\end{equation}
Furthermore, we approximate the transition domain $\Omega^h _T$ as
%and porous domain $\Omega^h_2$, respectively in
\begin{equation}
    \Omega^h _T = \{ x \in \Omega | -\epsilon \leq {\phi}^h _2 (x) \leq 
     \epsilon \}.
    \label{Eq:Transition}
\end{equation}

In (\ref{Eq:physDfem1}) and (\ref{Eq:physDfem2}), $\phi^h _1 (x) \in {\mathcal{Q}}_h$ is the linear approximation of the level set function $\phi_1$ and ${\phi}^h _2 (x) \in {\mathcal{Q}}_h$ is the linear approximation of level set function ${\phi}_2$. By using these level set functions, we define the position of the microscale features and pores over a single fixed mesh arbitrarily (in a nonconforming manner). Now we can present the approximate interface $\Gamma^h _1$

\begin{equation}
  \Gamma^h _1 = \{ x \in \Omega_\mathcal{M} ^h \, | \, \phi^h _1 (x) =0 \}. 
\end{equation}

The pores with arbitrary geometries can have non-zero intersection with either macro or micro domains, where all the elements of $\mathcal{T}$ intersected by $\Gamma^h _1$ will be grouped in set
\begin{equation}
    {T}^h _1:= \{ K \in \mathcal{T} : K \cap \Gamma^h _1 \neq \emptyset\}, 
\end{equation}
 where the corresponding domain is defined as ${T}^H _1 = \bigcup_{K\in {T}^{h} _1} K$. \\
 Furthermore, let $T^H_p$ denote the set of all elements, which are fully inside the pores, i.e. $\phi_1^h >0$ in all vertices of the element.

\begin{figure}
\centering
  \includegraphics[scale=0.4]{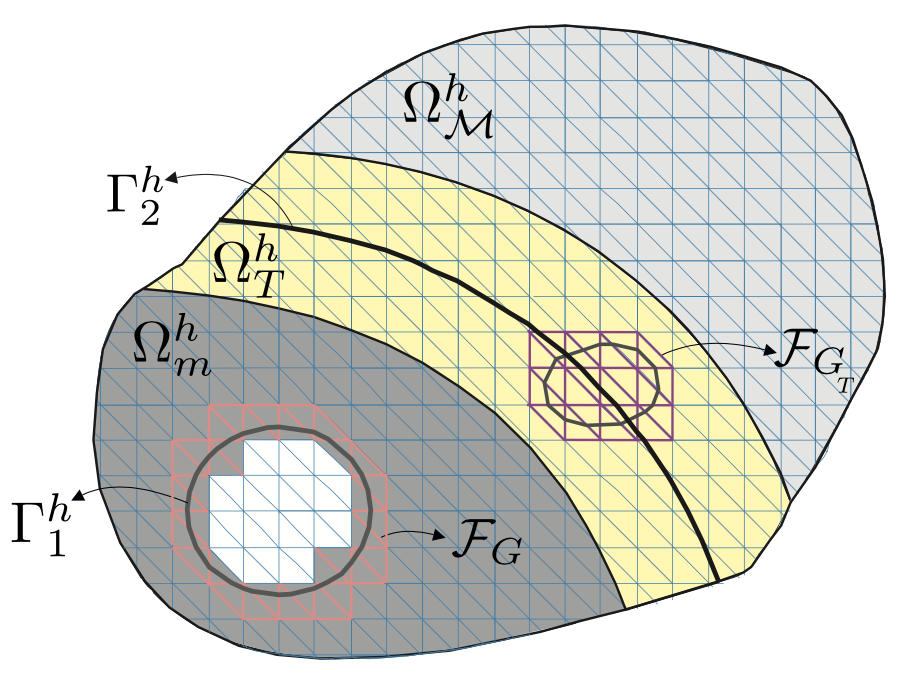}
  \caption{Schematic presentation of the discretized domain for the mixed multiscale method.}
  \label{fig:DiscretizedDomain}
\end{figure}

\subsection{Fictitious domain}
First, we define a set of all elements in the background mesh $\mathcal{T}$ which have a non-zero intersection with  $\Omega_\mathcal{M}^h$ or ${\Omega}_{m}^h$
\begin{equation}
      {\mathcal{T}_h}:= \{ K \in \mathcal{T} : K \cap (\Omega_\mathcal{M} ^h \cup {\Omega}_{m} ^h) \neq \emptyset\} .
\end{equation}
Note that, this fictitious domain mesh consists of all elements in the background mesh except for the elements fully inside the pores outside of the transition region (elements shown in white in Figure~\ref{fig:DiscretizedDomain}). 
The domain associated with this set of elements is called fictitious domain and is denoted by $ \Omega_{\mathcal{T}}:=\bigcup_{K\in (\Omega_\mathcal{M} ^h \cup {\Omega}_{m} ^h)} K$.

Notably, all elements inside the pores in the transition region are contained in the fictitious domain mesh. These elements are not integrated over in the micro-domain and therefore yield a zero contribution in the system matrix. Therefore, these elements rely mainly on the stiffness of the macro-domain. However, if almost the full weight is on the micro-domain, i.e. $\alpha \approx 0$, in the transition region there is very little contribution from the macro-domain inside the pores and this yields ill-conditioning. \\
In addition to this source of ill-conditioning, we can obtain ill-conditioned matrix entries through the integration of elements which lie almost entirely in the pores in the micro-region and therefore contain very little material from the micro-domain. \\
We address these two sources of ill-conditioning by introducing two ghost penalty regularization terms.
The first one is used for the elements intersected by $\Gamma^h_1$ in the microscale region $\Omega^h_m$, and is applied to the elements edges (shown in red in Figure \ref{fig:DiscretizedDomain}) given by 

\begin{equation}
{\mathcal{F}}_G:= \{ F = K \cap K'  : K  \in \mathcal{T}_h \mbox{ and } K' \in \mathcal{T}_h, \, F \cap {T}^H _1 \neq \emptyset  \}.
\end{equation}

The second ghost penalty regularization term is applied to the edges of elements in the transition region $\Omega^h _T$ that are intersected by $\Gamma^h _1$ or are inside the pores. These ghost penalty terms extend the solution of the micro-domain into the pores. It gives the elements in the pores a stiffness, which alleviates ill-conditioning, while maintaining the consistency and accuracy of the solution (i.e. terms vanish with optimal rate with mesh refinement and continuity of the solution). The corresponding edges are shown schematically in Figure \ref{fig:DiscretizedDomain} in purple and are defined as

\begin{equation}
{\mathcal{F}}_{G_{T}}:= \{ F = K \cap K'  : K  \in \mathcal{T}_h \mbox{ and } K' \in \mathcal{T}_h, \, F \cap ({T}^H _1 \cup {T}^H _2) \neq \emptyset  \},
\end{equation}
where ${T}^H _2$ is the domain related to the set of all elements of $\mathcal{T}$ intersected by pores in the transition domain defined as ${T}^H _2 = \bigcup_{K\in {T}^{h} _2} K$, where the set of elements $T^h _2$ is given by,

\begin{equation}
    {T}^h _2:= \{ K \in \mathcal{T} : K \cap \Omega^h _T \neq \emptyset \mbox{ and }  ( K \in T^H_1 \mbox{ or } K \in T^H_p )\}. 
\end{equation}

In this paper, since we use one adapted background mesh for the multiscale problem, the displacement field is continuous throughout the whole domain. For its discretiztion, we choose the vector-valued continuous piecewise linear space \\
%Then we look for the displacement approximation $u_h$ in the following space,
\begin{equation}
    {\mathcal{U}}_h:=\{ u \in \mathcal{C}^0 (\Omega_{\mathcal{T}}):  u|_K \in \mathcal{P}^{d,1} (K) \ \ \forall K \in {\mathcal{T}}_h \},
\end{equation}
where $d$ denotes the spatial dimension, $d=2,3$.

\subsection{Stabilized mixed finite element formulation}
The mixed finite element formulation for the proposed multiscale method is the following: find ${u}_h \in {\mathcal{U}}_h$, such that

\begin{equation}
  \alpha_h \,  {a}_{\mathcal{M}}({u}_h,v) + (1-\alpha_h)  \, {a}_{m}({u}_h,v)
=  \, \alpha_h {l}_{\mathcal{M}}(v)+  (1-\alpha_h) \, {l}_{m}(v)
\end{equation}
for any ${v_h} \in {\mathcal{U}}_h$ satisfying homogeneous Dirichlet boundary conditions. The bilinear form $a_\mathcal{M}$ and linear form ${l}_\mathcal{M}$ of the macro model are given by 

\begin{equation}
    {a_{\mathcal{M}}}({u}_h,v)=\int_{\Omega_{\mathcal{M}} ^h} {D}_{\mathcal{M}}\nabla_s{{u}_h}\nabla_s{v} \ dx ,
\end{equation}

\begin{equation}
    {l_\mathcal{M}}(v)=\int_{\Omega_{\mathcal{M}} ^h} f_\mathcal{M} \cdot v \ dx  + \int_{\partial \Omega_t ^h} t_d \cdot v \, ds. 
\end{equation}

In the previous problem statement, the regularized bilinear form $a_m$ is defined for the micro scale model as 
\begin{equation}
    {a}_m({u}_h,v)=\int_{{\Omega}_m ^h}  {D}_m \nabla_s{{u}_h}\nabla_s{v_h} \ dx \ + \sum_{F\in {\mathcal{F}}_G} \bigg(\int_F \frac{\beta h}{E_m}  \llbracket {D}_{m}\nabla_s{{u}_h} \rrbracket \llbracket {D}_{m}\nabla_s{v_h} \rrbracket \ dS \bigg)  . 
\end{equation}
Here, the second term, called ghost-penalty, ensures a uniformly bounded condition number for the system matrix and $\llbracket x \rrbracket$ denotes the normal jump of quantity $x$ over the facet $F$, and $\beta$ denotes the ghost penalty stabilization parameter that needs to be large enough to guarantee the coerciveness of bilinear form $a_m$ \cite{Burman15, BURMAN2012328} on the fictitious domain. The linear form of the microscale model is given by   
\begin{equation}
    {l_m}(v)=\int_{{\Omega}_m ^h}  f_m \cdot v \ dx.
\end{equation}

\clearpage

% --------------------------------------------------------------
% ----------------- Numerical Results --------------------------
% --------------------------------------------------------------

\section{Numerical results}

In this Section, we first test the projection of an analytical signed distance function over an adaptive background mesh using the CutFEM technique. Next, we investigate the performance of the proposed smooth mixing approach in a simplified multiscale problem. Eventually, we adopt a homogenized medium in the coarse domain of the mixed multiscale model and demonstrate the efficacy and robustness for 2D and 3D elasticity problems. In our 2D simulations, we use an analytical signed distance function to define the geometry. In contrast, in our 3D case study, we use a mesh surface derived from micro-CT image data to describe the geometry of a trabecular bone with a complex microstructure. All the numerical results are produced by the CutFEM library \cite{Burman15} developed in FEniCS \cite{fenics}.\\
   
\subsection{Adaptive CutFEM technique}
\begin{figure}[h]
\centering
  \includegraphics[scale=.3]{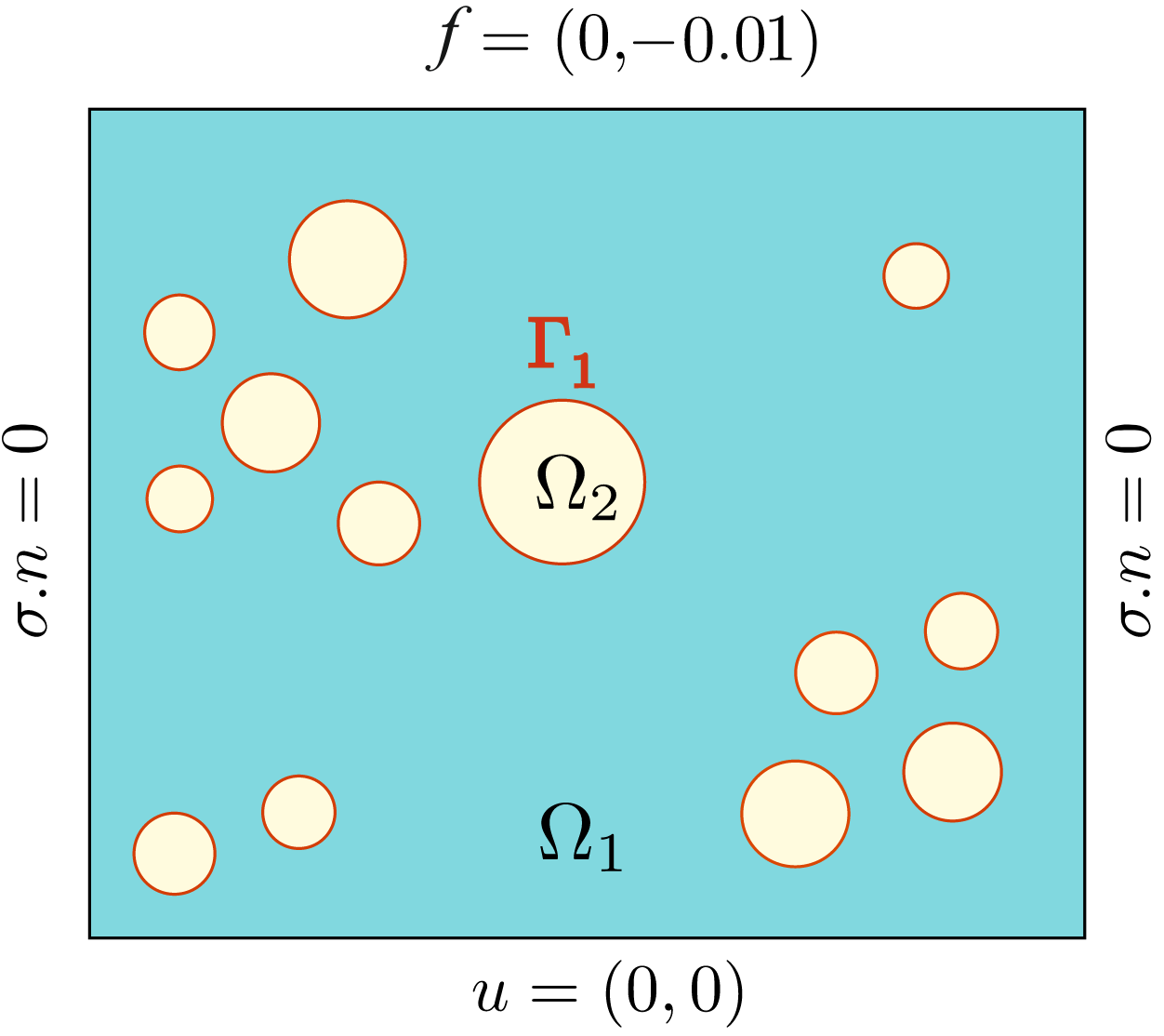}
  \caption{Schematic presentation of 2D rectangular domain with a quasi-uniform distribution of micro-pores.}
  \label{fig:ModelQuasiSchm}
\end{figure}

Let us consider the heterogeneous structure shown in Figure \ref{fig:ModelQuasiSchm}, comprised of a matrix and pores which are distributed all over the domain. The matrix domain is defined as $\Omega_1 = \Omega \backslash \Omega_2$, where $\Omega = [0,12]\times[0,10]$ is the rectangular background domain and $\Omega_2$ are the pores. We restrict the displacement at the bottom edge and prescribe the force $f= (0, -0.01)$ along the top edge of the domain. We assume the corresponding mechanical properties as following: $E=1$ and $\nu=0.3$.

\begin{figure}[h]
   \centering

     \subfloat[ \label{CutFEM1}]{%
       \includegraphics[scale=0.27]{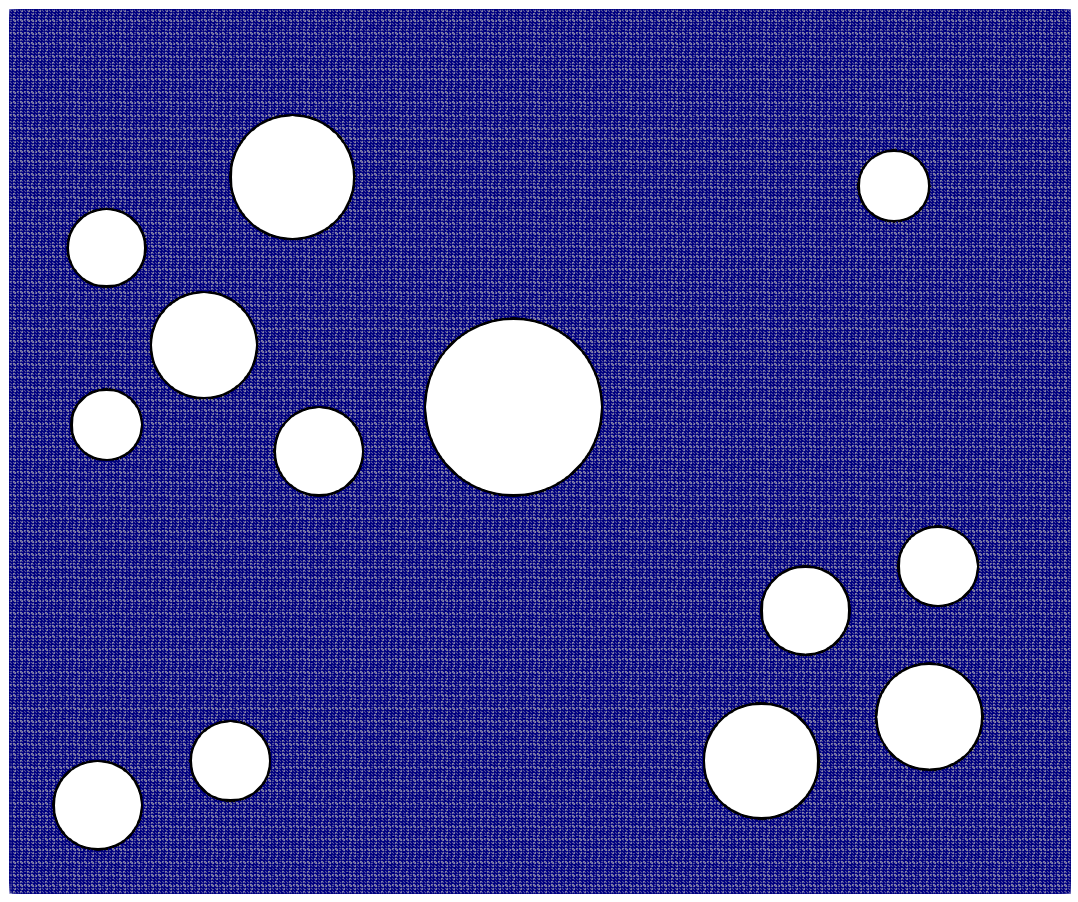}
     }
    \hfill
     \subfloat[ \label{CutFEM2}]{%
      \includegraphics[scale=0.27]{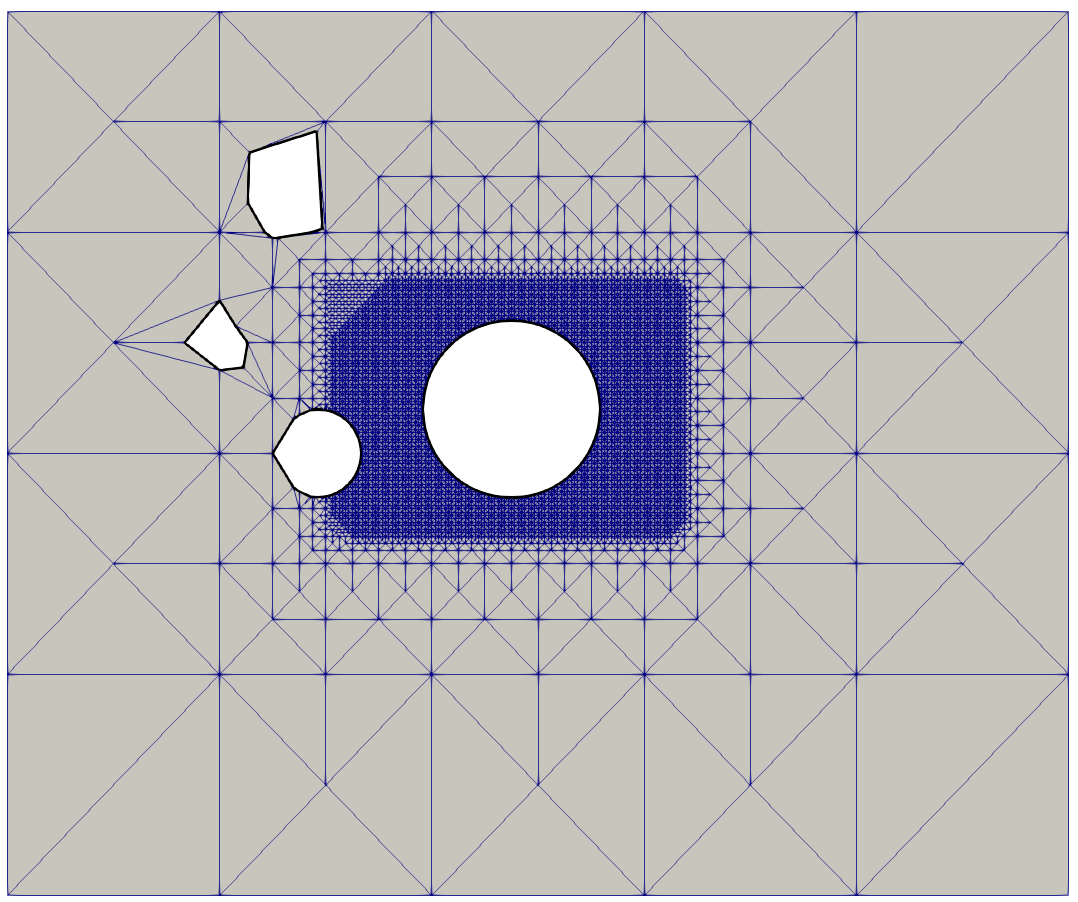}
     }
    \hfill
     \caption{ CutFEM method with a) high-resolution uniform mesh $h_{min}=0.054$ and b) CutFEM method with adaptive background mesh $h_{min}=0.054$ and $h_{min}/h_{max}=0.015$.}
     \label{fig:MeshesModelBAdaptive}
\end{figure}

Here, we define the geometry by a piecewise linear signed distance function over two types of background meshes. As depicted in Figure \ref{fig:MeshesModelBAdaptive}a,b, the first background mesh is uniform and fine everywhere; however, the second background mesh is fine only in regions of interest and coarse elsewhere, and the corresponding adaptive mesh refinement scale is $s=1/4$. The zero level set function $\Gamma_1 ^h$  represents the pore interfaces which intersect the background mesh arbitrarily. For both mesh configurations, the mesh size is defined as $h=h_x =h_y$ with $h_{min}=0.005$ and the regularization parameter is set to $\beta =0.005$.  

\begin{figure}[h]
   \centering
    \subfloat[ \label{CutFEM}]{%
     \includegraphics[scale=0.22]{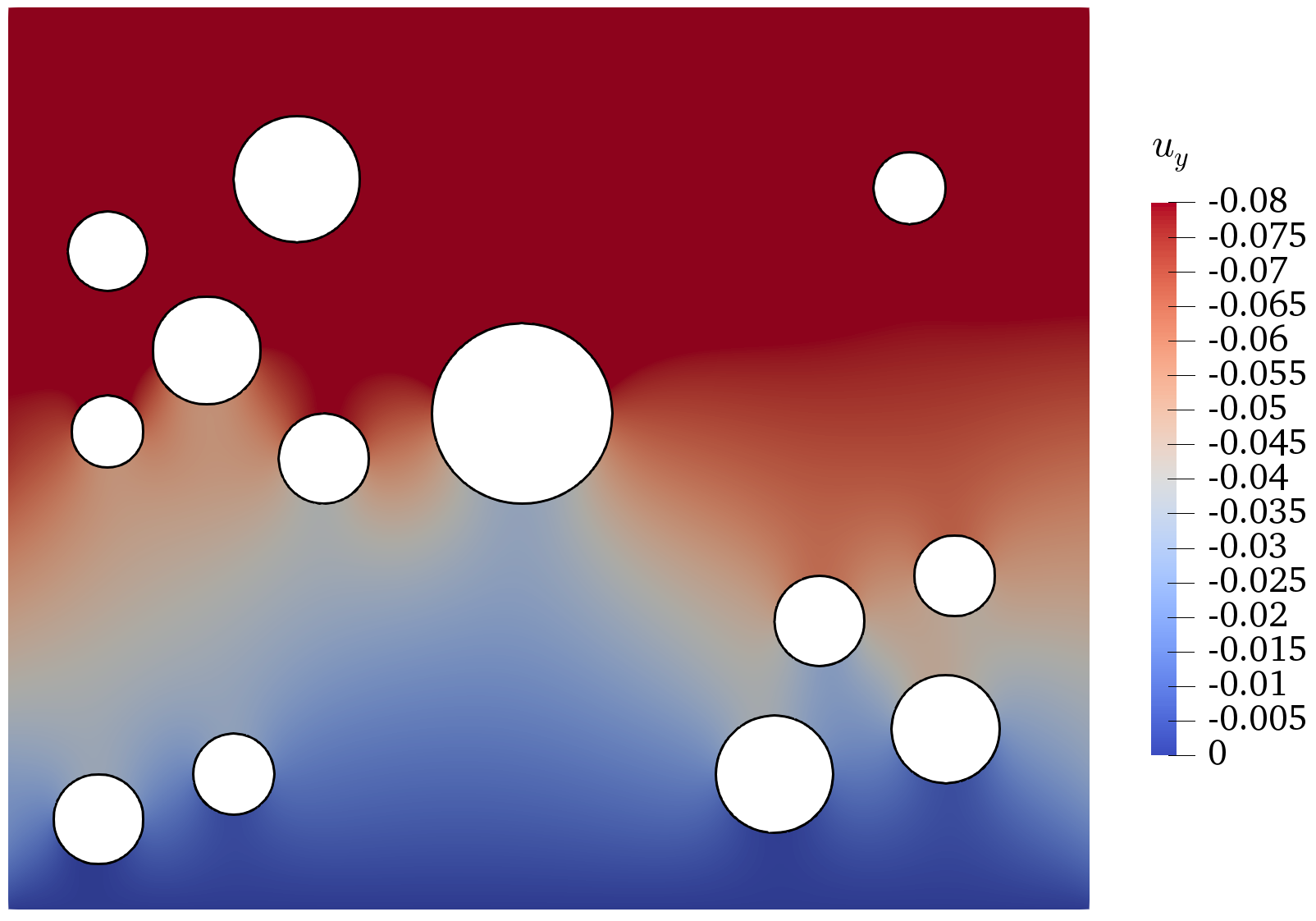}
     }
    \hfill
     \subfloat[ \label{AdaptiveCutFEM}]{%
       \includegraphics[scale=0.22]{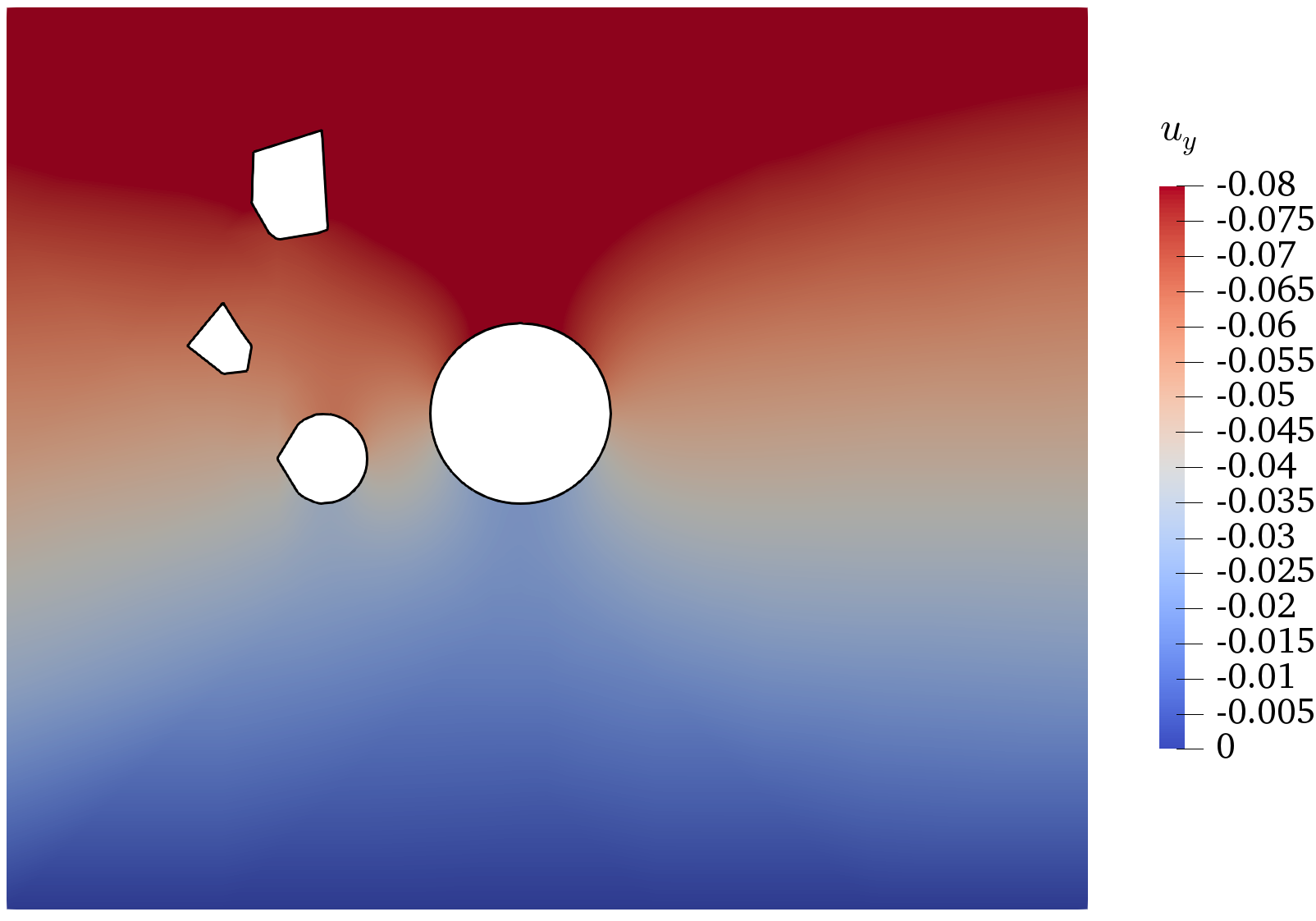}
     }
    \hfill
    % \subfloat[ \label{Mixed01}]{%
    %  \includegraphics[scale=0.22]{./Figs/Results/U_AdapBMix01.png}
    % }
   % \hfill
    %\subfloat[ \label{Mixed08}]{%
     %\includegraphics[scale=0.22]{./Figs/Results/U_AdapBMix08.png}
     %}
    %\hfill
     \caption{Displacement component $u_y$ for a) uniformly refined background mesh and b) adaptive background mesh.}
     %, c) mixed multiscale, $2\epsilon =0.2$ and d) mixed multiscale, $2\epsilon =0.8$.}
     \label{fig:ModelBAdaptiveDisps}
\end{figure}

We perform a mechanical compression test and consider the model with the uniform fine mesh as a reference. As shown in Figure \ref{fig:MeshesModelBAdaptive}b, using the signed distance function in the coarse domain leads to the random appearance of geometrical artifacts. The comparison of the displacement field component $u_y$ in Figure \ref{fig:ModelBAdaptiveDisps} shows the response in the fine mesh region of CutFEM is precise; however, in the coarse mesh region, the geometrical artifacts impose unrealistic additional stiffness. To address this limitation of the CutFEM technique for very coarse meshes, we will employ our mixed multiscale framework in Section 4.3, whereby instead of using a coarse signed distance function in the coarse domain, we replace it with a homogenized medium.

%\subsection{2D LOCALLY POROUS MEDIUM}
\subsection{Smooth mixing approach adopted for a 2D locally porous medium}
%Paragraph 1. Schematic geometry. Material Properties.\\

\begin{figure}[h]
\centering
  \includegraphics[scale=.4]{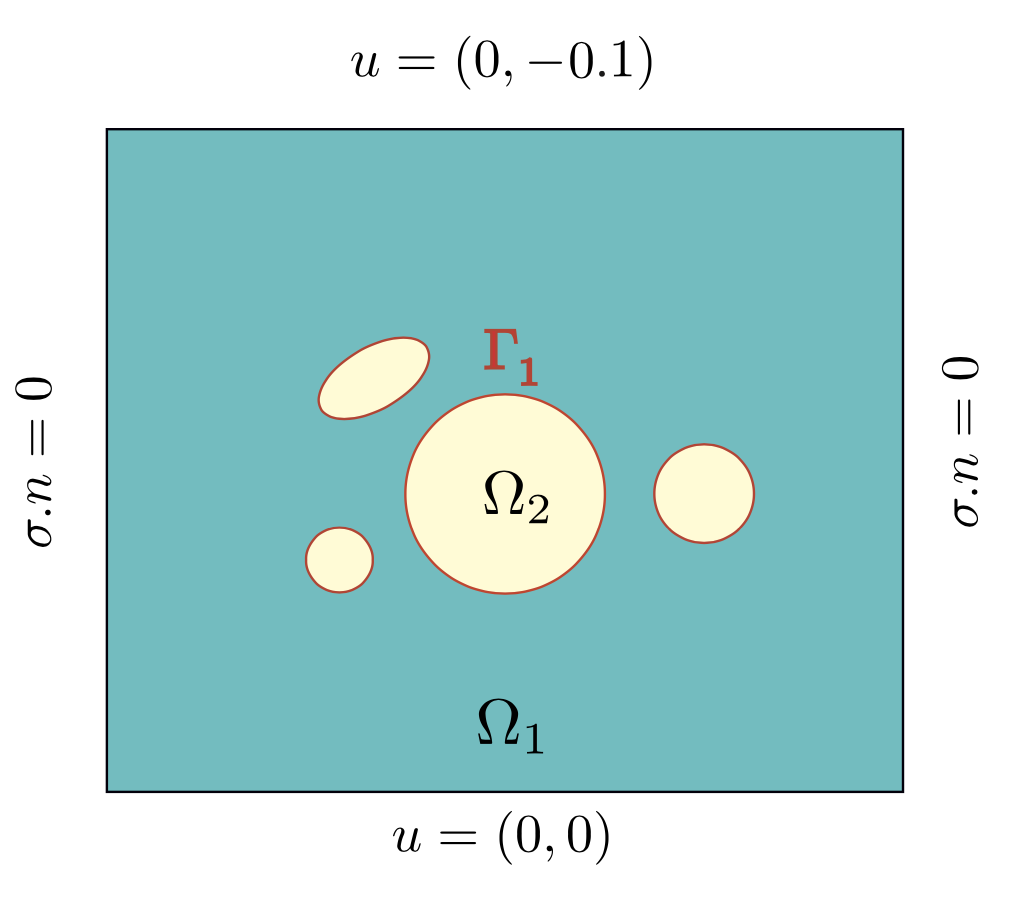}
  \caption{Schematic presentation of 2D rectangular domain with locally distributed pores.}
  \label{fig:ModelASchem}
\end{figure}

Here, we investigate the performance of the proposed smooth mixing approach in a 2D locally porous medium, shown schematically in Figure \ref{fig:ModelASchem}. This structure is a simple case for multiscale modelling, as homogenization is not essential in the coarse domain due to the local distribution of micro-pores.

We define the rectangular domain as $\Omega = [0,12]\times[0,10]$, comprised of matrix domain $\Omega_1 = \Omega \backslash \Omega_2$ and pore domain $\Omega_2$. We block the displacement at the bottom edge and insert displacement $u= (0, -0.1)$ along the top edge of the domain. Then, we set the macro and microscale mechanical properties to $E_{\mathcal{M}}=E_m=1$ and $\nu_{\mathcal{M}}=\nu_{m}=0.3$.

\begin{figure}[h]
   \centering
    \subfloat[ \label{Mesh1}]{%
     \includegraphics[scale=0.15]{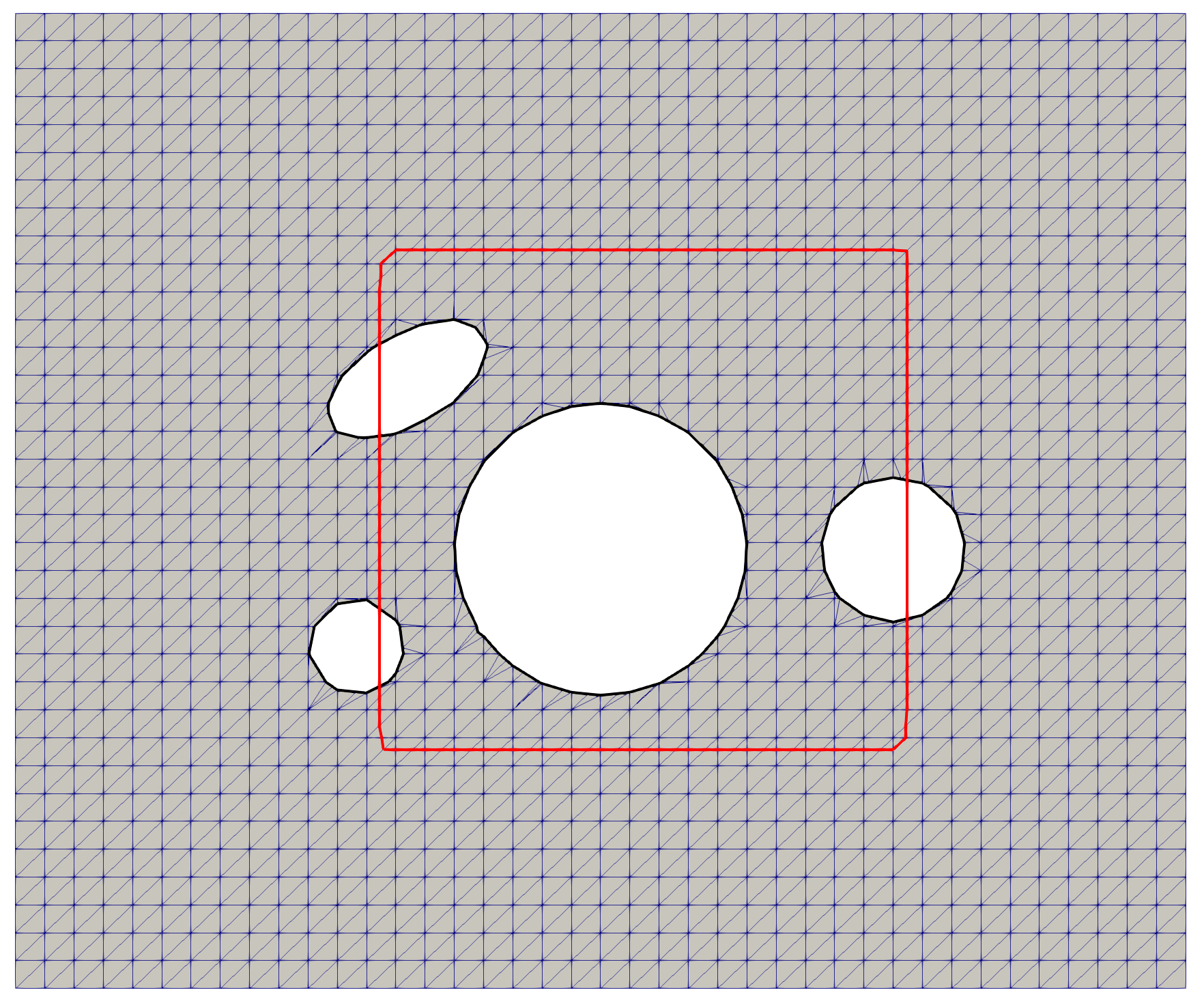}
     }
    \hfill
     \subfloat[ \label{Mesh2}]{%
       \includegraphics[scale=0.15]{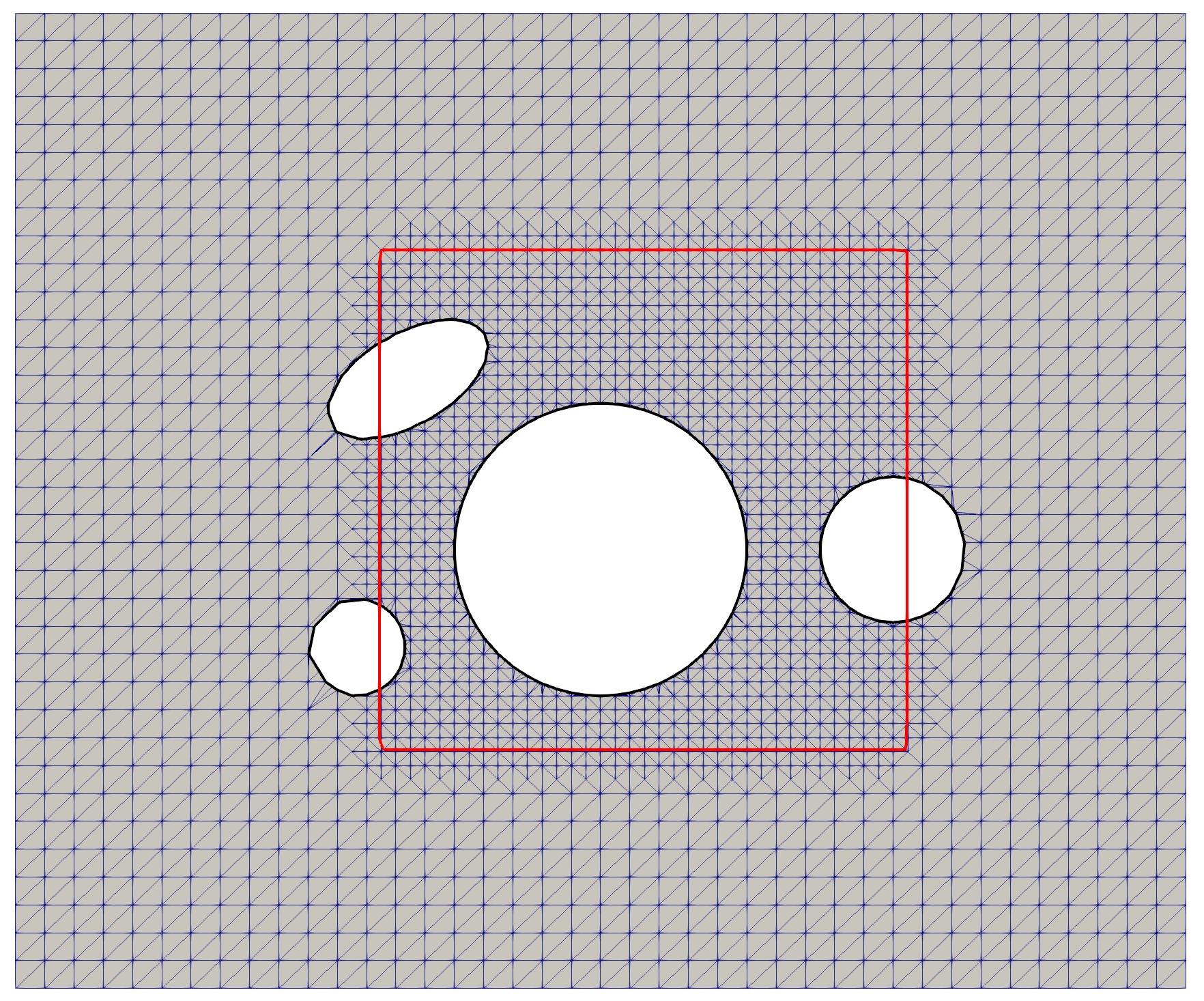}
     }
    \hfill
     \subfloat[ \label{Mesh3}]{%
      \includegraphics[scale=0.15]{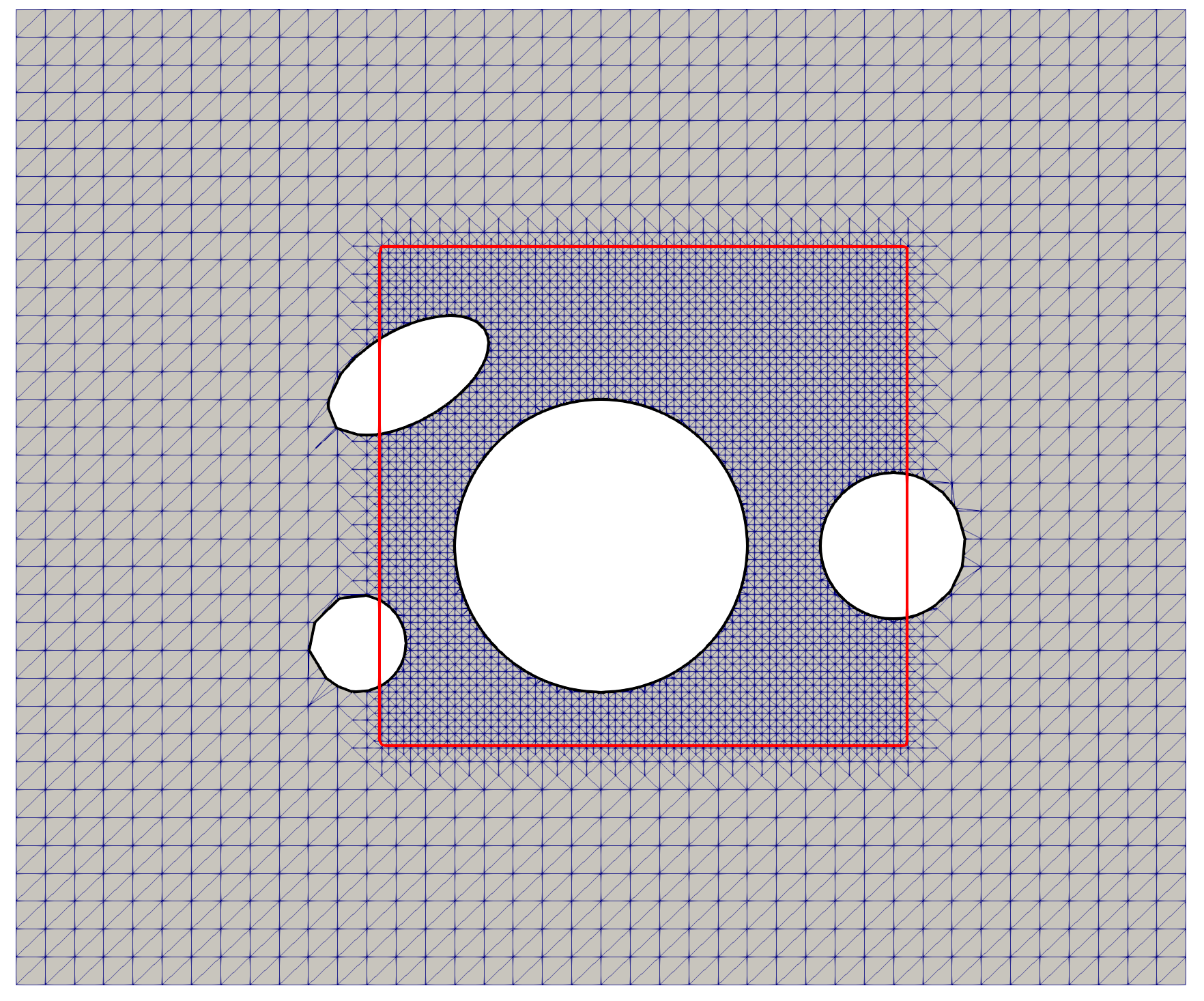}
     }
    \hfill
     \caption{Computational mesh for the physical domain of the 2D model with locally distributed pores a) uniform meshing, b) adaptive meshing type-1, c) adaptive meshing type-2.}
     \label{fig:MeshesA}
\end{figure}

We test three structured background meshes consisting of one uniform and two adaptively refined meshes generated independently of the pore and zoom interfaces. We employ linear Langrangian elements, with a uniform background mesh size $h=h_x =h_y$ and the regularization parameter set to $\beta =0.005$. The corresponding discretizations of the physical domain $\Omega_1$ are shown in Figure \ref{fig:MeshesA}. The zero level set functions of $\Gamma_1 ^h$ (shown as black lines) and $\Gamma_2 ^h$ (shown as red line) represent the micro-pore and the zooming regions, respectively. The corresponding discretized domains in Figure \ref{fig:MeshesA} show the arbitrary intersection of the interfaces with the elements, where the zooming interface determines the middle of the transition region $\Omega_T$ and the mesh is refined inside the zoom.

In this study, we choose the following smooth weight function to mix the two models inside transition region $\Omega^T$,

\begin{equation}
    \mathcal{S}=\frac{1}{2}(1+\sin(\frac{\pi}{2\epsilon}\xi(x))) .
    \label{Eq:Weight}
\end{equation}

\begin{figure}[h]
   \centering
    \subfloat[ \label{Mix01w}]{%
     \includegraphics[scale=0.17]{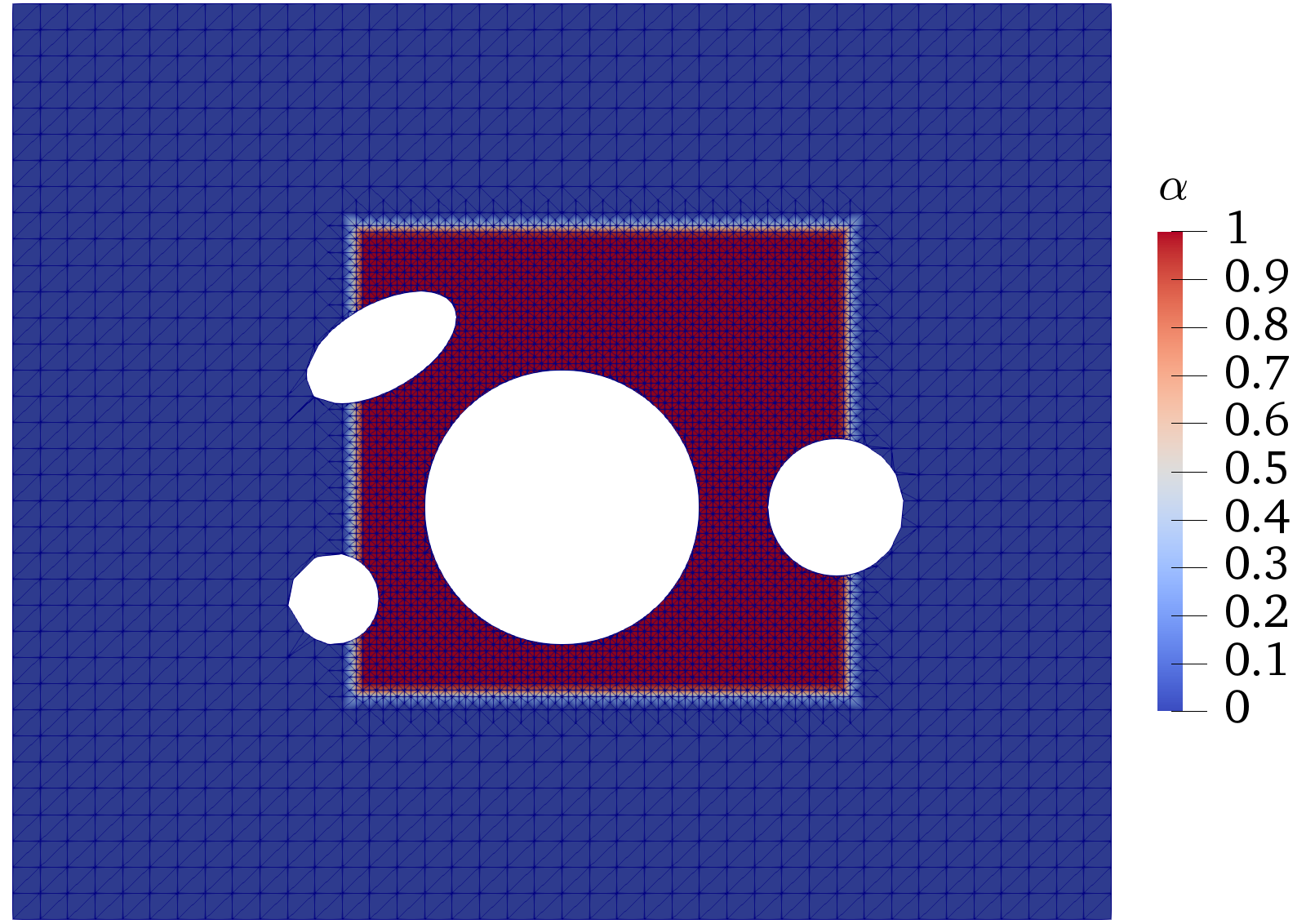}
     }
    \hfill
     \subfloat[ \label{Mix04w}]{%
           \includegraphics[scale=0.17]{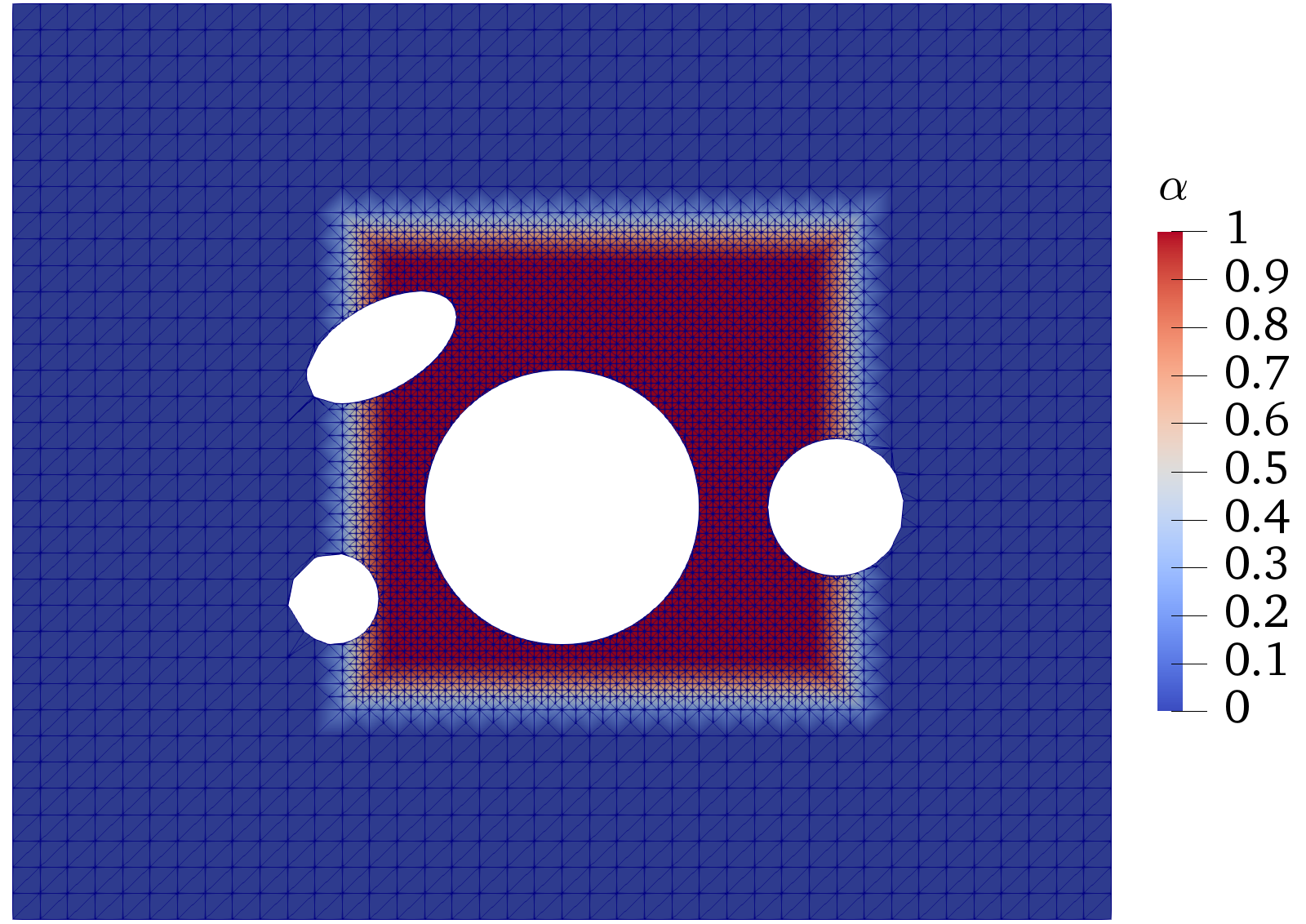}
     }
    \hfill
     \subfloat[ \label{Mix1w}]{%
      \includegraphics[scale=0.17]{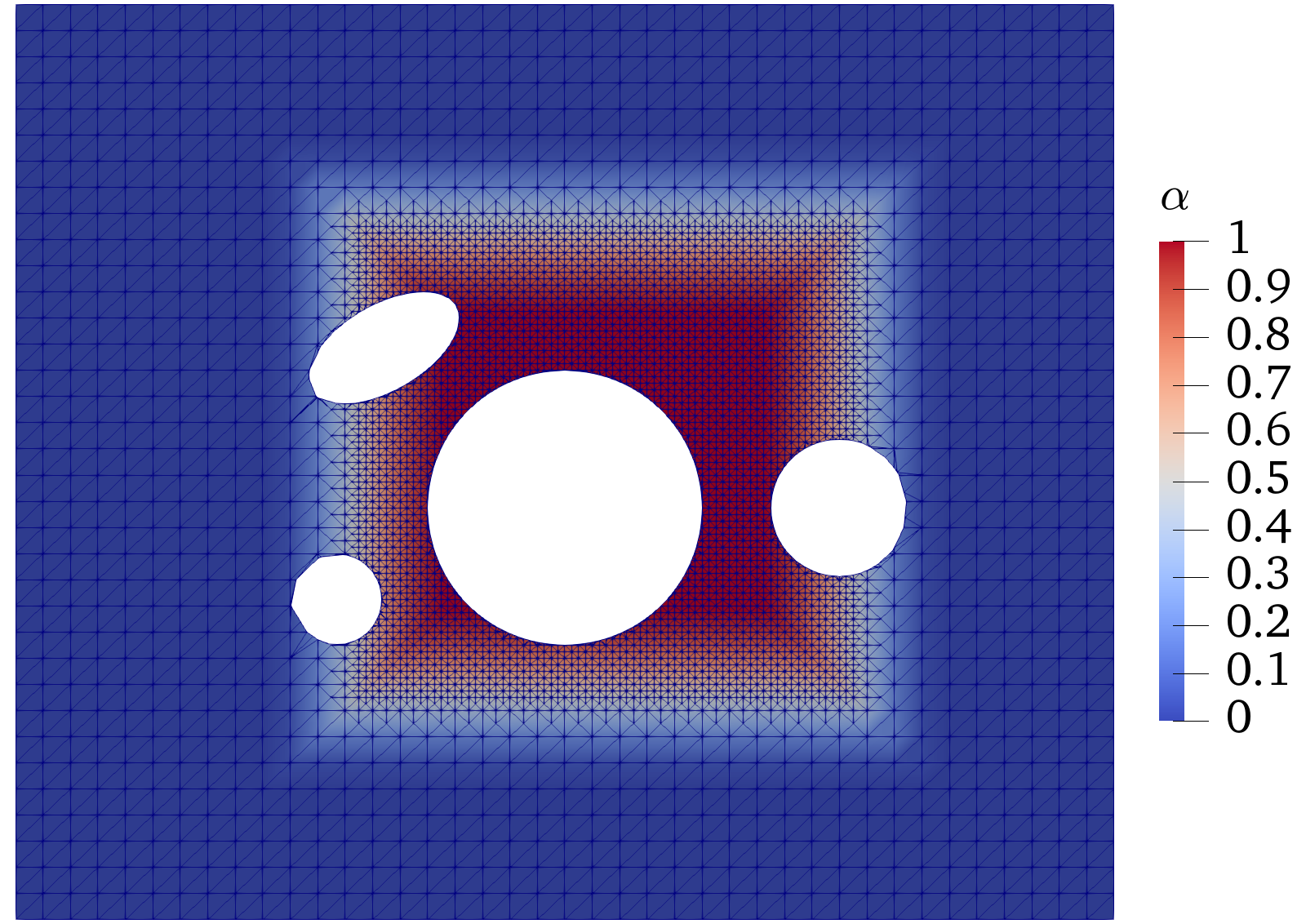}
     }
    \hfill
     \caption{Smoothing weight function $\alpha$ contour over finest adaptive mesh, a) $\epsilon=0.1$, b) $\epsilon=0.4$ and c) $\epsilon=1$.}
     \label{fig:IdicA}
\end{figure}

\begin{figure}[h]
   \centering
    \subfloat[ \label{FEMd}]{%
     \includegraphics[scale=0.21]{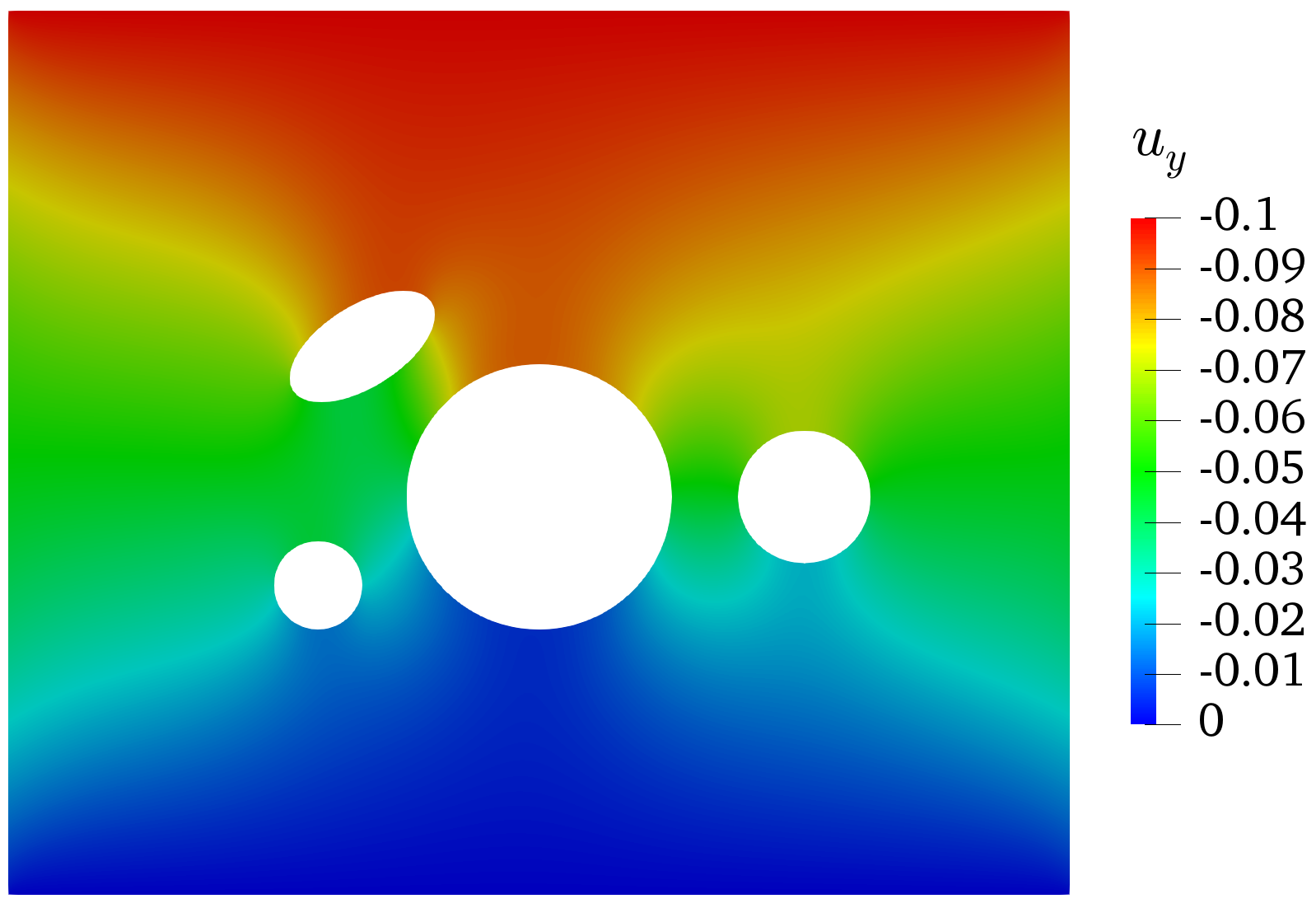}
     }
    \hfill
     \subfloat[ \label{CutFEMd}]{%
       \includegraphics[scale=0.21]{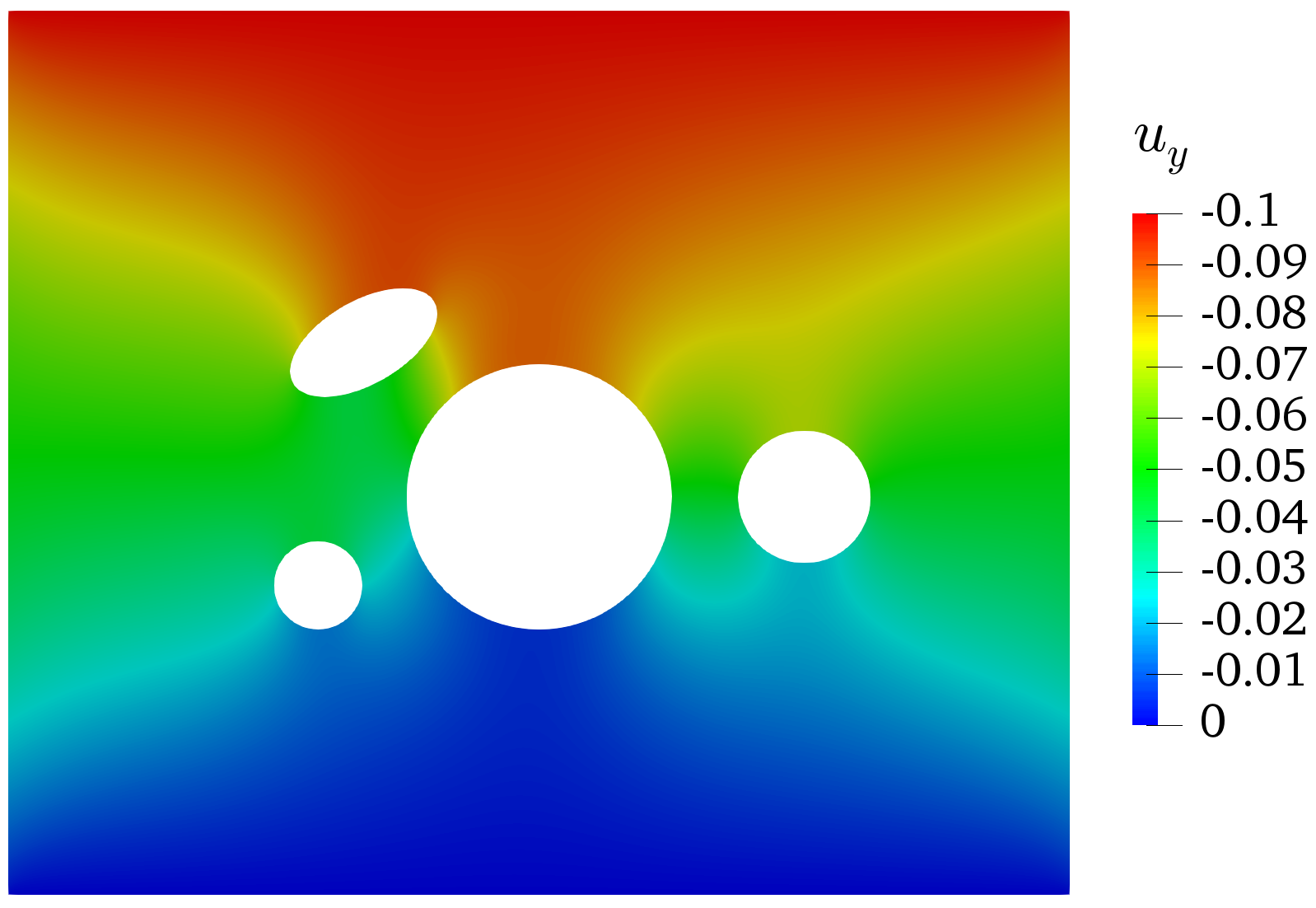}
     }
    \hfill
     \subfloat[ \label{Mix01d}]{%
      \includegraphics[scale=0.21]{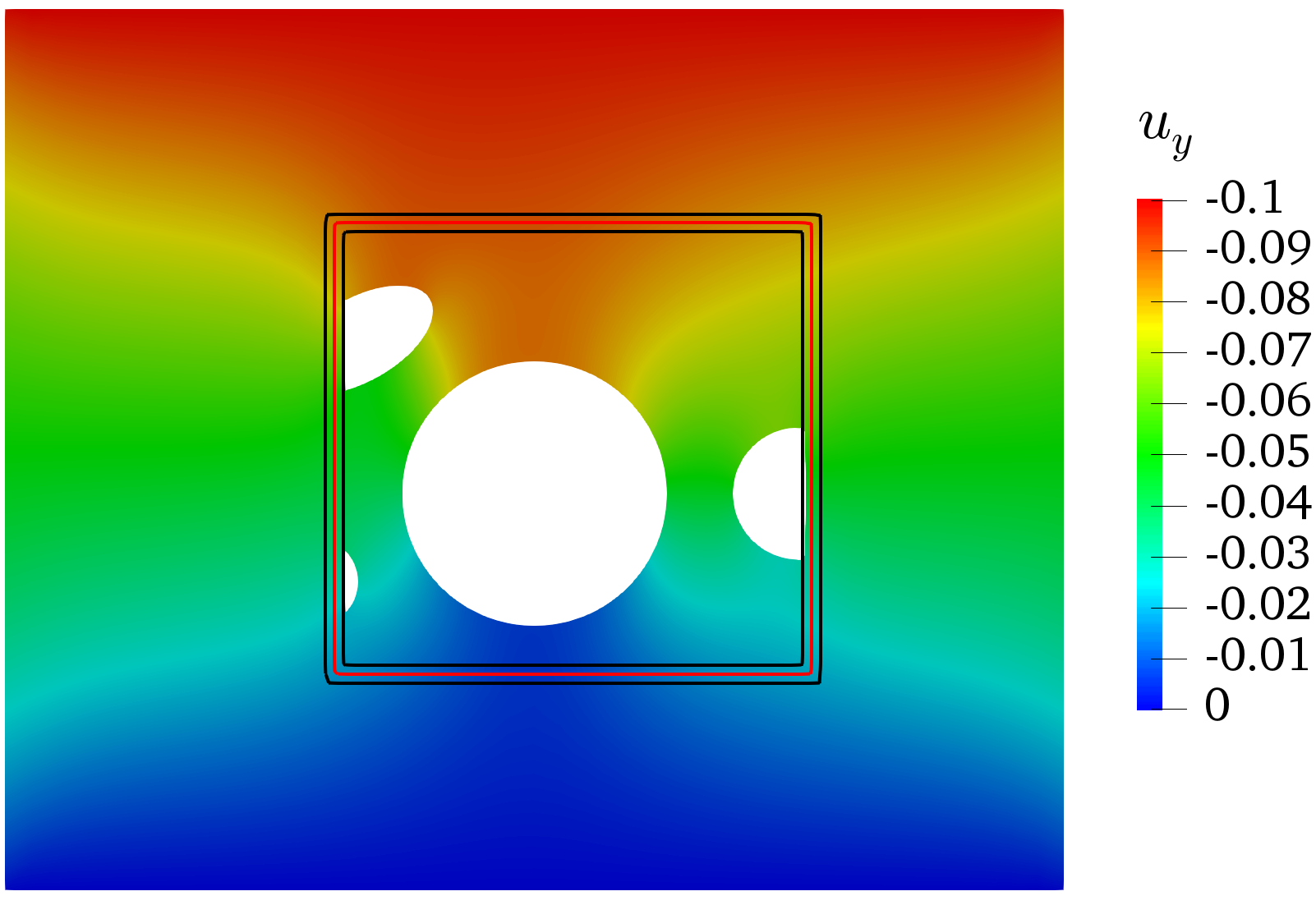}
     }
    \hfill
     \subfloat[ \label{Mix1d}]{%
      \includegraphics[scale=0.21]{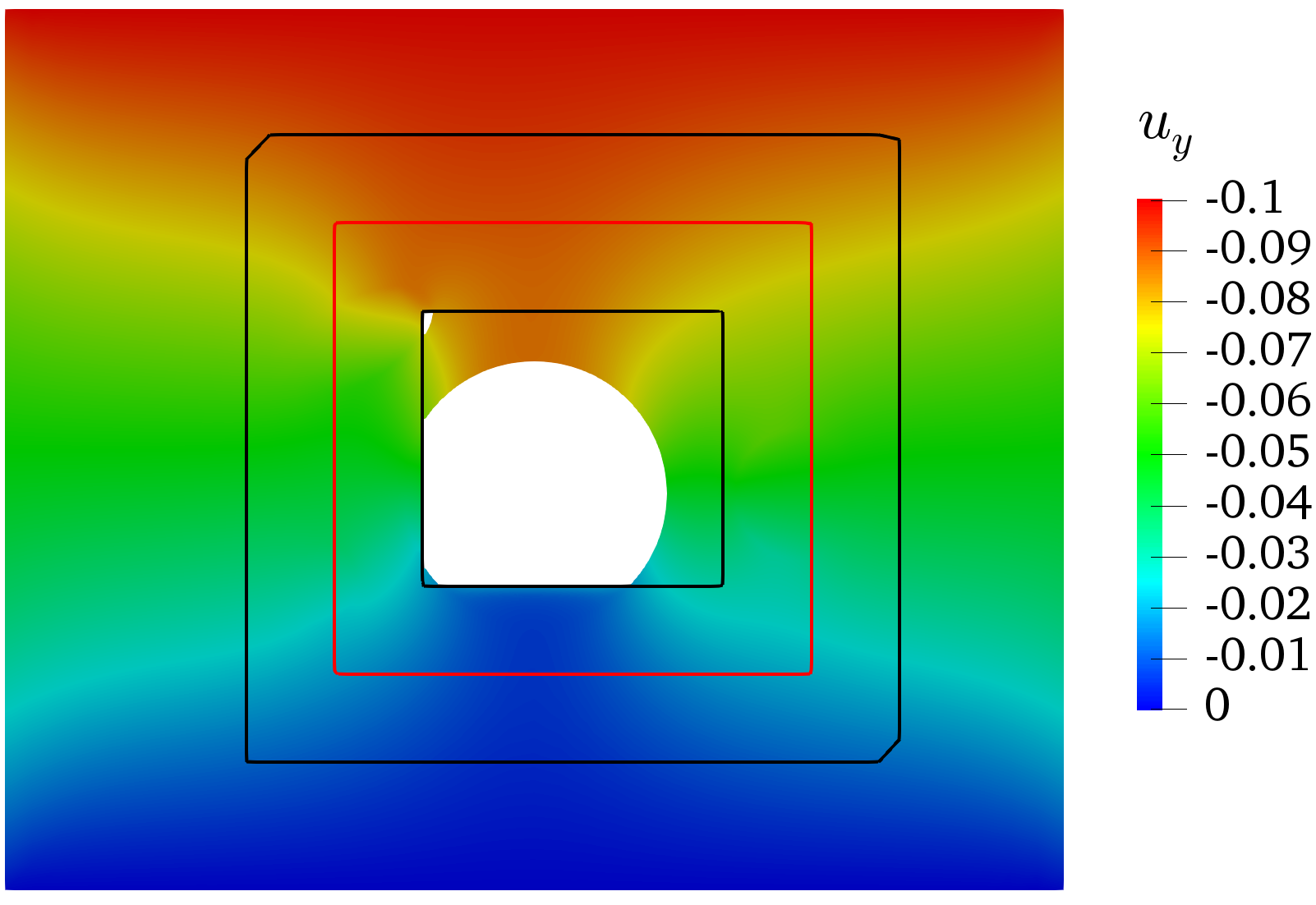}
     }
    \hfill
     \caption{Displacement component $u_y$ contours for different methods: a) FEM, b) CutFEM, c) mixed multiscale model with $2\epsilon=0.1$, d) mixed multiscale model with $2\epsilon=1$.}
     \label{fig:ModelAdisp}
\end{figure}

Figure \ref{fig:IdicA} illustrates how the scalar function $\alpha_h$ is distributed in the discretized physical domain with different mixing lengths. Note that our multiscale mixing approach operates over a single mesh, and its mixing length is defined in a mesh-independent manner. 
The displacement field component $u_y$ for two smooth mixing lengths $2\epsilon =0.1, 1$ and the finest adaptive mesh with $h_{min} = 0.2$ are shown in Figure~\ref{fig:ModelAdisp}c, d. We choose standard FEM and unfitted CutFEM as reference models and present the corresponding $u_y$ in Figure \ref{fig:ModelAdisp}a, b. As expected, we find that our CutFEM displacement field converges to the FEM displacement field, verifying our single-scale unfitted method. For the mixed multiscale model, $u_y$ inside the zoom is similar to the corresponding references and exhibits smooth behaviour in the transition domain $\Omega_T$.
%Paragraph 4. Stress and Strain fields. Fictitious domains. Physical domains\\

The energy distribution inside $\Omega_T$ is the average of the FEM macroscale and the CutFEM microscale model. Next, we will investigate how the mixing approach via the weight function (\ref{Eq:Weight}) impacts the stress field in the physical and the fictitious domains. The stress field is given by
\begin{equation}
    \sigma_{\text{mix}} (x) = \Bigg\{ \begin{array}{ll}
         \sigma_m \qquad \qquad \qquad \qquad \ \text{in} \ \ \Omega_m \backslash \Omega_T  , \\
         \sigma_{\mathcal{M}}  \qquad \qquad \qquad  \qquad  \text{in} \ \ \Omega_{\mathcal{M}} \backslash \Omega_T,   \\
        (1-\alpha_h) \sigma_m  + \alpha_{h} \sigma_{\mathcal{M}}  \ \ \ \ \ \ \ \text{in} \ \  \Omega_T . 
    \end{array}
    \label{MixStress}
\end{equation}

\begin{figure}[!h]
   \centering
    \subfloat[ \label{FEMs}]{%
     \includegraphics[scale=0.2]{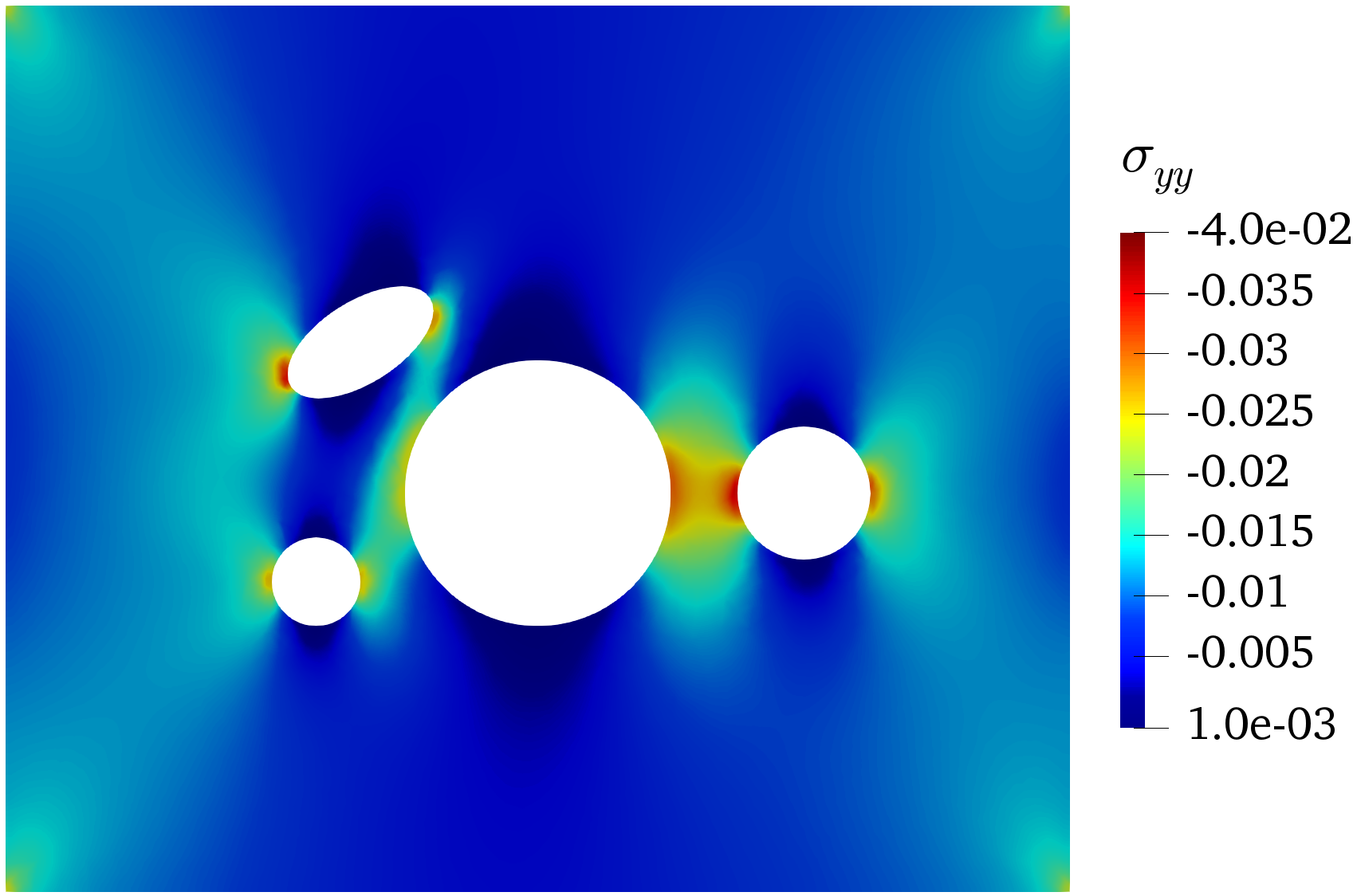}
     }
    \hfill
     \subfloat[ \label{CutFEMs}]{%
       \includegraphics[scale=0.21]{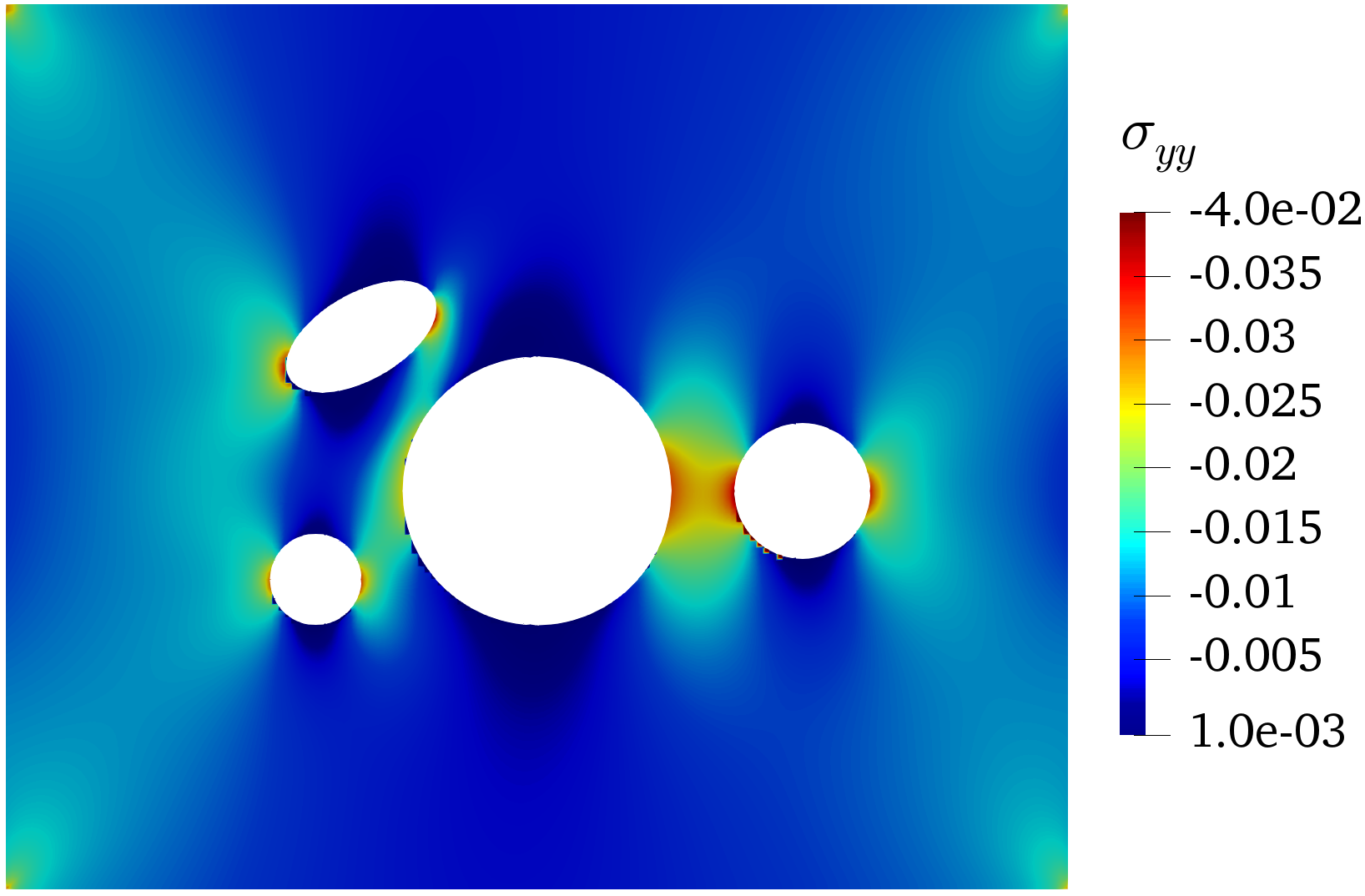}
     }
    \hfill
     \subfloat[ \label{Mix01s}]{%
      \includegraphics[scale=0.2]{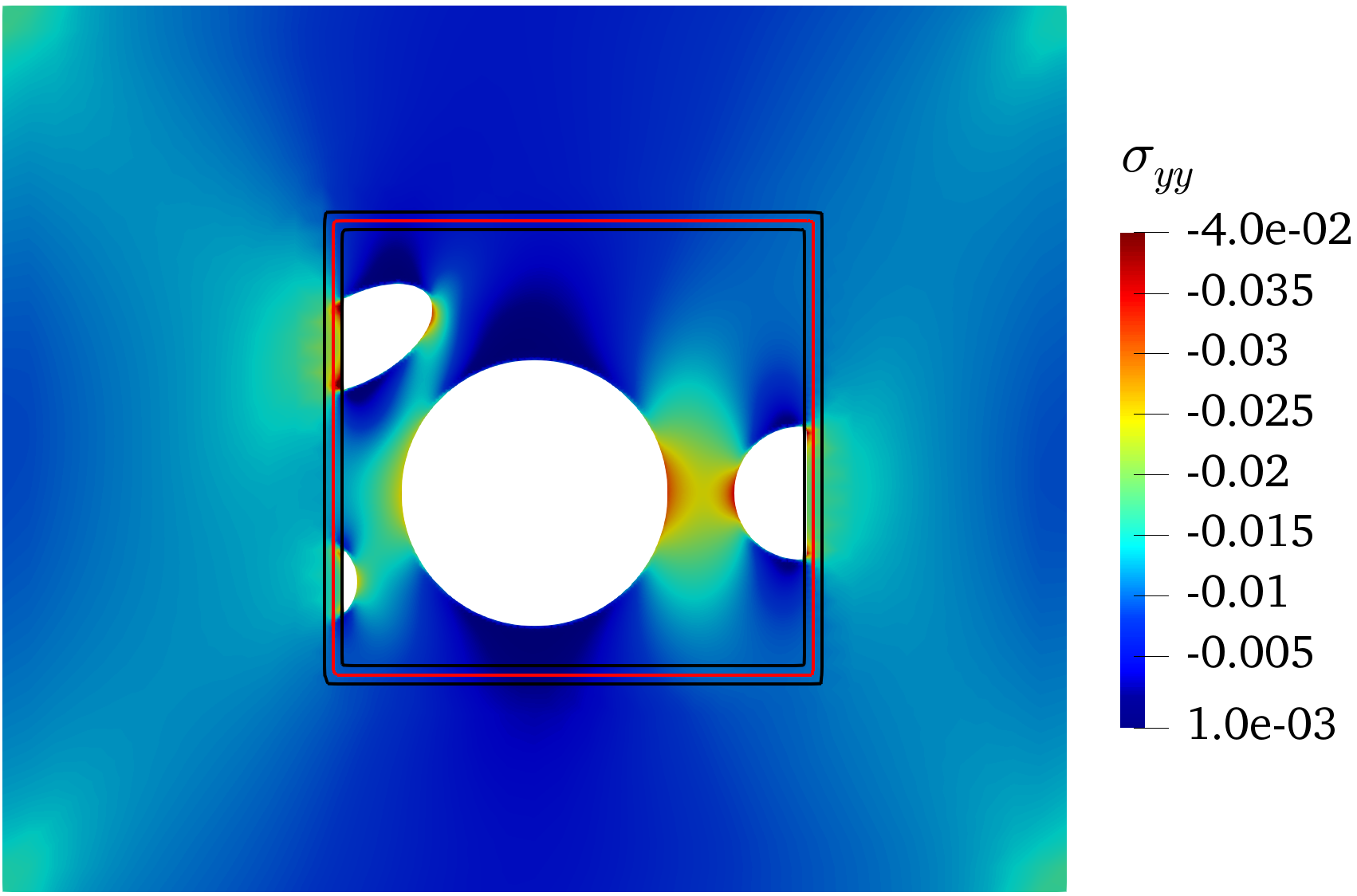}
     }
    \hfill
         \subfloat[ \label{Mix1s}]{%
      \includegraphics[scale=0.2]{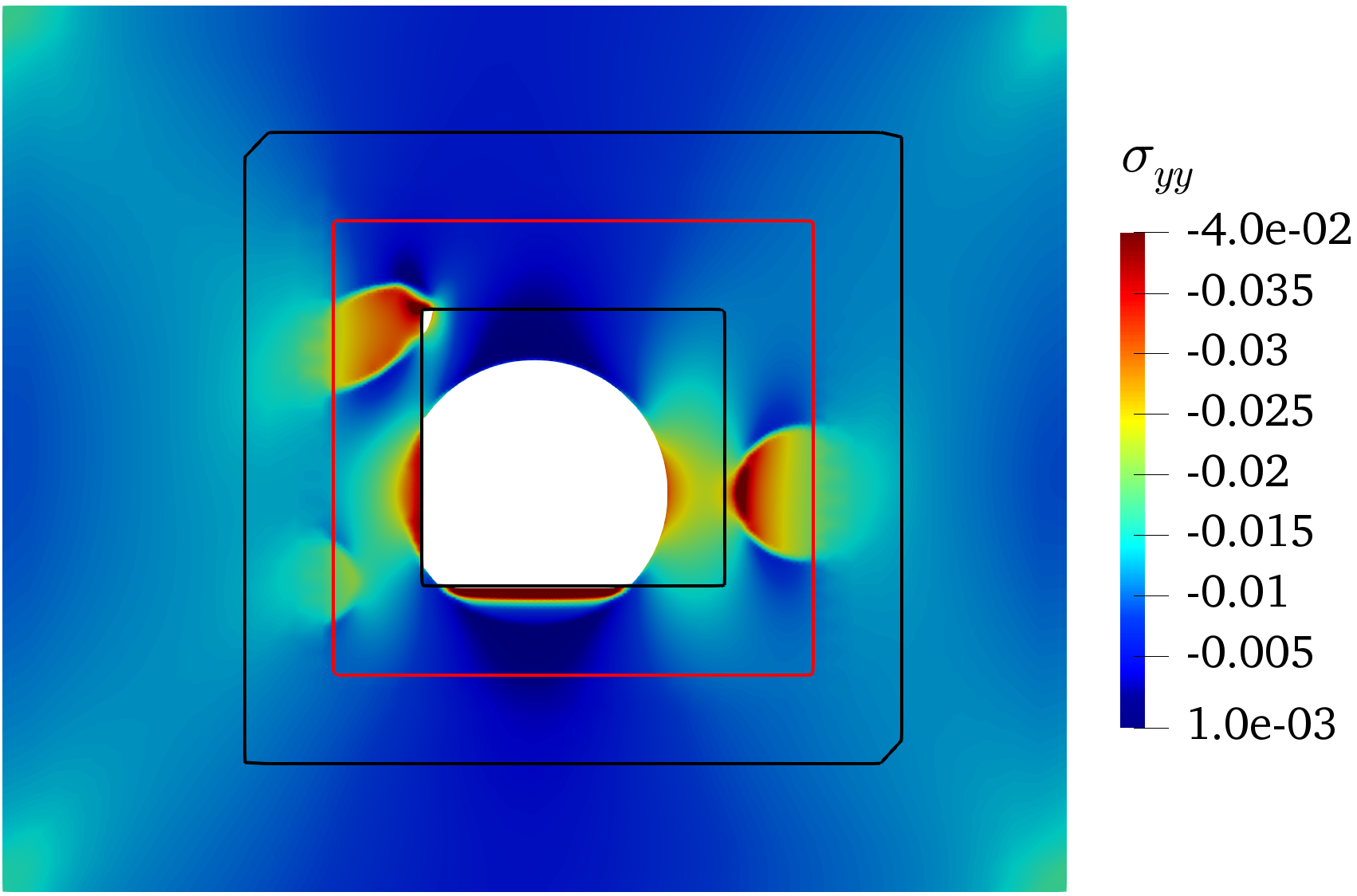}
     }
    \hfill
         \subfloat[ \label{Mix01Fict}]{%
      \includegraphics[scale=0.2]{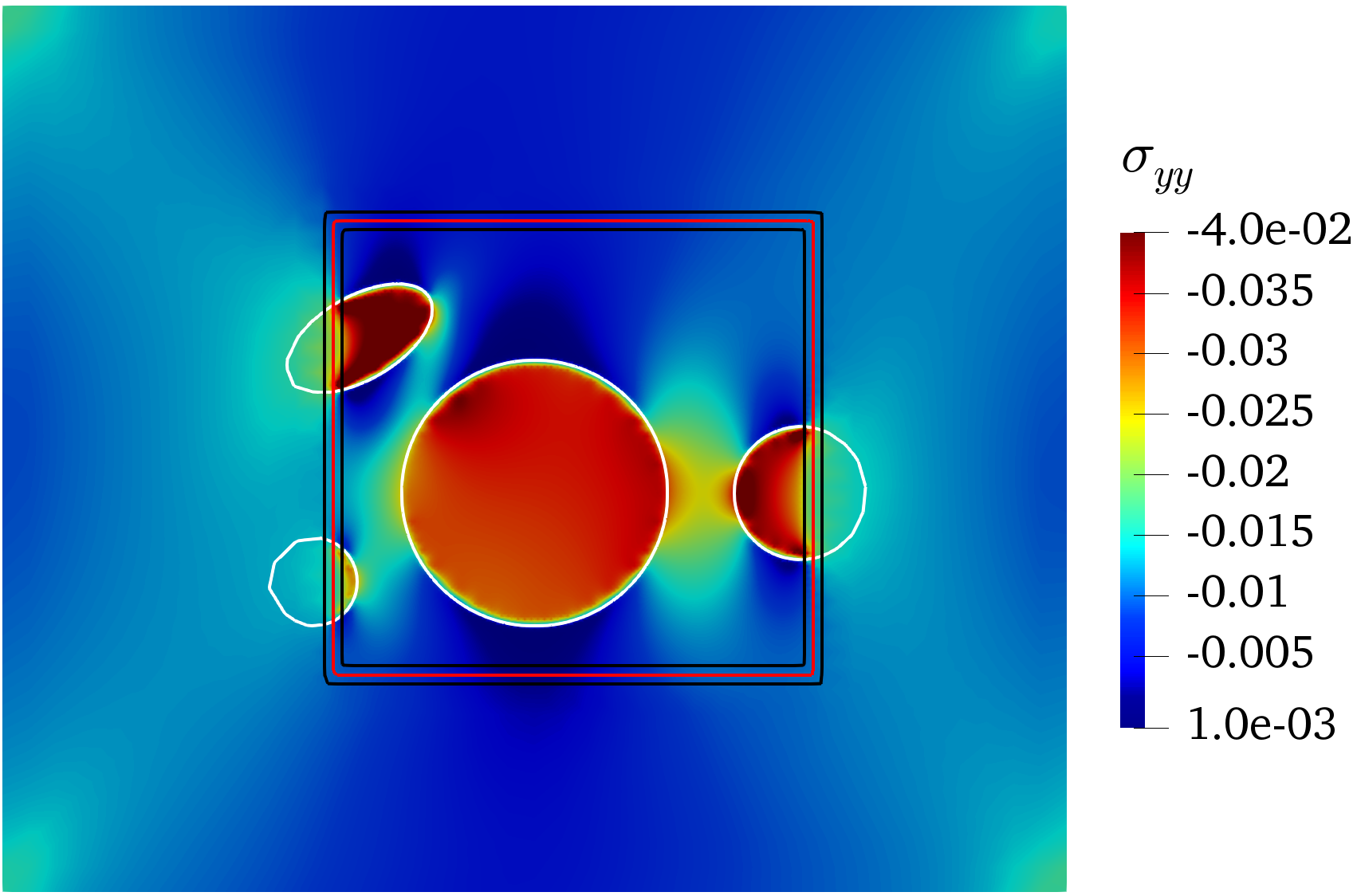}
     }
    \hfill
         \subfloat[ \label{Mix1Fict}]{%
      \includegraphics[scale=0.2]{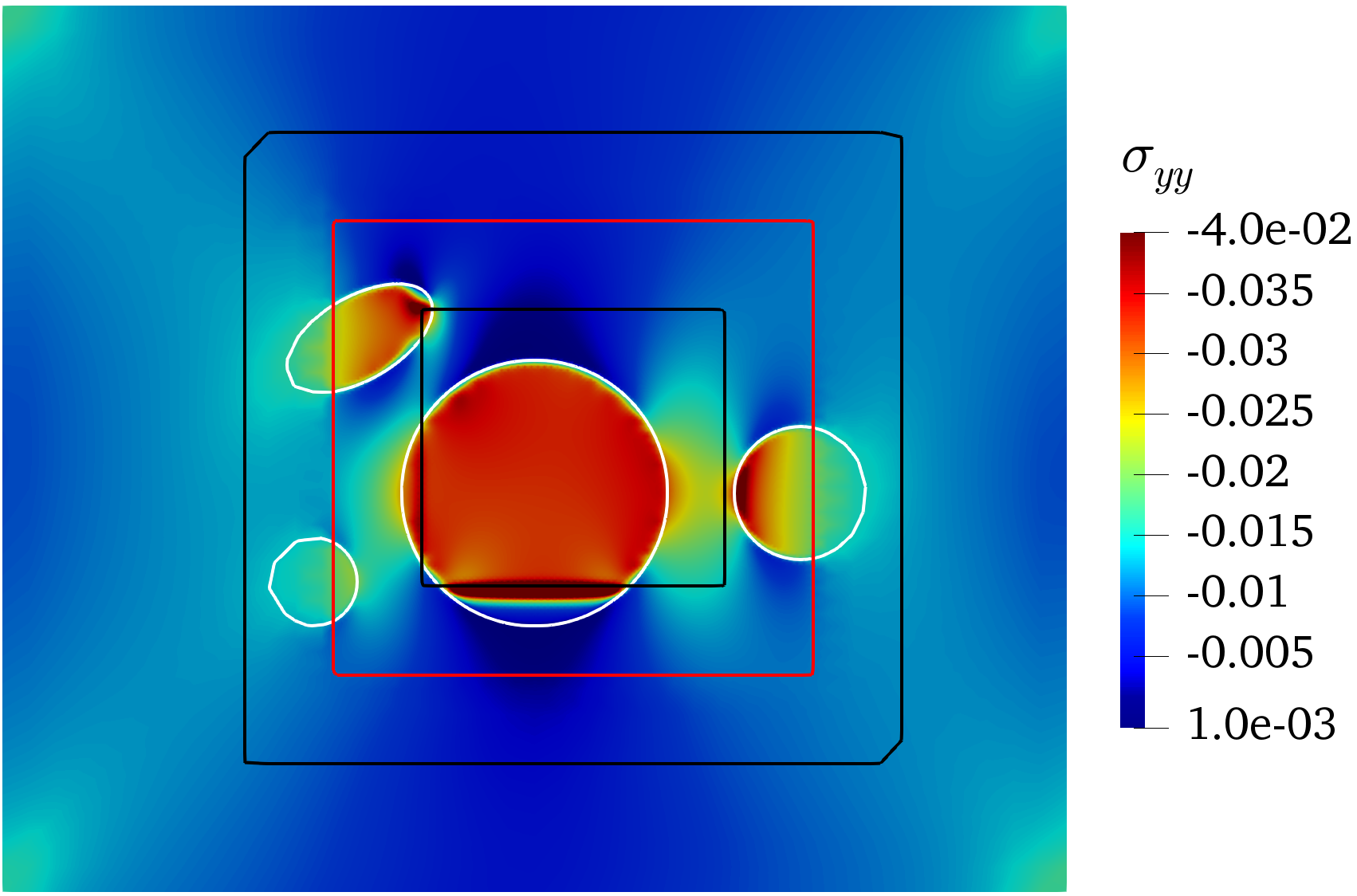}
     }
    \hfill
     \caption{Stress component $\sigma_{yy}$ contours, a) FEM model, b) CutFEM model, c) mixed multiscale model in physical domain with $2\epsilon=0.1$, d) mixed multiscale model in physical domain with $2\epsilon=1$, e) mixed multiscale model in fictitious domain with $2\epsilon=0.1$, f) mixed multiscale model in fictitious domain with $2\epsilon=1$.}
     \label{fig:ModelASig}
\end{figure}

As shown in Figure \ref{fig:ModelASig}a,b, the stress component $\sigma_{yy}$ in CutFEM converges to its FEM counterpart. We compute $\sigma_{\text{mix}}$ given in (\ref{MixStress}) for two smoothing lengths over the physical and fictitious domains in Figure \ref{fig:ModelASig}c-\ref{fig:ModelASig}f. Our results show that $\sigma_{\text{mix}}$ in $\Omega_T$ is smooth and without oscillations.

\begin{figure}[h]
   \centering
    \subfloat[ \label{GPCut}]{%
      \includegraphics[scale=0.26]{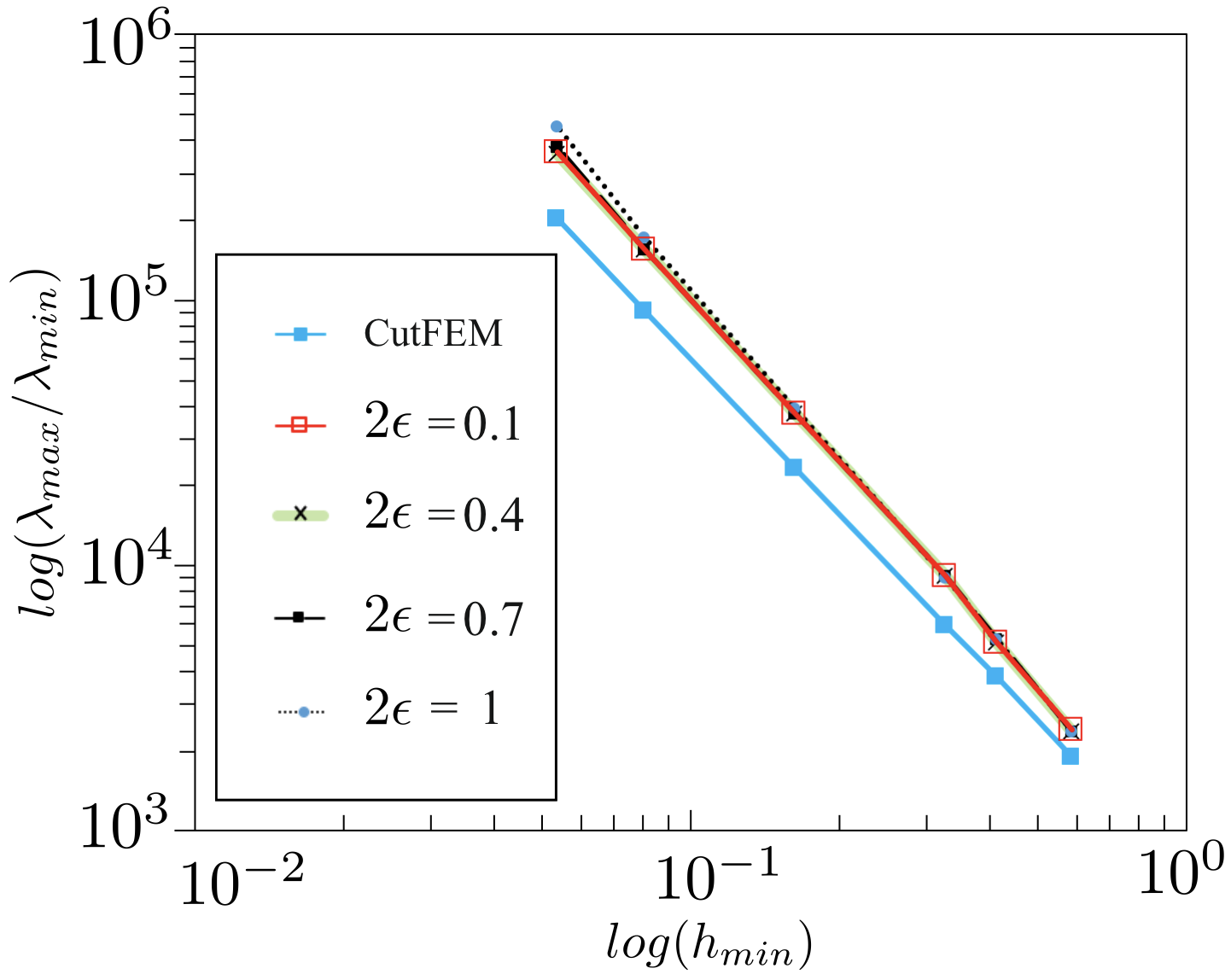}
     }
    \hfill
         \subfloat[ \label{GPeverywhere}]{%
      \includegraphics[scale=0.26]{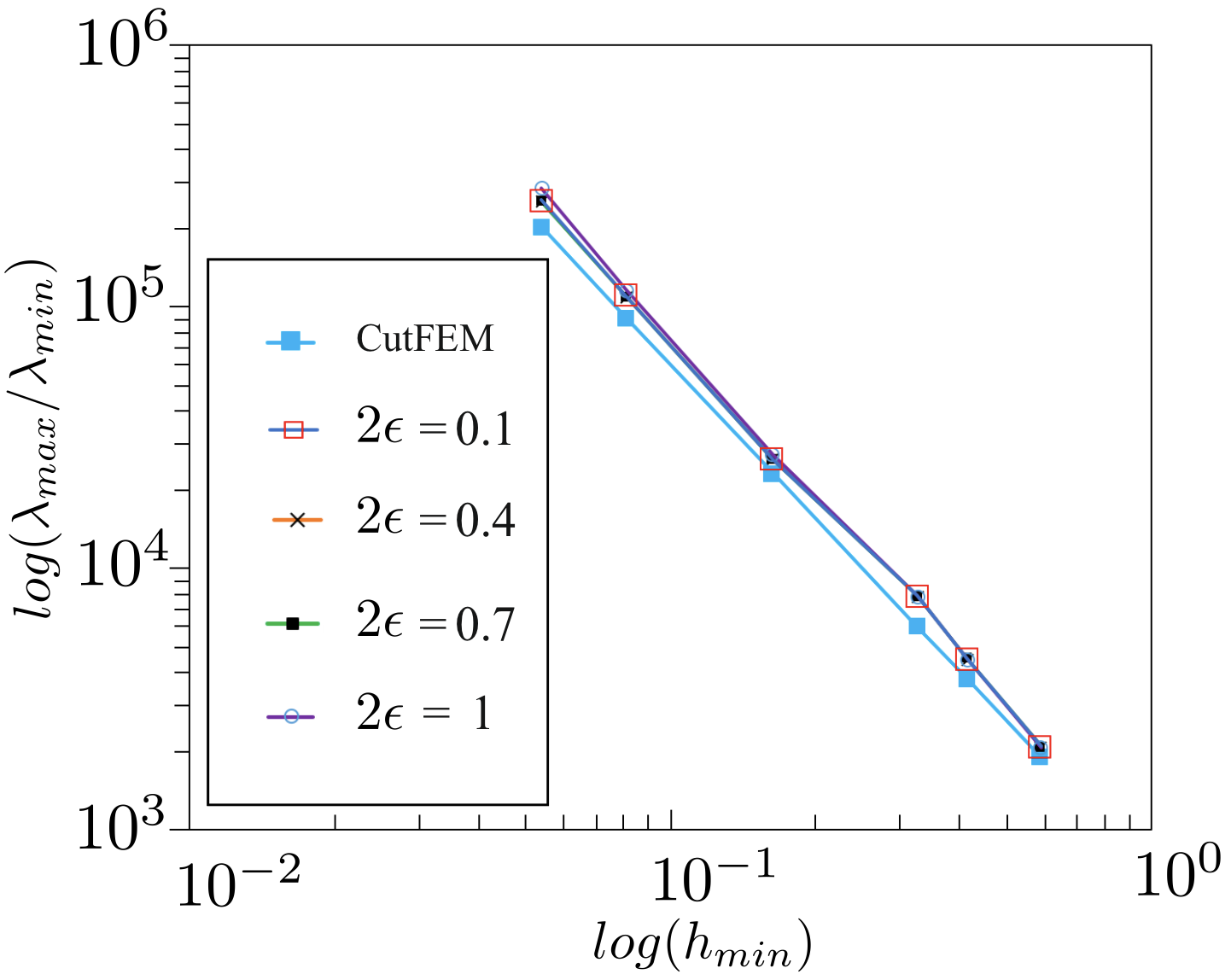}
     }
    \hfill
  \caption{The condition number of the system matrix versus mesh size, for different mixing lengths: a) ghost penalty regularization is applied to cut elements only,  b) ghost penalty regularization is applied to every element inside the porous domain in addition to cut elements. In both cases, the regularization parameter is chosen as $\beta =0.005$.}
  \label{fig:ModelACond}
\end{figure}

To enhance the stability of our multiscale framework, in the microscale model, we regularize the elements inside the porous domain in addition to the intersected elements by $\Gamma^h _1$. Then we compute the condition number of the multiscale system matrix to investigate the stability by using SLEPc \cite{Hernandez05} which finds the ratio of the maximum to the minimum eigenvalue of the system matrix (i.e. $\lambda_{max}/ \lambda_{min}$). We use a sequence of uniform and adaptive meshes with different mixing lengths and then compare them with the CutFEM reference model. In Figure \ref{fig:ModelACond}a, we find that the behaviour of our mixed approach with different mixing lengths is well conditioned and similar to the standard CutFEM approach. In Figure \ref{fig:ModelACond}b, we investigate the impact of extending the ghost-penalty regularization to the inside of the pores (in addition to the cut elements by the pore interfaces) on the condition number of the multiscale system matrix. As expected, this technique improves the condition number effectively. Also, we find that the corresponding behaviour with respect to mesh refinement for our mixed approach is proportional to $h^{-2}$ for both regularization approaches, which are very close to the "pure" CutFEM results.

\clearpage

%---------------------------- Second 2D model -------------------------------------------------
%\subsection{2D mixed multiscale problem with two zooms}
\subsection{The mixed multiscale method for a 2D quasi-uniform porous medium}
In this Section, we consider the quasi-uniform porous domain given in Figure \ref{fig:ModelQuasiSchm} for our mixed multiscale analysis. As discussed in Section 4.1, structures with uniform heterogeneity require homogenization in the coarse domain to avoid geometrical artifacts which yield unrealistic stiffness  and stress singularities. Hence, here, we replace the signed distance function in the coarse domain with a homogenized domain and use the smooth mixing approach to couple the fine and coarse-scale domains. 

In our mixed multiscale framework, we construct the homogenized model by using the Modified Mori Tanaka (MMT) approach \cite{IMANI201816489} to reproduce the effects of micropores in the homogenized macro model. Employing the MMT homogenization approach for $\Omega_\mathcal{M}$ with $\textit{n}$ circular pores of different radii, the effective Young's modulus will be computed as follows.

\begin{equation}
    E_\mathcal{M} ^i = (1-\bar{\phi}_i)E_\mathcal{M} ^{i-1}(\bar{\phi}_i L_i+(1-\bar{\phi}_i)I)^{-1}, \ \ i=1,...,n,  
    \label{Eq.MMT}
\end{equation}
where $E_\mathcal{M} ^i$ and $E_\mathcal{M} ^{i-1}$ are the homogenized Young's modulus with inclusion of $i^{th}$ and $({i-1})^{th}$ circular pores, respectively, and $\bar{\phi}_i$ is the instantaneous porosity parameter defined as

\begin{equation}
    \bar{\phi}_i = V_v ^i /V_t , 
\end{equation}
where $V_v ^i$ is the void volume with $\textit{i}$ number of pores and $V_t$ is the total volume. $L_i$ is the Eshelby parameter given for circular inclusions in \cite{Mura87}. To calculate the effective elastic modulus of a domain with $\textit{n}$ pores, we add the inclusions one by one, and in each step number $\textit{i}$; we update Equation~\ref{Eq.MMT}. For more details regarding the MMT approach, see \cite{IMANI201816489}.

\subsubsection{The mixed multiscale method with one arbitrary zoom}
\begin{figure}[h]
\centering
  \includegraphics[scale=.32]{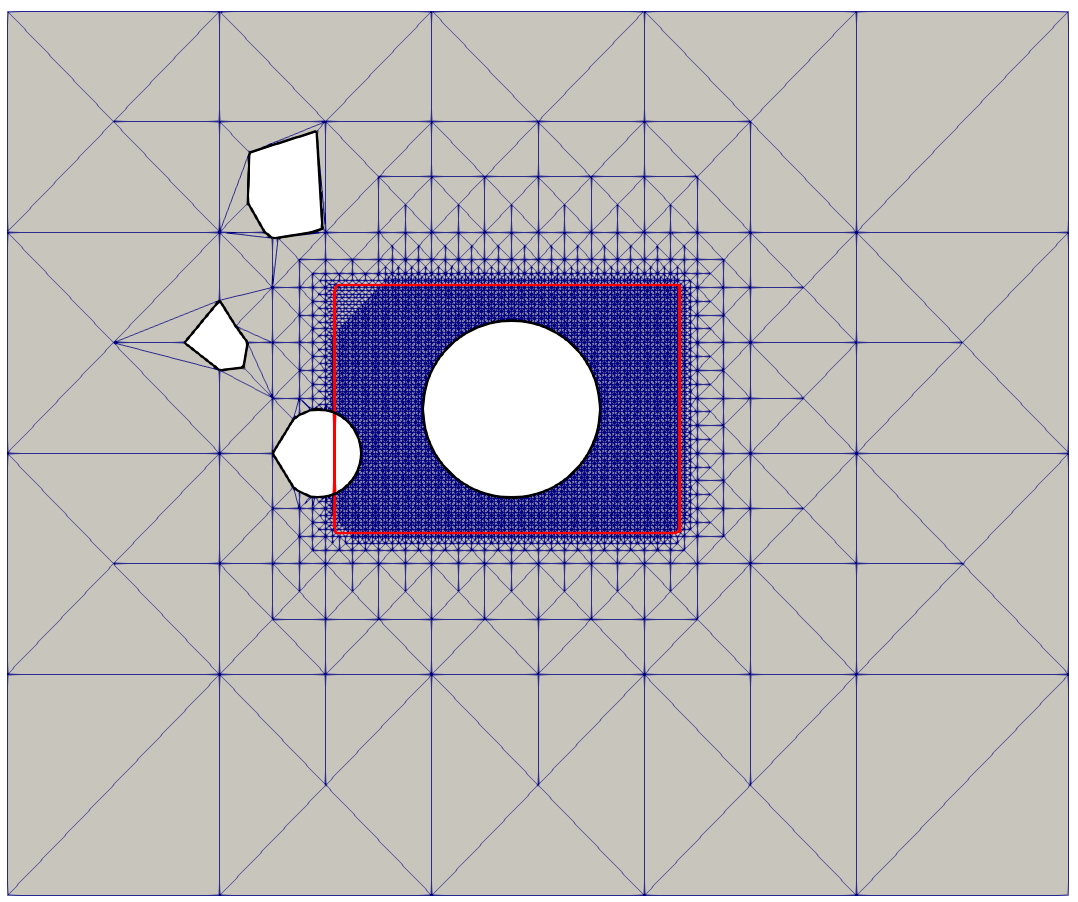}
  \caption{Background mesh with projected pores and zooming interfaces of the mixed multiscale method.}
  \label{fig:MeshZoomAdapt1}
\end{figure}

Here, we use the mixed multiscale framework with one zoom and compare it with the equivalent adaptive CutFEM approach discussed in Section 4.1. The zooming interface is projected over the background mesh and shown with a red line in Figure \ref{fig:MeshZoomAdapt1}. The material properties and boundary conditions are the same as in the adaptive CutFEM model from Section 4.1. To compute the homogenized material properties, we consider the pores in the entire domain $\Omega$. We use Equation \ref{Eq.MMT} for this purpose and then calculate the corresponding effective Young's modulus as $E_{\mathcal{M}}=0.78$.

\begin{figure}[h]
   \centering
     \subfloat[ \label{Mixed01}]{%
      \includegraphics[scale=0.22]{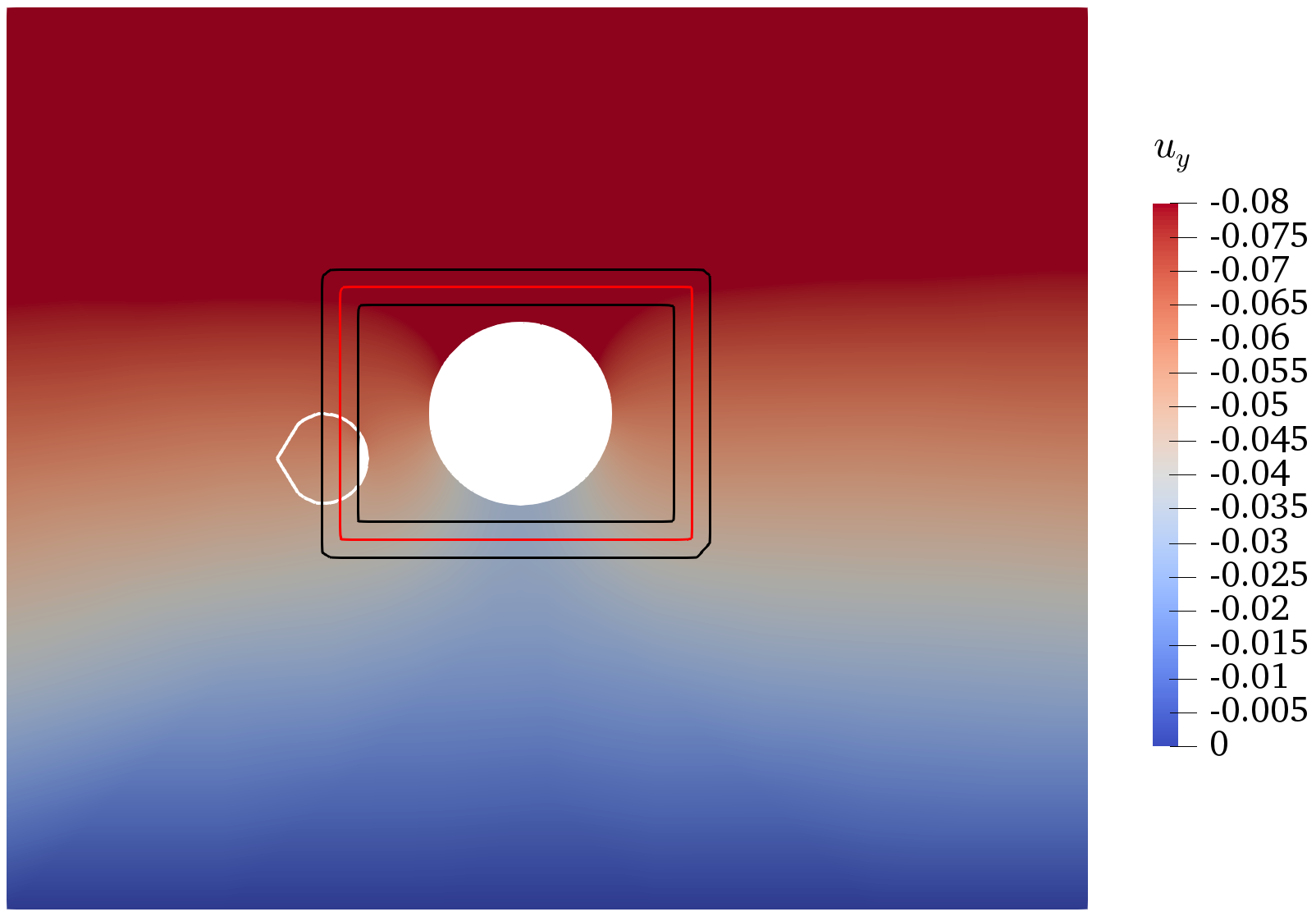}
     }
    \hfill
    \subfloat[ \label{Mixed08}]{%
     \includegraphics[scale=0.22]{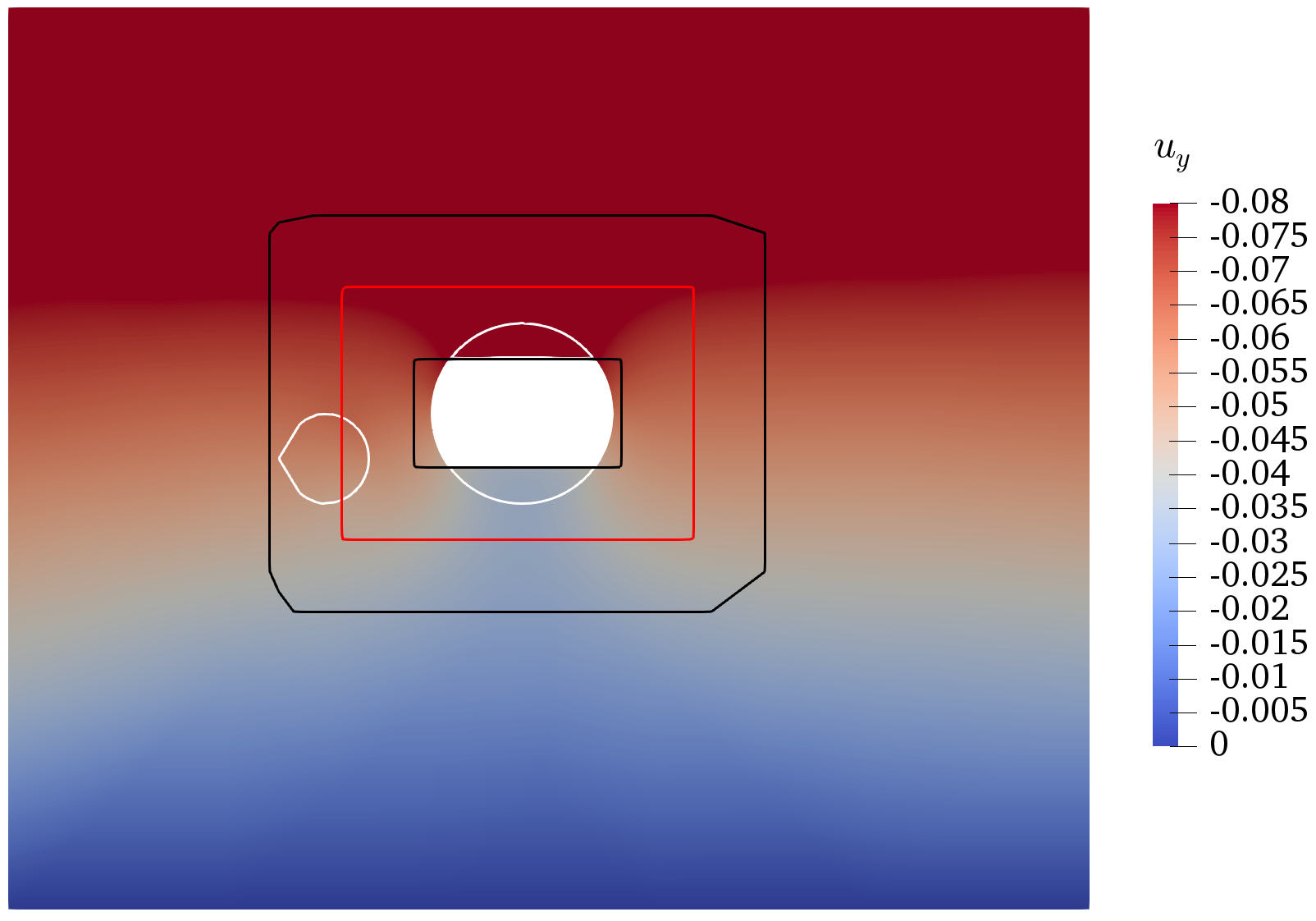}
     }
    \hfill
     \caption{Displacement component $u_y$ for a) mixed multiscale, $2\epsilon =0.2$, and b) mixed multiscale, $2\epsilon =0.8$.}
     \label{fig:ModelBAdapZoomDisps}
\end{figure}

We test two length sizes for the transition region, $2\epsilon ={0.2,0.8}$. The displacement field component $u_y$ for both $\epsilon$ is shown in Figure \ref{fig:ModelBAdapZoomDisps}. When compared to the full microscale model as a reference, shown in Figure \ref{fig:ModelBAdaptiveDisps}a, the mixed multiscale method with homogenization is much closer to the reference solution in comparison to the adaptive CutFEM approach, shown in Figure \ref{fig:ModelBAdaptiveDisps}b. Therefore, using homogenized models in the coarse domains is necessary when the signed distance functions fail to detect the microstructure precisely.

\clearpage
%, while assuming the RVE being restricted inside the zooms, the density of pores increases and as expected we achieve a smaller value of $E_{\mathcal{M}} ^n= $0.58.
\subsubsection{The mixed multiscale method with two arbitrary zooms}
%Paragraph 1. Schematic geometry. Meshing.\\

Next, we investigate the efficiency of our mixed multiscale approach for the same quasi-uniform porous domain (see Figure \ref{fig:ModelQuasiSchm}) using two separate zooms. The displacement at the bottom edge is blocked and $u= (0, -0.1)$ is applied along the top edge of the domain. We consider the following microscale material properties: $E_m =1$ and $\nu_m =0.3$, while for the macro scale, we derive effective material properties by using homogenization Equation \ref{Eq.MMT}. Like in the previous Section, we compute the effective Young's modulus based on the pores in the entire domain $\Omega$ as $E_{\mathcal{M}}=0.78$.

\begin{figure}[h]
   \centering
    \subfloat[ \label{FEMm2}]{%
     \includegraphics[scale=0.22]{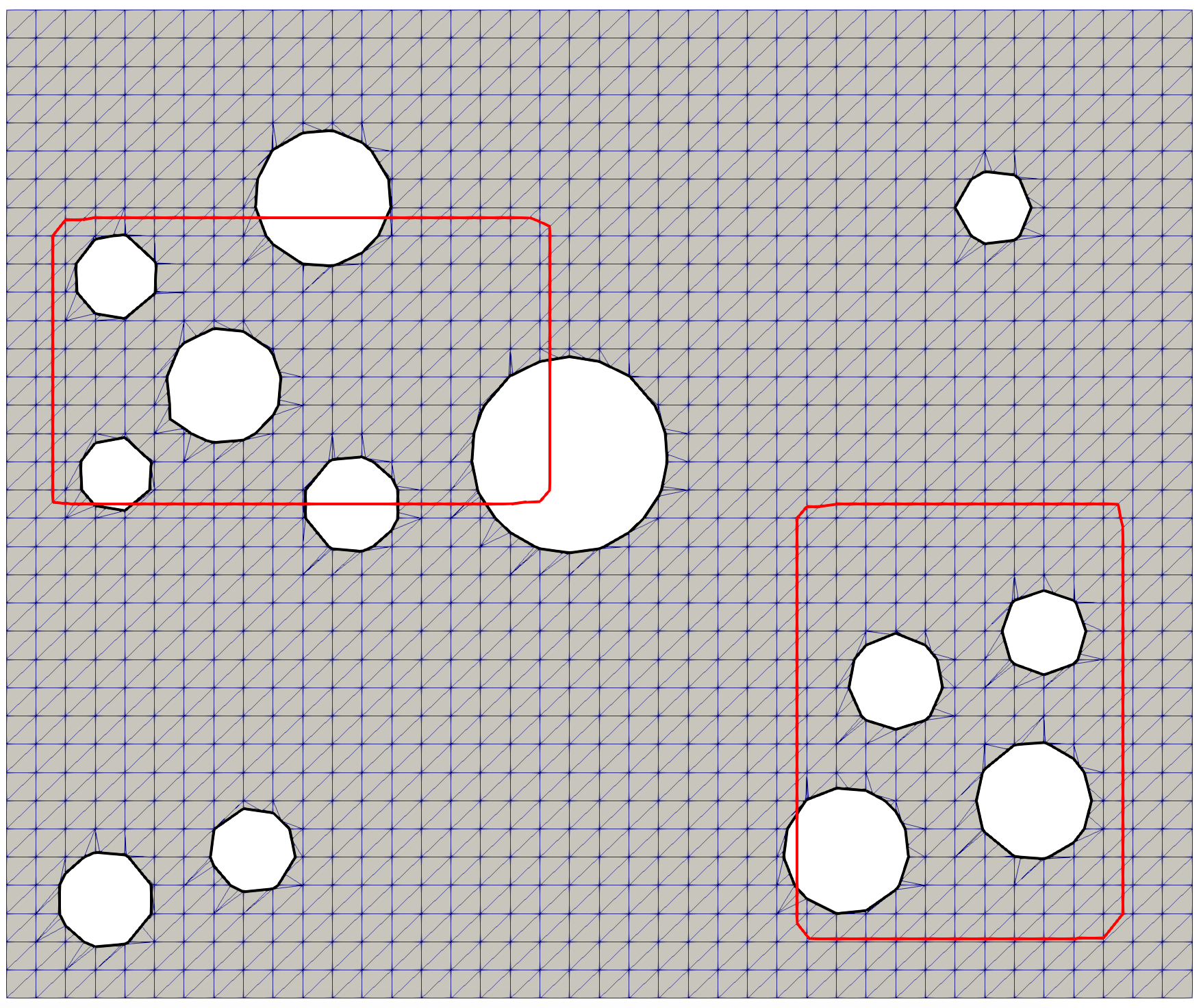}
     }
    \hfill
     \subfloat[ \label{CutFEMm2}]{%
       \includegraphics[scale=0.22]{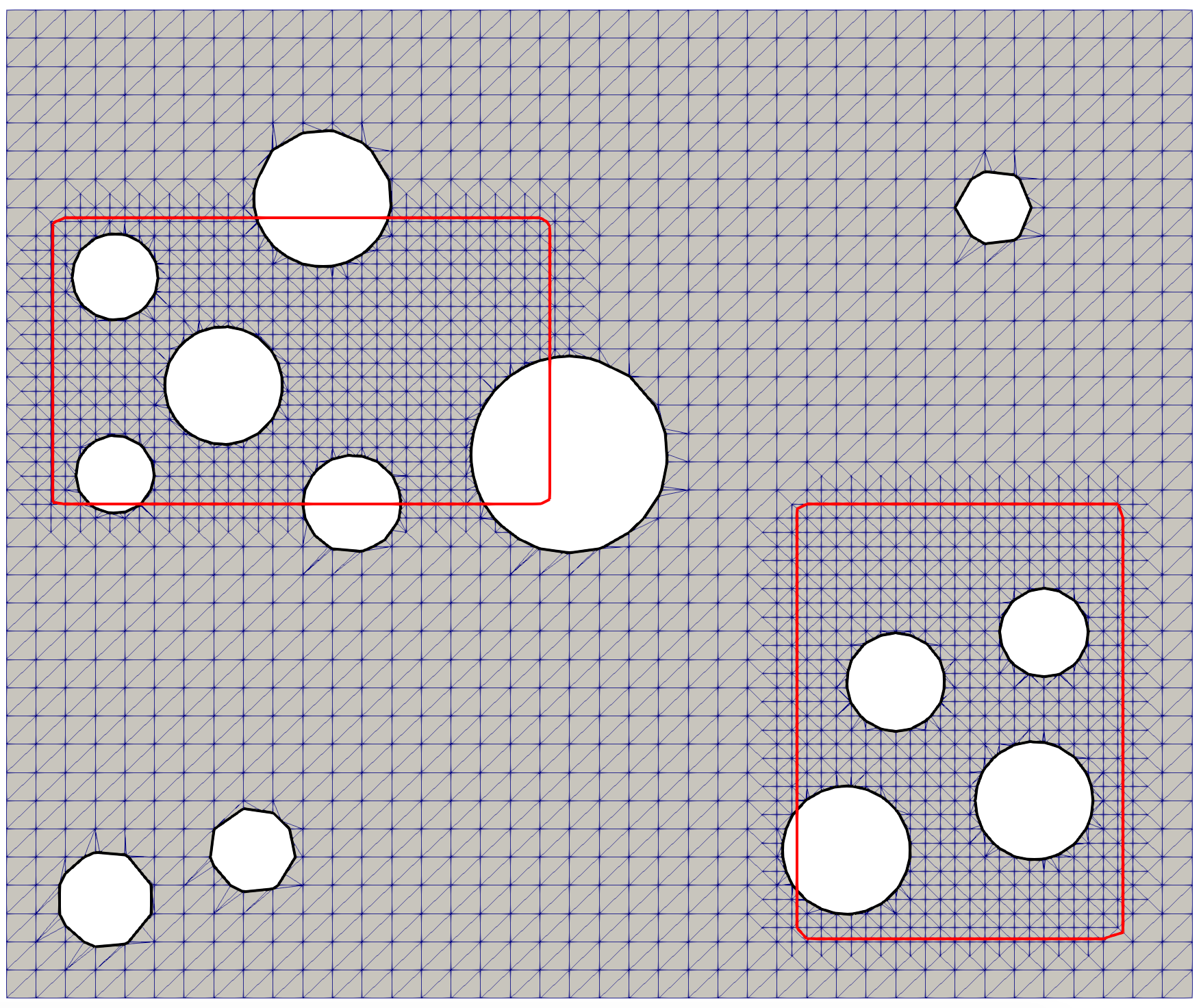}
     }
    \hfill
     \subfloat[ \label{CutFEMm2b}]{%
      \includegraphics[scale=0.22]{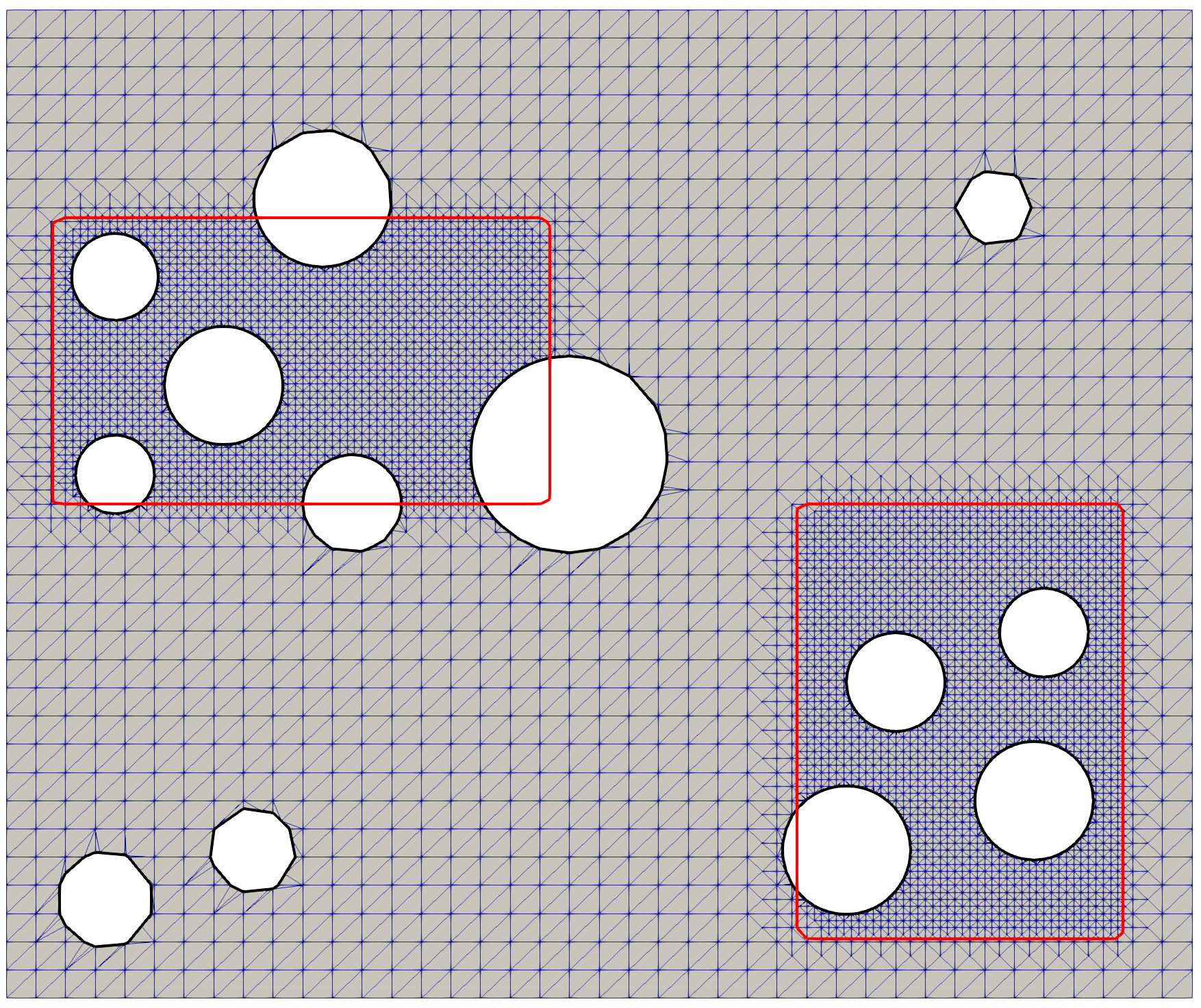}
     }
    \hfill
     \caption{Computational mesh for physical domain of 2D model with quasi-uniform distributed pores, a) uniform meshing, b) adaptive meshing type 1, c) adaptive meshing type 2.}
     \label{fig:MeshesModelB}
\end{figure}

\begin{figure}[h]
   \centering
    \subfloat[ \label{FEM2b}]{%
     \includegraphics[scale=0.135]{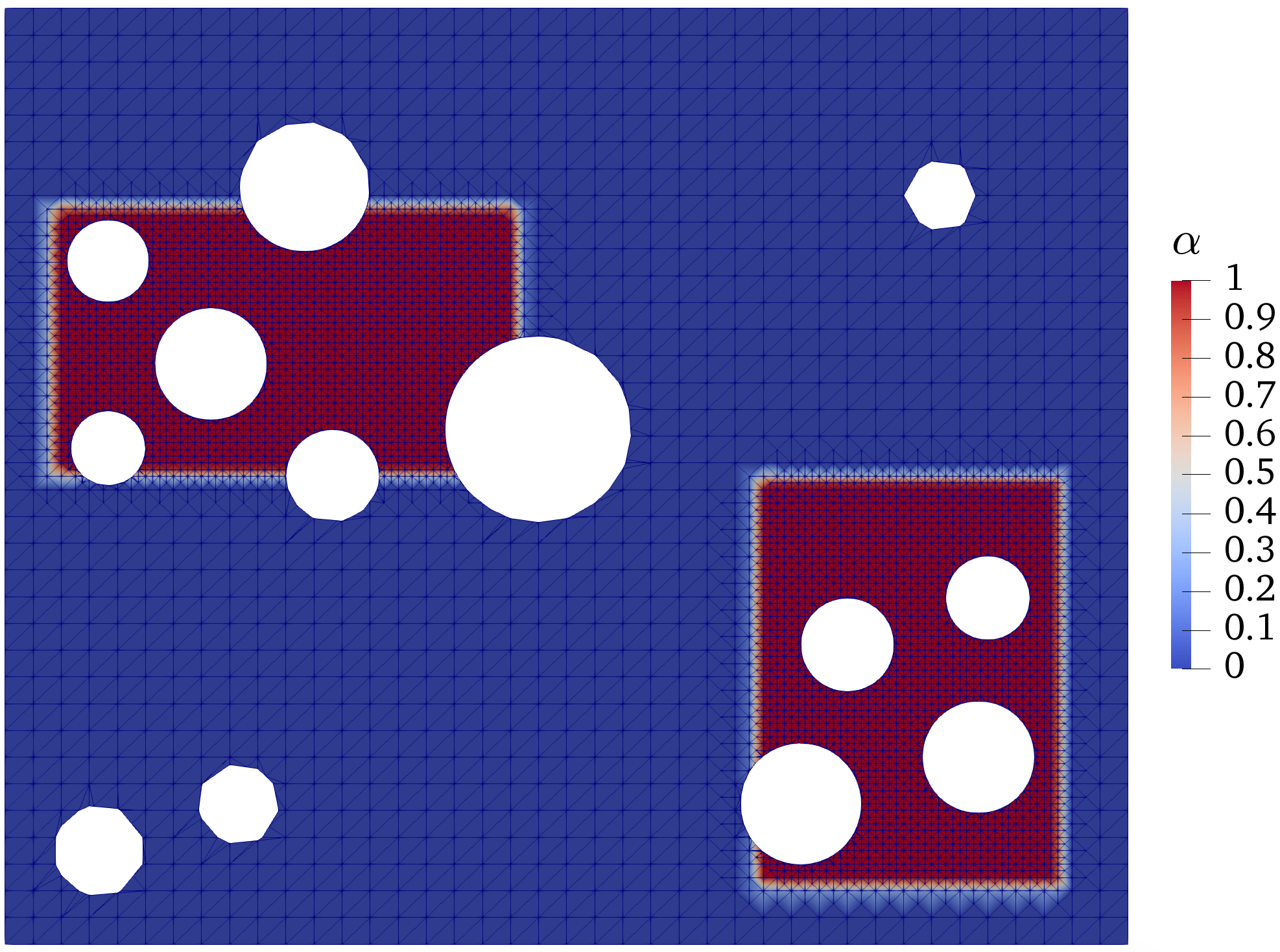}
     }
    \hfill
     \subfloat[ \label{CutFEMw2}]{%
       \includegraphics[scale=0.135]{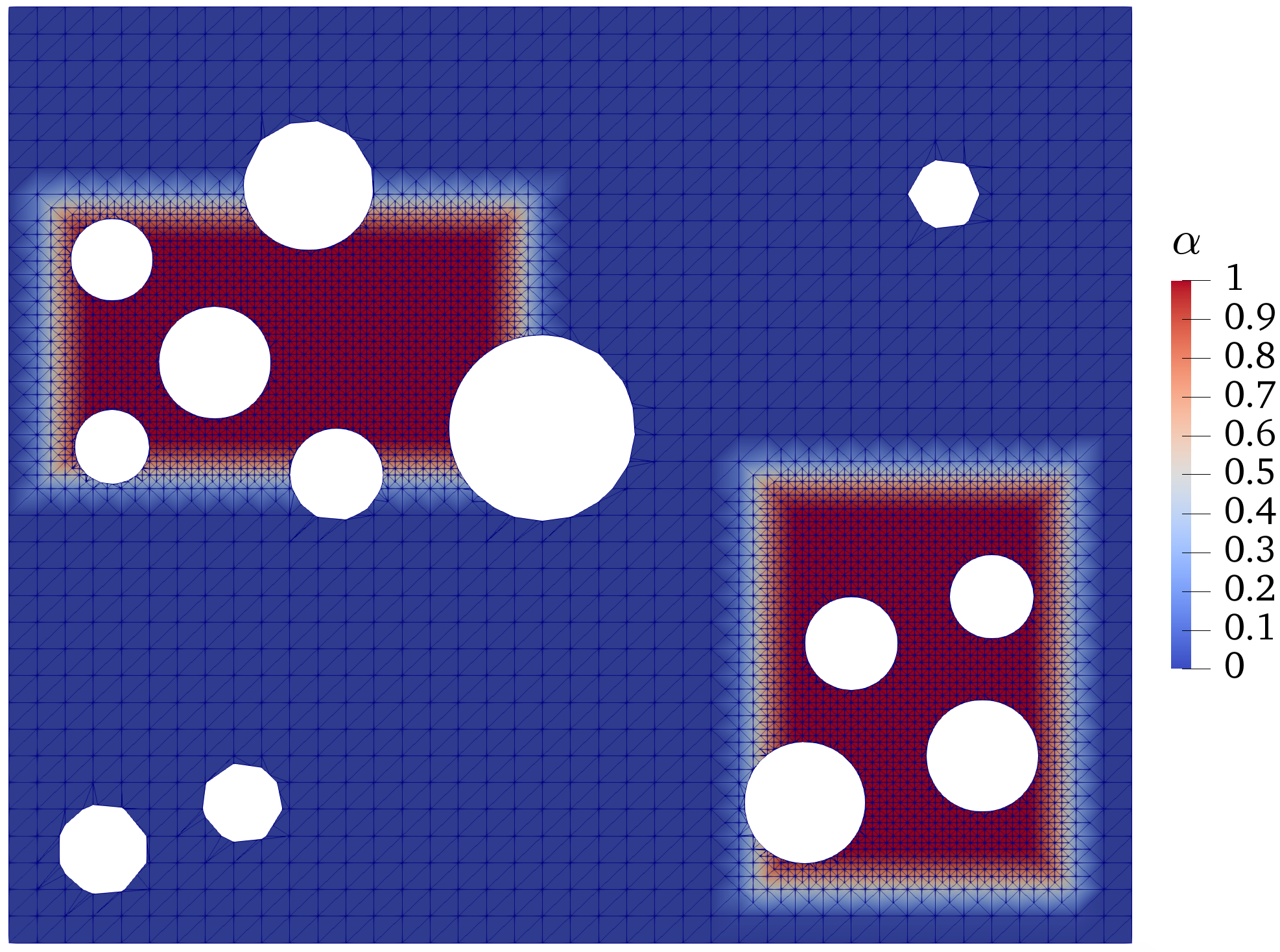}
     }
    \hfill
     \subfloat[ \label{CutFEMw2b}]{%
      \includegraphics[scale=0.135]{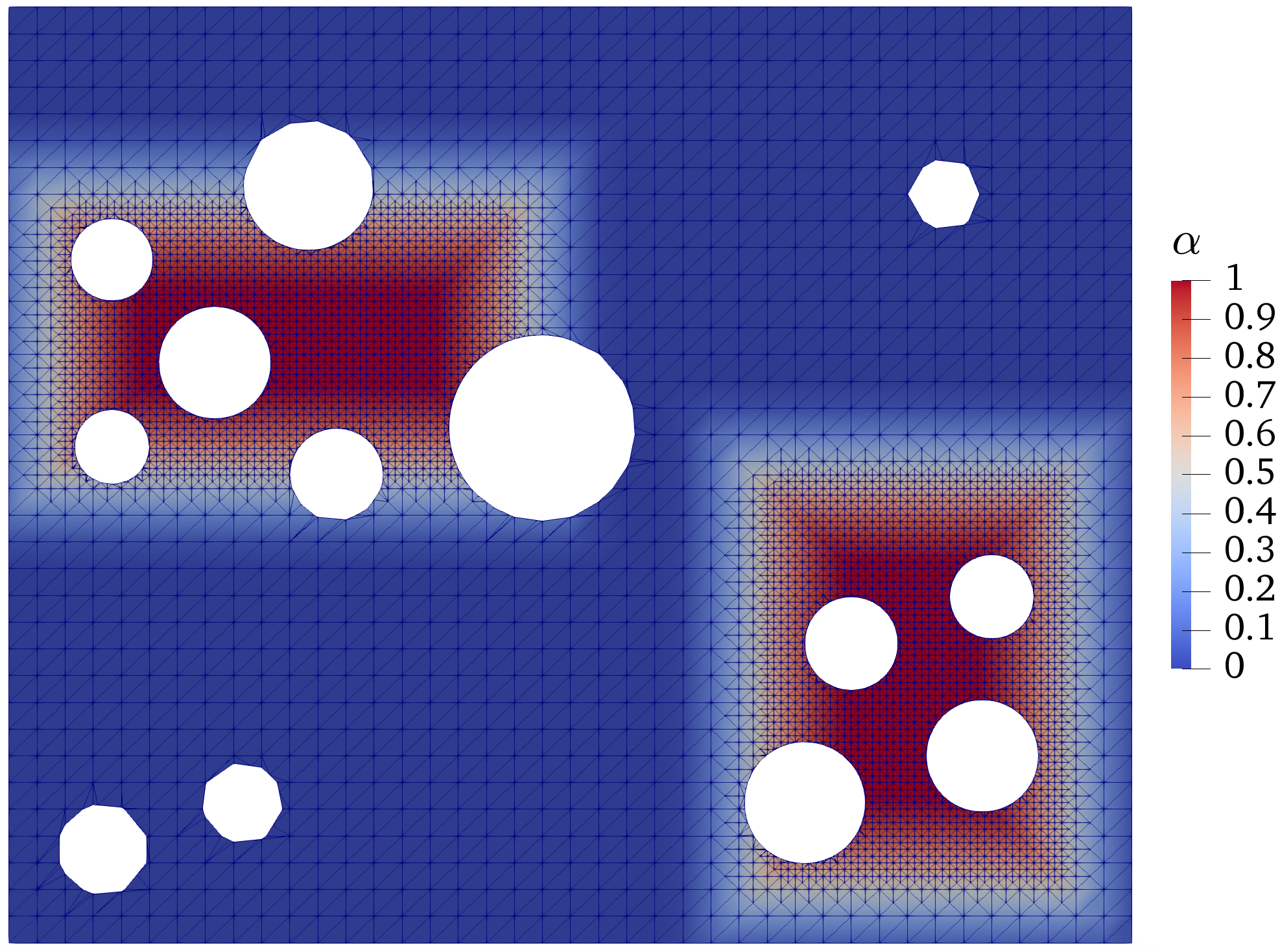}
     }
    \hfill
     \caption{Smooth weight function field $\alpha_h$ over finest adaptive mesh with a) $\epsilon=0.1$ b) $\epsilon=0.4$ c) $\epsilon=1$.}
     \label{fig:IdicB}
\end{figure}

For this example, we employ the same background meshes as for the locally porous domain (Figure \ref{fig:ModelASchem}) and show the corresponding discretized domain and generated interfaces in Figure \ref{fig:MeshesModelB}. The smooth indicator function with three lengths is computed for the finest adaptive mesh in Figure \ref{fig:IdicB}. In Figures \ref{fig:MeshesModelB} and \ref{fig:IdicB}, we observe the independency of the microstructure, zooming geometry and mixing length to the computational mesh, which creates a straightforward preprocessing pipeline and saves mesh regeneration costs.
%Our framework allow us to locate our zooming interfaces arbitrarily at different parts of domain requiring accurate micro scale analysis

\begin{figure}[!h]
   \centering
    \subfloat[ \label{FEMd2}]{%
     \includegraphics[scale=0.23]{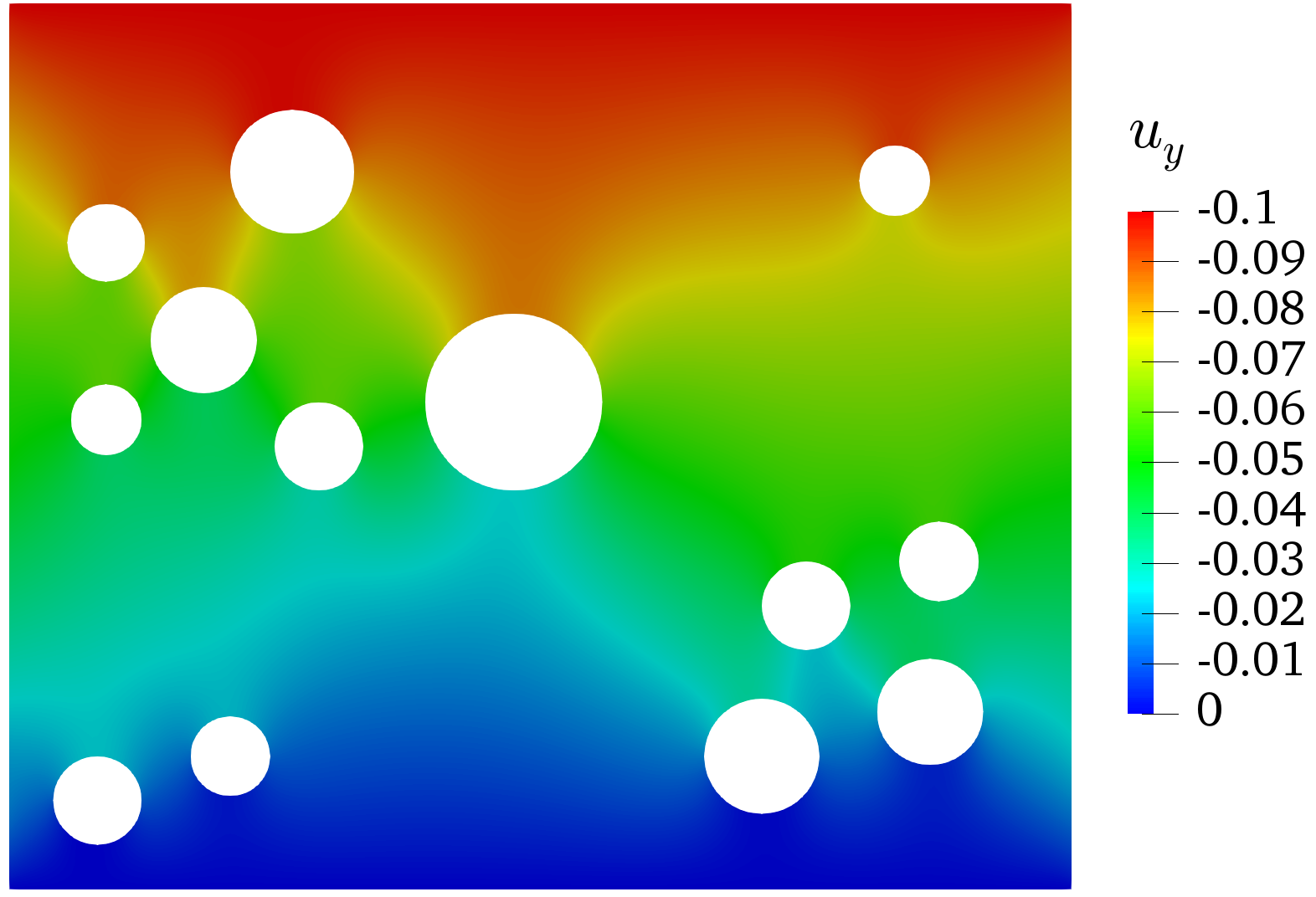}
     }
    \hfill
     \subfloat[ \label{CutFEMd2b}]{%
       \includegraphics[scale=0.23]{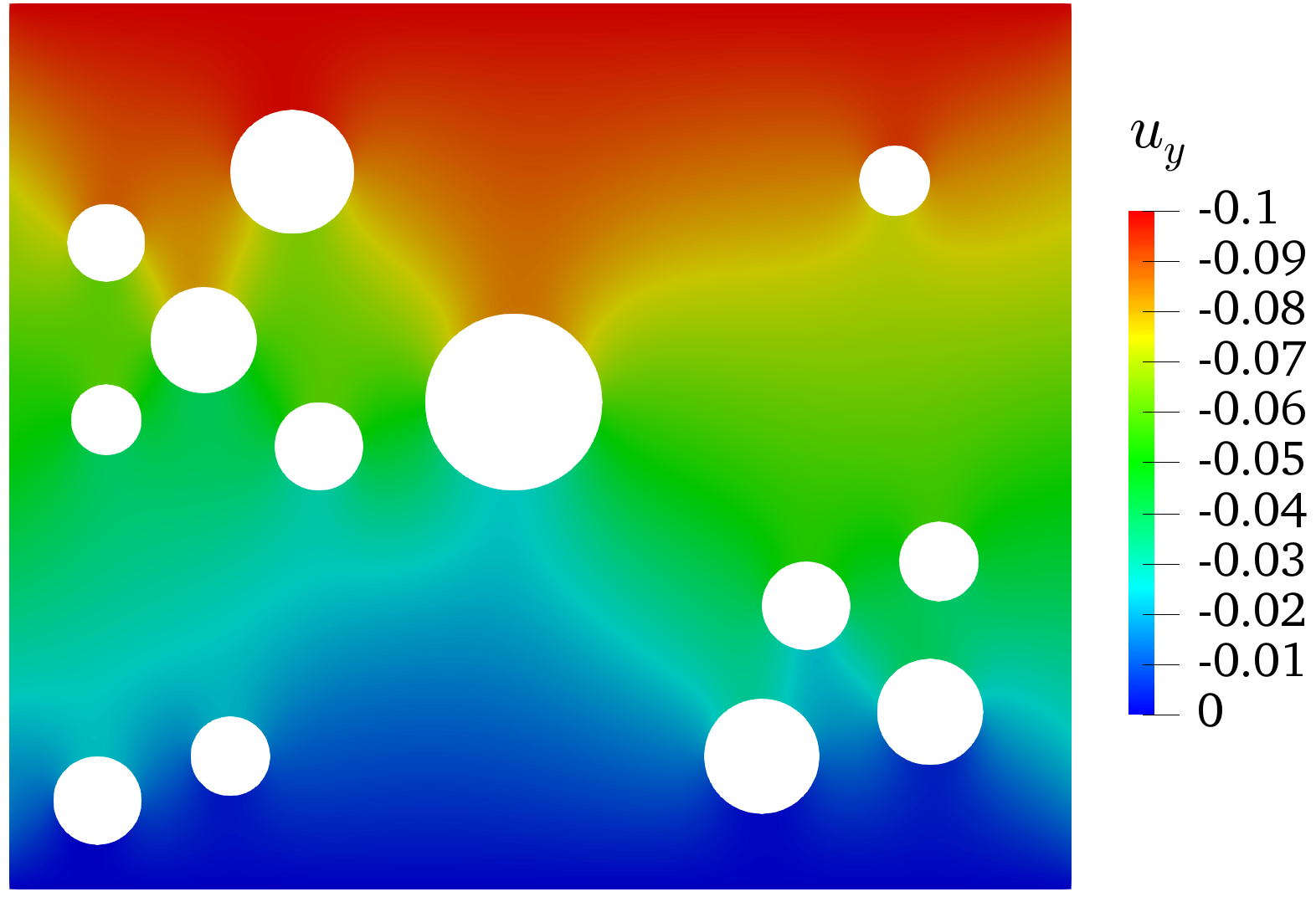}
     }
    \hfill
     \subfloat[ \label{CutFEMd2m}]{%
      \includegraphics[scale=0.23]{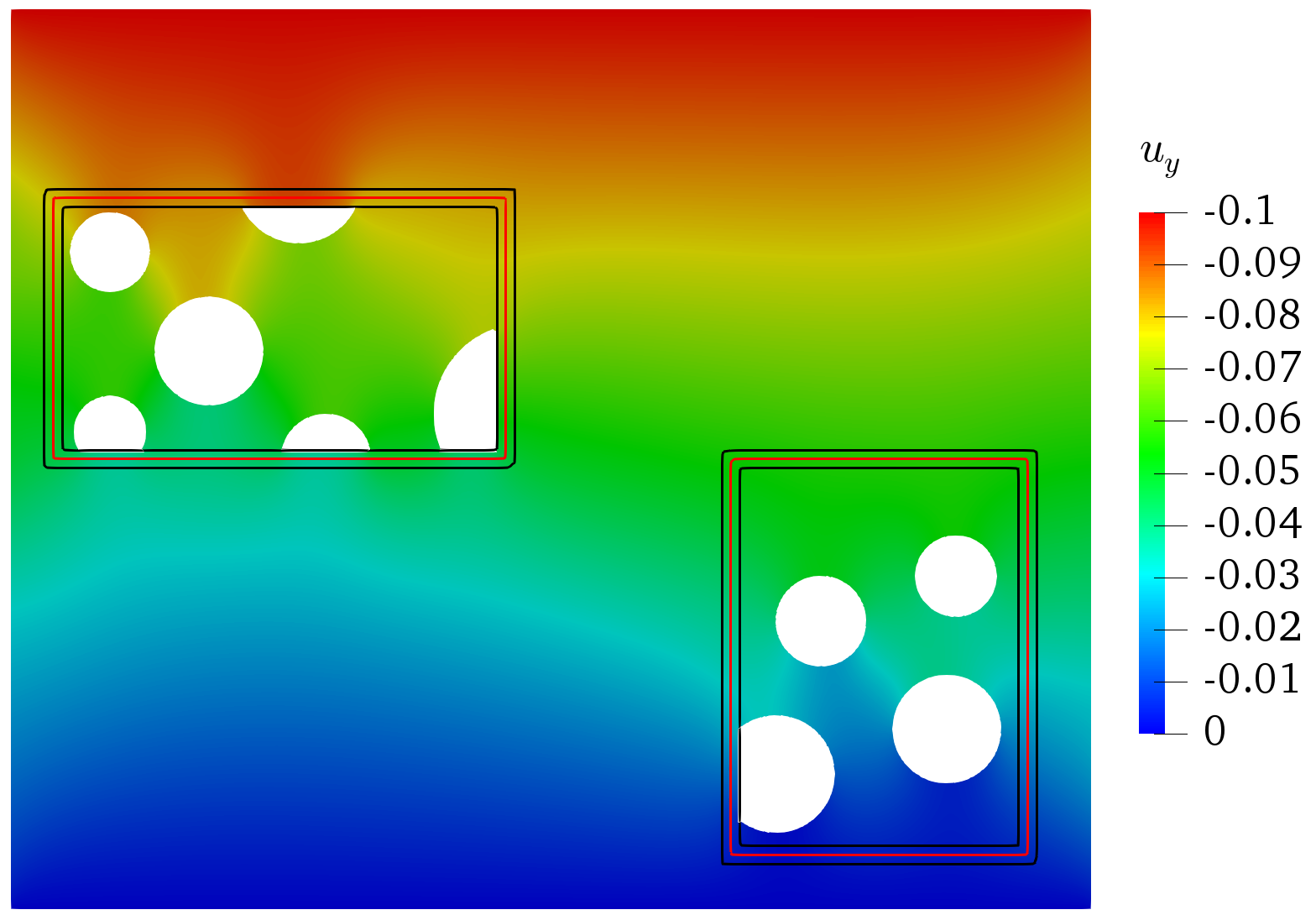}
     }
    \hfill
     \subfloat[ \label{CutFEMd2m2}]{%
      \includegraphics[scale=0.23]{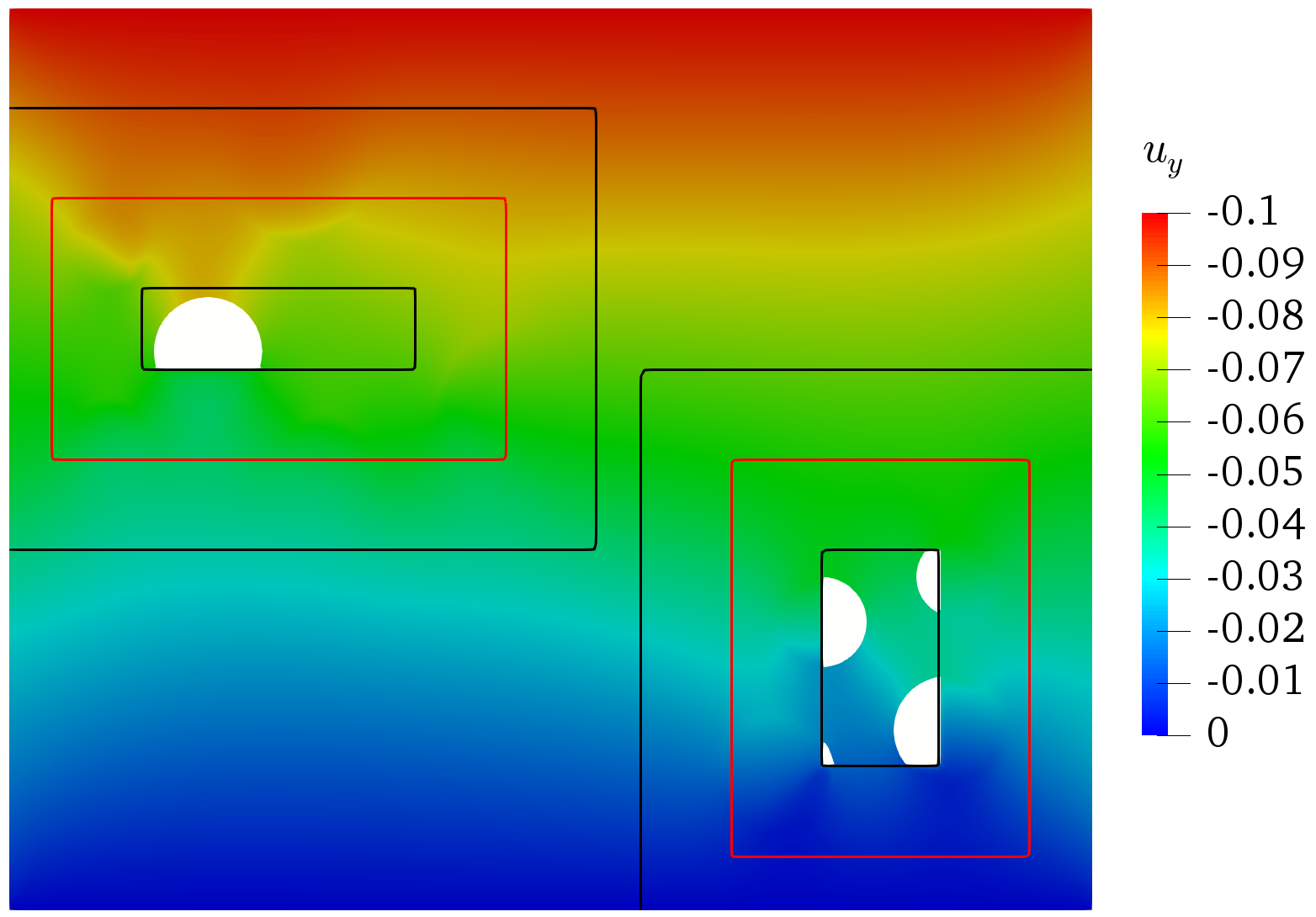}
     }
      \hfill
     \caption{Displacement field component $u_y$, a) FEM model, b) CutFEM model, c) mixed multiscale model with $\epsilon=0.1$, d) mixed multiscale model with $\epsilon=1$.}
     \label{fig:ModelBDisp}
\end{figure}

We compute displacement field component $u_y$ for two smoothing lengths $2\epsilon=\{0.1, 1\}$, and show the corresponding results over the physical and the fictitious domains in Figure \ref{fig:ModelBDisp}c-f. The results prove a high relevance of the multiscale framework in the microscale domain to the corresponding reference models (depicted in Figure \ref{fig:ModelBDisp}a,b). In the transition regions, $u_y$ as a global response is smooth for both mixing lengths and outside the zooms (homogenized domain) the trend is similar to the references.

\begin{figure}[h]
   \centering
    \subfloat[ \label{FEMs2}]{%
     \includegraphics[scale=0.21]{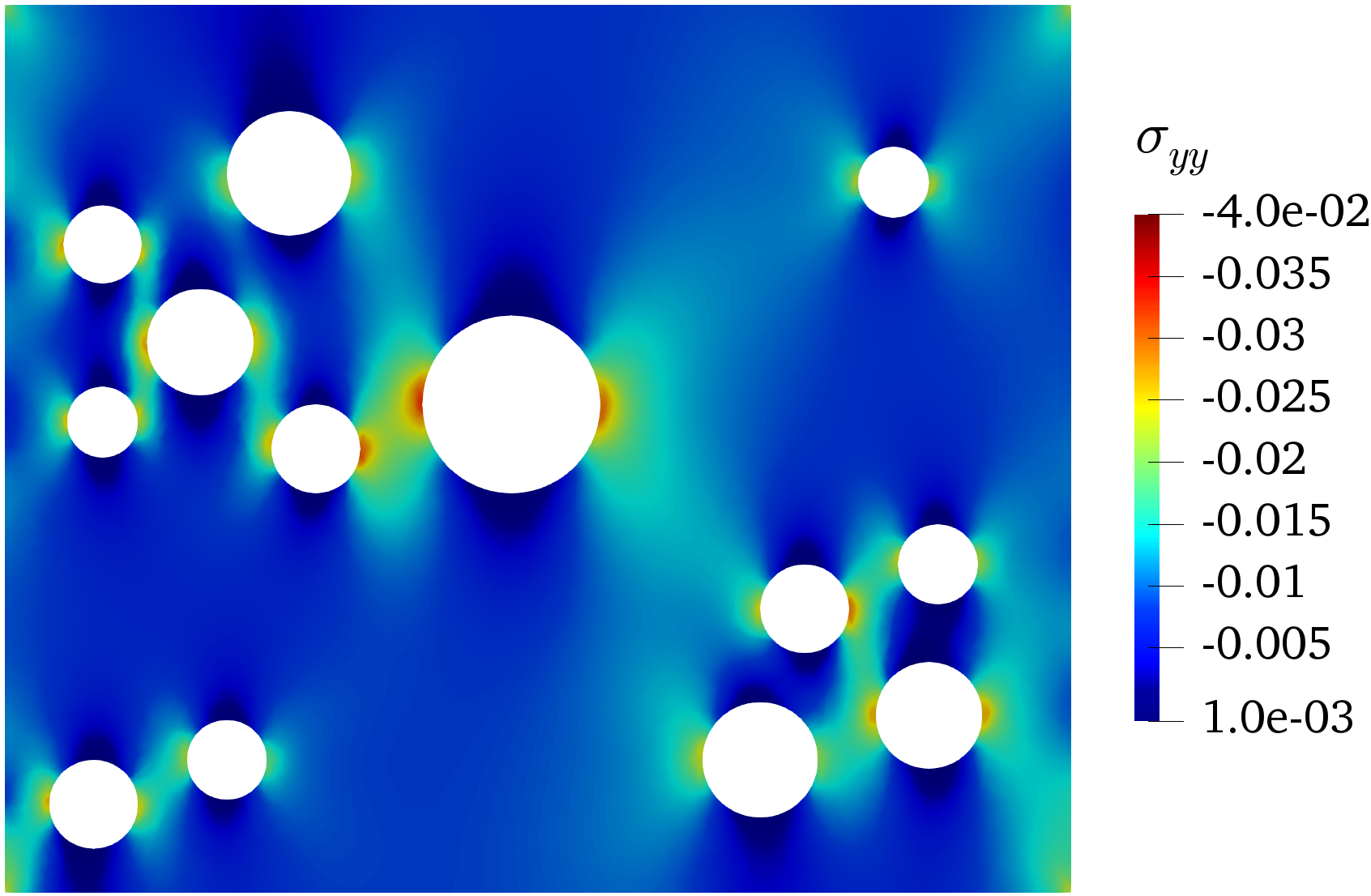}
     }
    \hfill
     \subfloat[ \label{CutFEMs2c}]{%
       \includegraphics[scale=0.21]{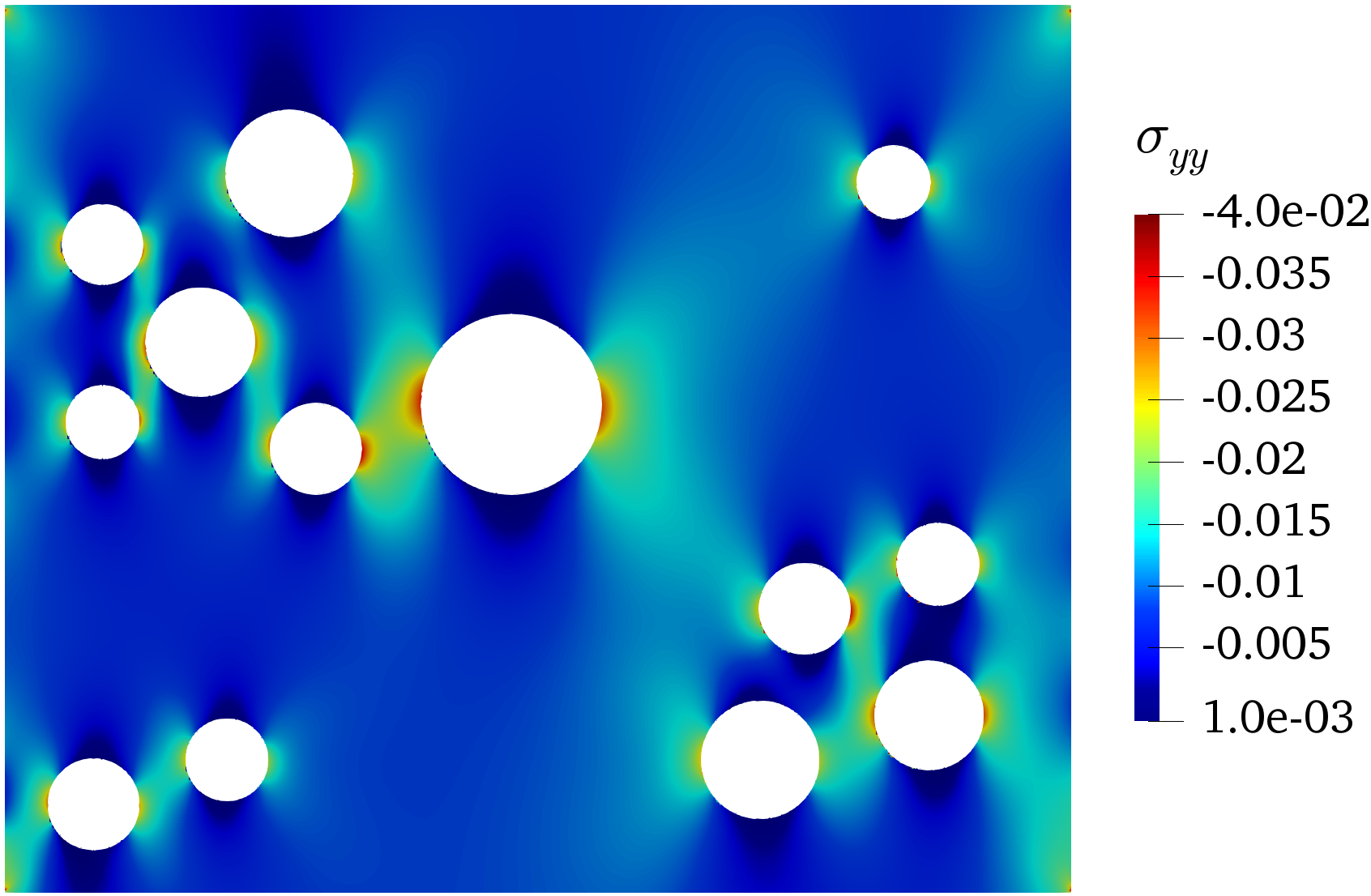}
     }
    \hfill
     \subfloat[ \label{smix2}]{%
      \includegraphics[scale=0.22]{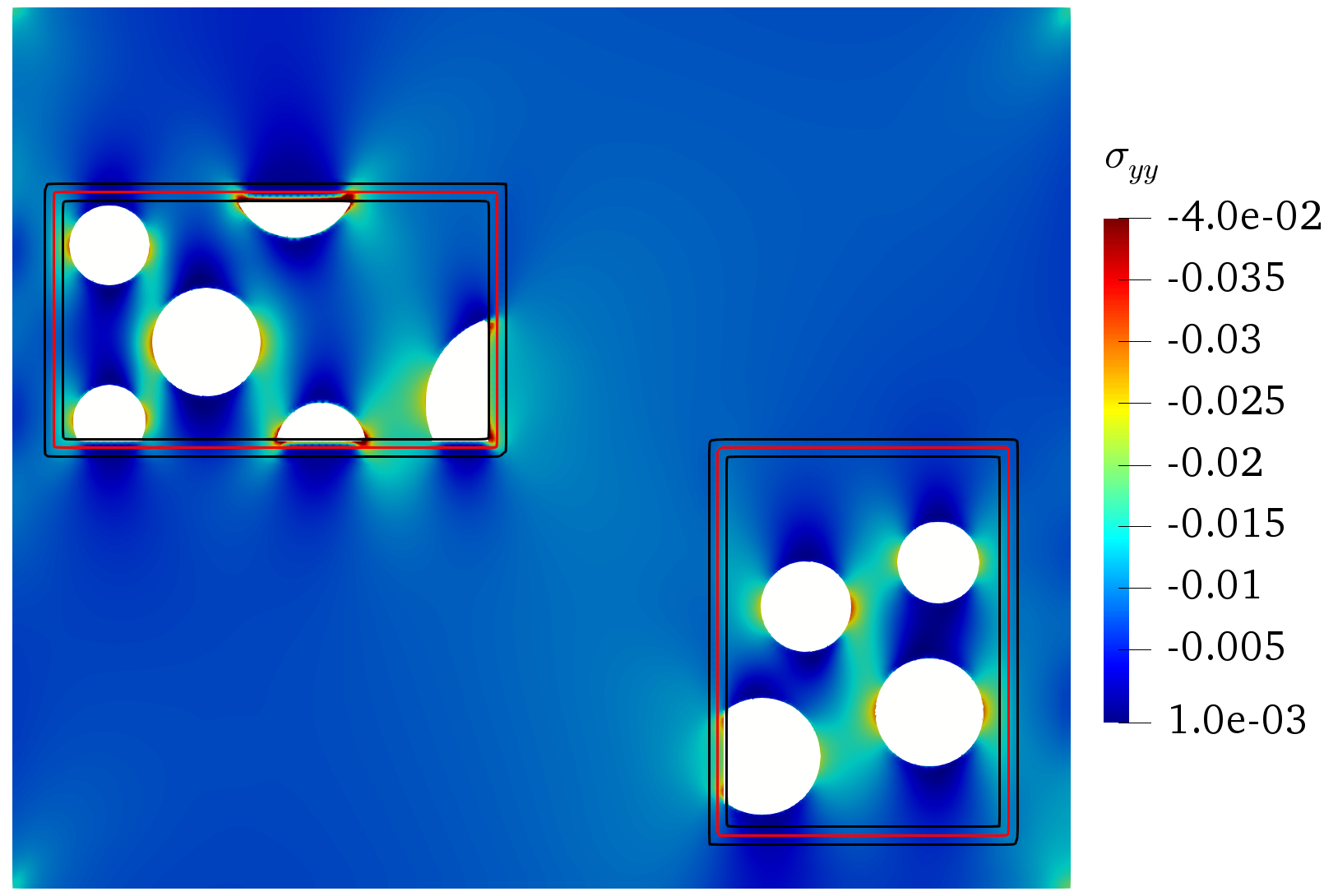}
     }
    \hfill
        \subfloat[ \label{smix2b}]{%
     \includegraphics[scale=0.22]{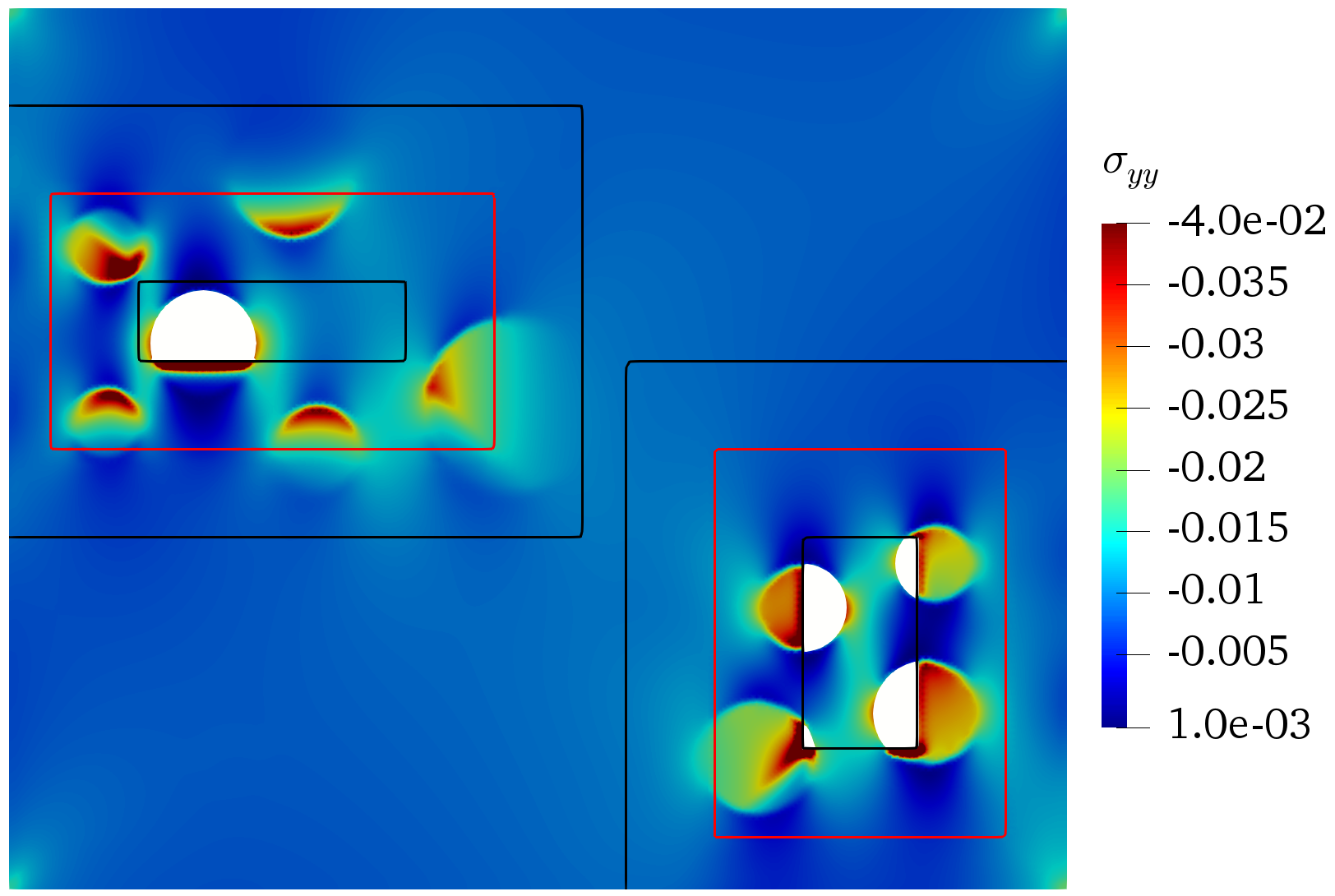}
     }
    \hfill
     \subfloat[ \label{smix2c}]{%
       \includegraphics[scale=0.22]{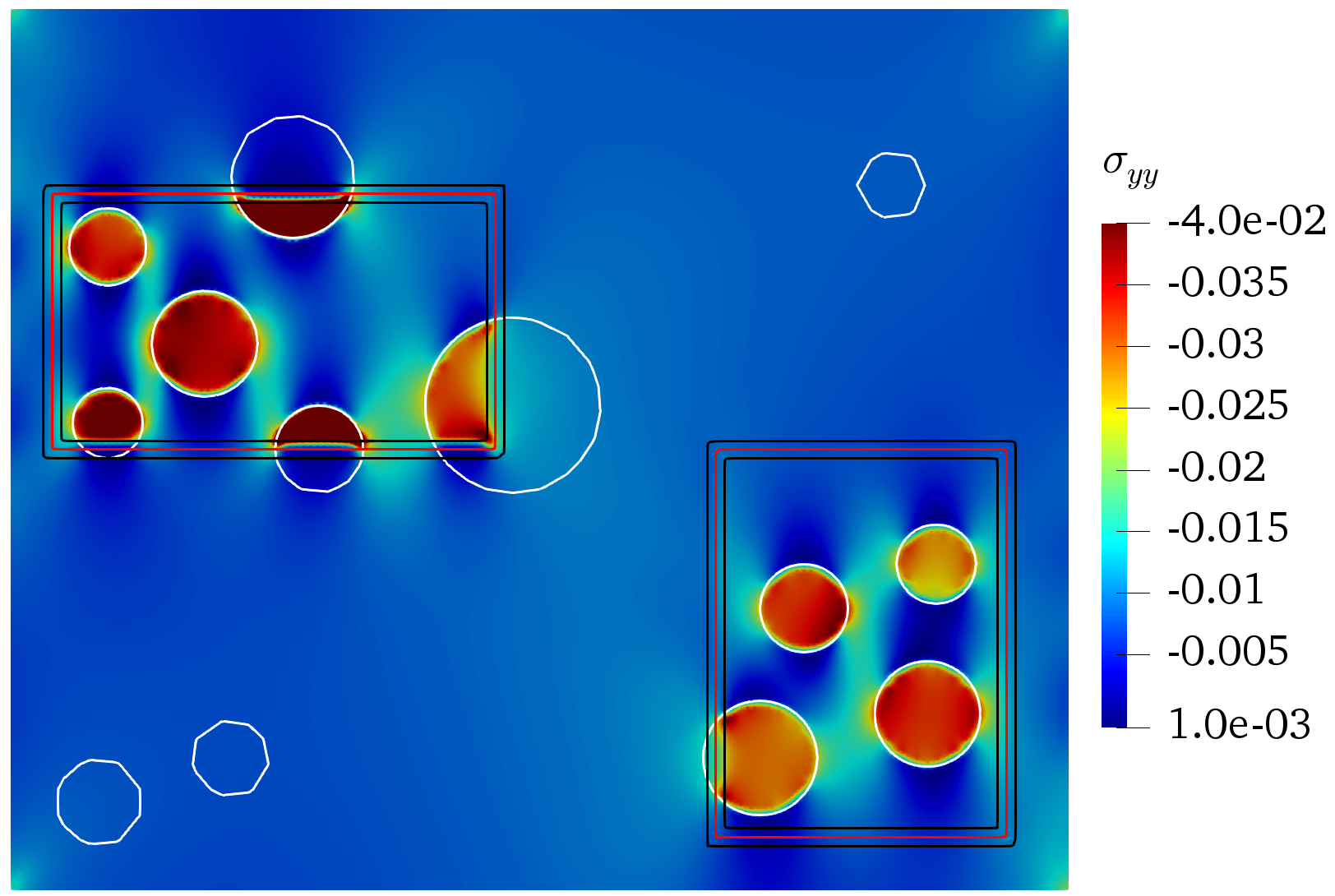}
     }
    \hfill
     \subfloat[ \label{smix2d}]{%
      \includegraphics[scale=0.22]{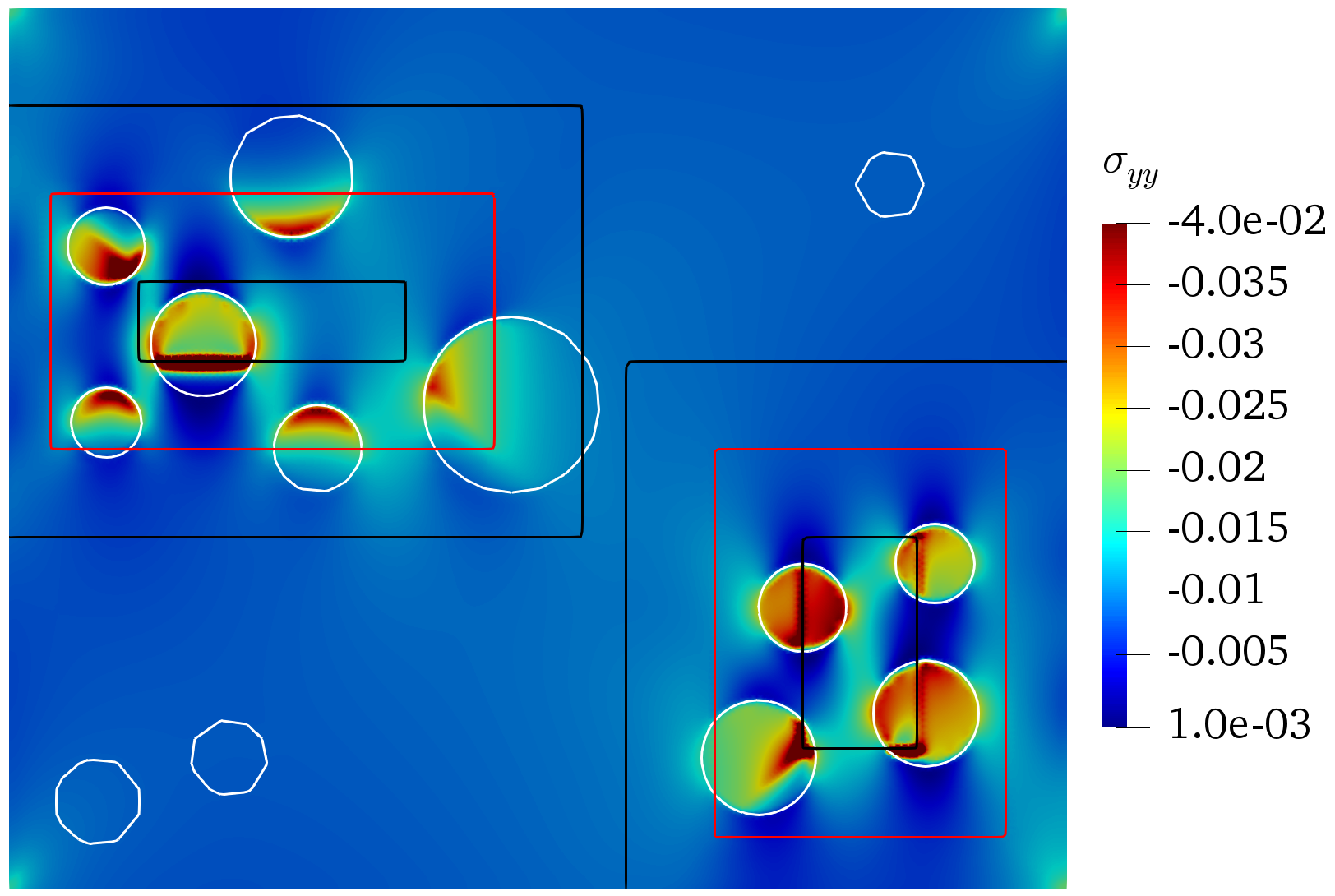}
     }
    \hfill
     \caption{Stress component $\sigma_{yy}$ contours, a) FEM model, b) CutFEM model, c) physical domain for the mixed multiscale model with $\epsilon=0.1$, d) physical domain for the mixed multiscale model with $\epsilon=1$, e) fictitious domain for the mixed multiscale model with $\epsilon=0.1$ and f) fictitious domain for the mixed multiscale model with $\epsilon=1$.}
     \label{fig:ModelBStress}
\end{figure}

Next, we inspect the distribution of the mixed stress field for two zooming problems. The results obtained for stress field component $\sigma_{yy}$ for two smoothing lengths and over physical and fictitious domains are given in Figure \ref{fig:ModelBStress}. The comparison with the full fine-scale reference models (see Figure \ref{fig:ModelBStress}) shows that the stress does not suffer from any oscillations neither in cut elements nor in the transition area. The ghost penalty regularization, which extends the solution from the physical domain to the fictitious domain alleviates the oscillations successfully while ensuring an accurate stress solution.

\begin{figure}[!h]
   \centering
    \includegraphics[scale=0.25]{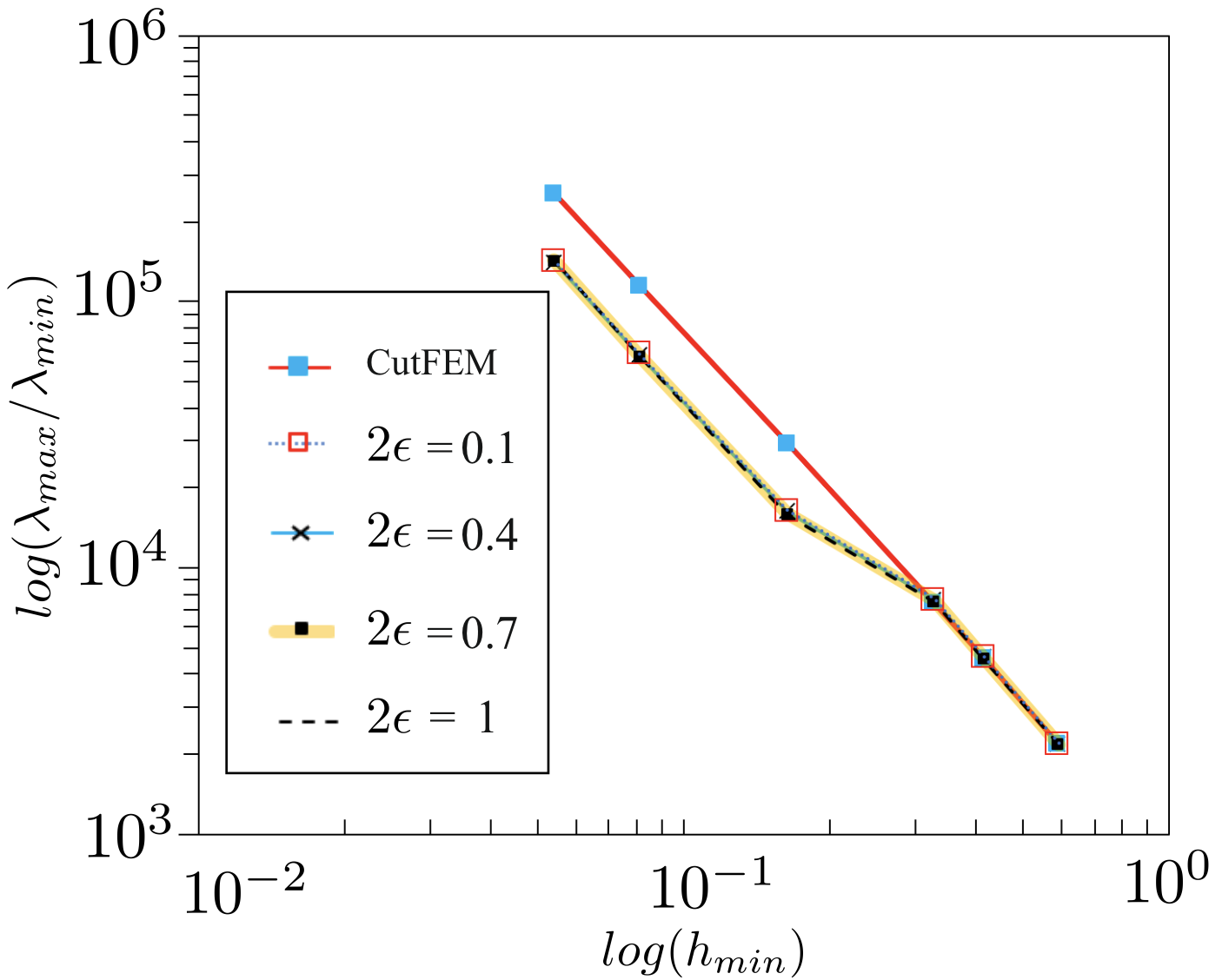}
    % \subfloat[ \label{fig:ModelBCondMixa}]{%
    %   \includegraphics[scale=0.25]{./Figs/Results/CondMixBa.png}
    %  }
    % \hfill
    %      \subfloat[ \label{fig:ModelBCondMixb}]{%
    %   \includegraphics[scale=0.25]{./Figs/Results/CondMixBb.png}
    %  }
    % \hfill
  \caption{Condition numbers for CutFEM model and mixed multiscale method for different smoothing lengths and mesh configurations. 
  %and with, a) homogenization RVE all over domain,  b) homogenization RVE inside the zooms. 
  The regularization parameter is chosen as $\beta =0.005$.}
  \label{fig:ModelBConds}
\end{figure}

\clearpage
The condition number of the multiscale system matrix for different mesh configurations and smoothing lengths are compared with the counterpart CutFEM microscale model in Figure \ref{fig:ModelBConds}a.  The comparison shows that our multiscale assembled matrix is well-conditioned under various smoothing lengths and mesh sizes and converges proportional to $h^{-2}$ that is similar to the CutFEM convergence.

% ----------- 3rd Model --------------------
\subsection{3D mixed multiscale modelling of trabecular bone}

\begin{figure}[!ht]
   \centering
    \subfloat[ \label{BoneImage1}]{%
     \includegraphics[scale=0.24]{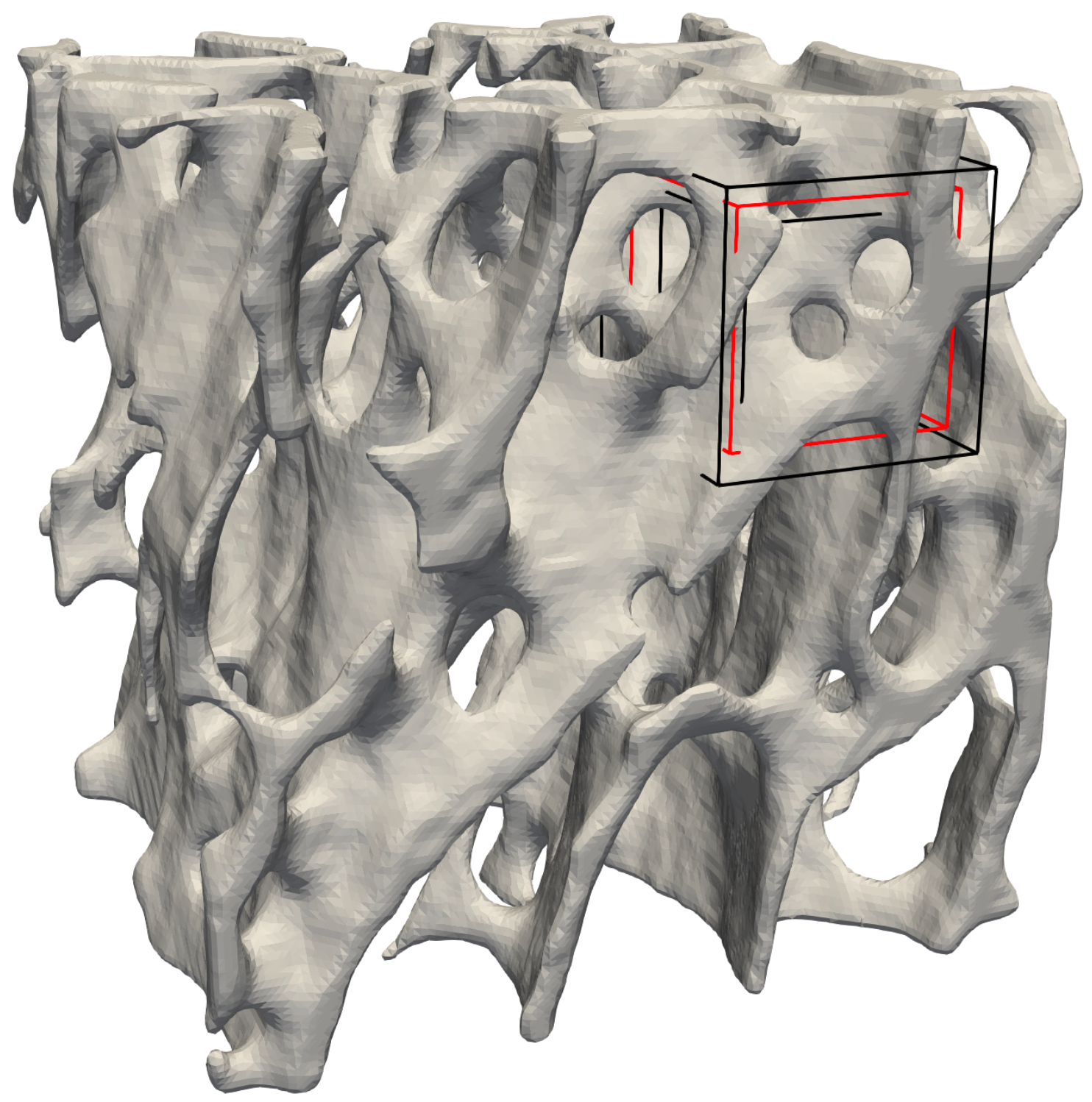}
     }
    \hfill
     \subfloat[ \label{BoneZoom1}]{%
       \includegraphics[scale=0.28]{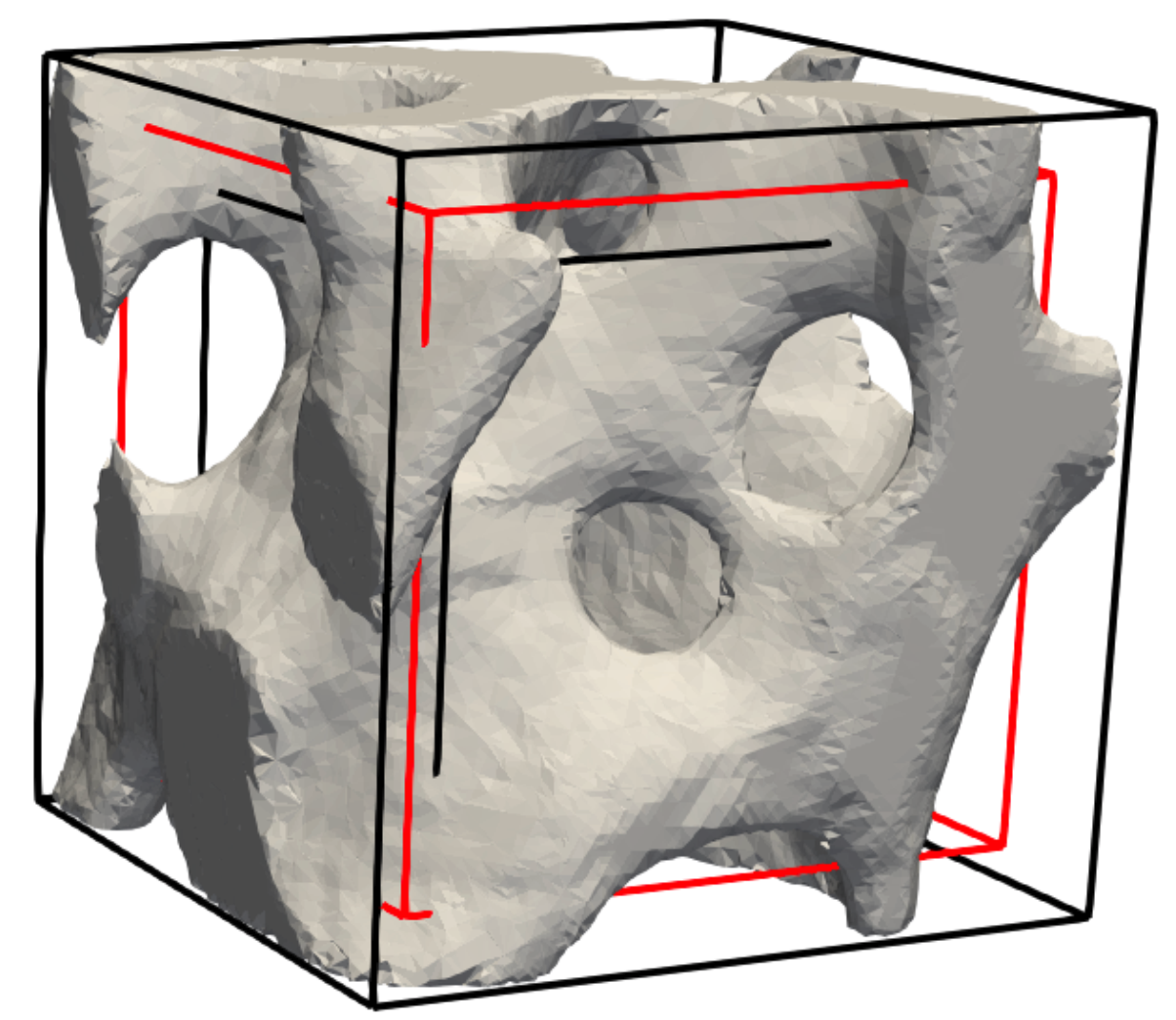}
     }
    \hfill
     \caption{3D trabecular bone with zoom: (a) Micro-CT image 3D reconstruction, (b) CutFEM interface.}
     \label{fig:BoneImage}
\end{figure}

\begin{figure}[!ht]
   \centering
    \subfloat[ \label{BoneImage2}]{%
     \includegraphics[scale=0.24]{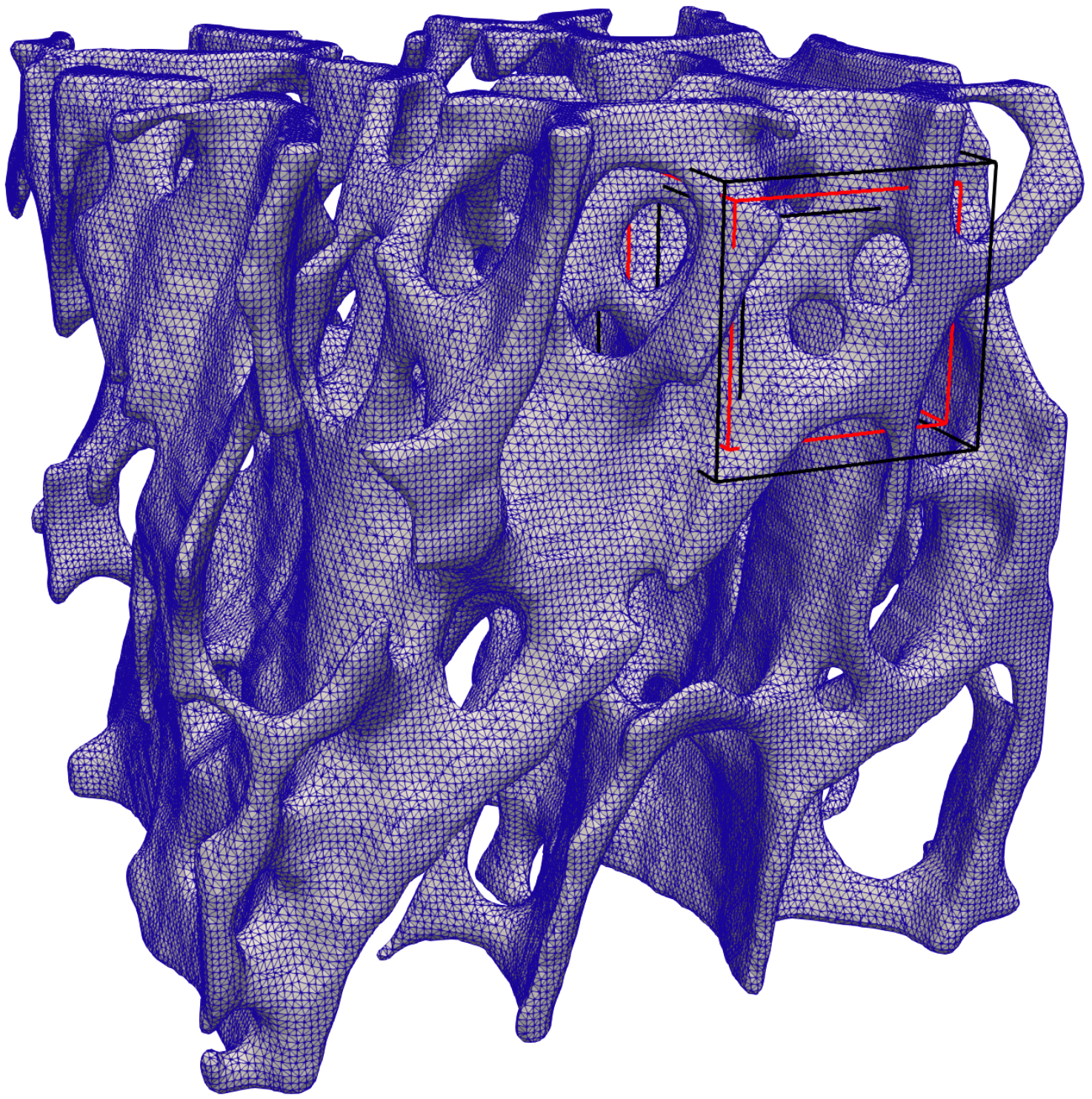}
     }
    \hfill
     \subfloat[ \label{BoneZoom2}]{%
       \includegraphics[scale=0.28]{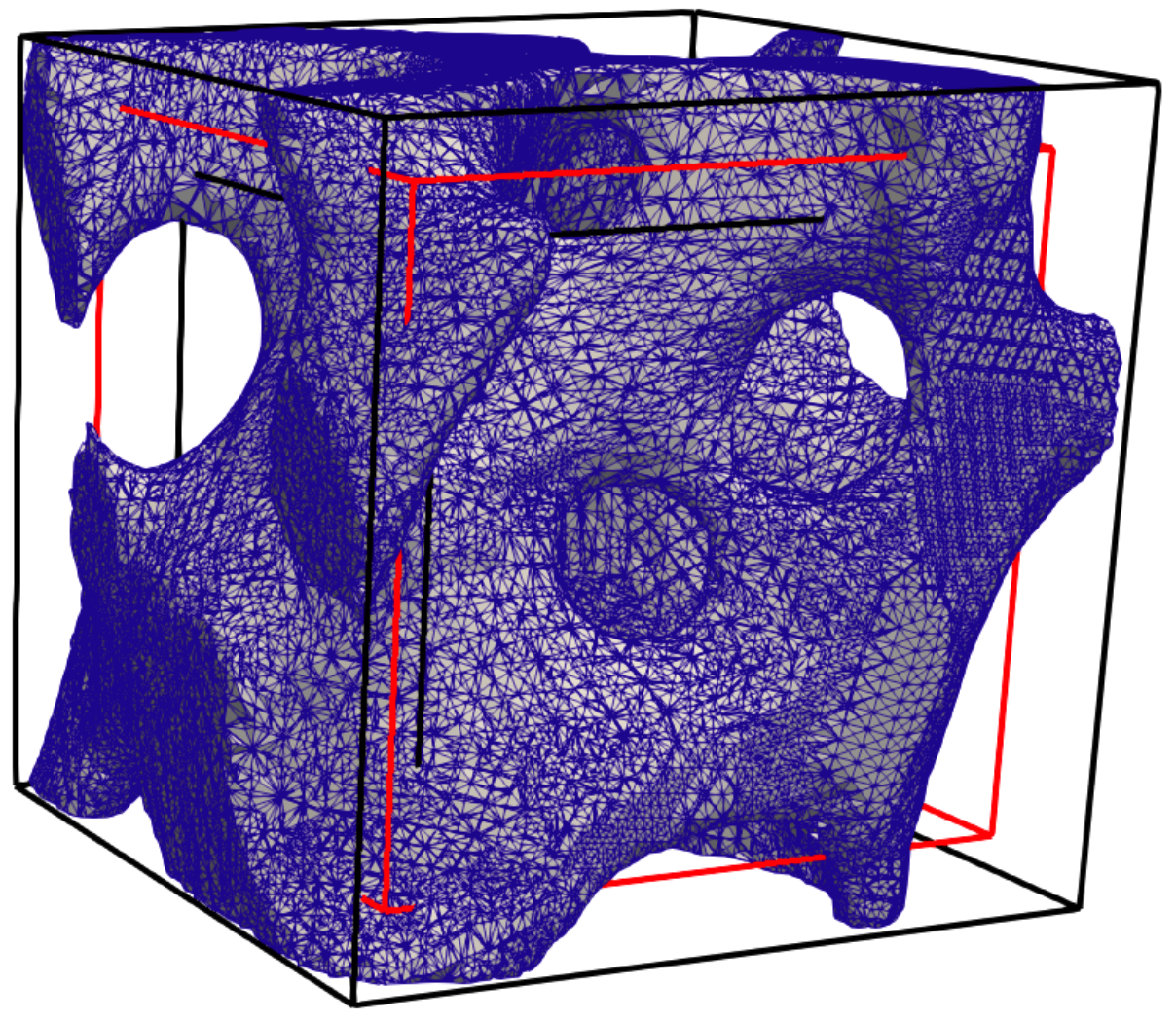}
     }
    \hfill
     \caption{3-D trabecular bone mesh with zoom: (a) surface mesh of micro-CT image, (b) CutFEM surface subtesselation for $h=0.036$.}
     \label{fig:BoneMesh}
\end{figure}

This numerical example illustrates the efficacy of the proposed mixed multiscale framework in 3D simulations. We use a 3D bone sample with a trabecular microstructure which is transferred directly from a micro-CT medical image. The corresponding 3D reconstructed micro-CT image is presented in Figure \ref{BoneImage1}. We use the 3D reconstructed image to compute a surface mesh (STL mesh data) which will be converted into a level set function. For more information on the digital pipeline that we have used to convert STL mesh data into a level set function, see \cite{Claus21b}. Using our proposed zooming technique, we select the zoom region and apply the mixing scheme to the bone as shown in Figure \ref{BoneZoom1}. The red and black lines represent the zoom surface and the upper/lower bounds of the mixing regions, respectively. The bone microstructure is defined by the zero level set function $\Gamma^h _1$ and the corresponding surface meshing and the CutFEM cell subtesselation are depicted in Figures \ref{BoneImage2} and \ref{BoneZoom2}, respectively.

%Paragraph 3. Stress component! Mori-Tanaka. Indicator function\\
We employ the MMT homogenization approach to compute the macroscale effective material properties. The mixing approach uses the level-set-based indicator function $\alpha_h$, defined in Equation~\eqref{Eq:Weight}, as shown in Figure \ref{Eq:Weight}. For the homogenization, we obtain the volume fractions of trabecular bone from \cite{perilli07.2}, where the bone volume fraction is reported as $B_v =0.192$. Assuming the microscale properties as $E_m =1$ and $\nu_{m}=0.3$, we derive the homogenized properties as $E_{\mathcal{M}}=0.15$ and $\nu_{\mathcal{M}}=0.3$.  

\begin{figure}[!ht]
   \centering
    \subfloat[ \label{FEMc}]{%
     \includegraphics[scale=0.25]{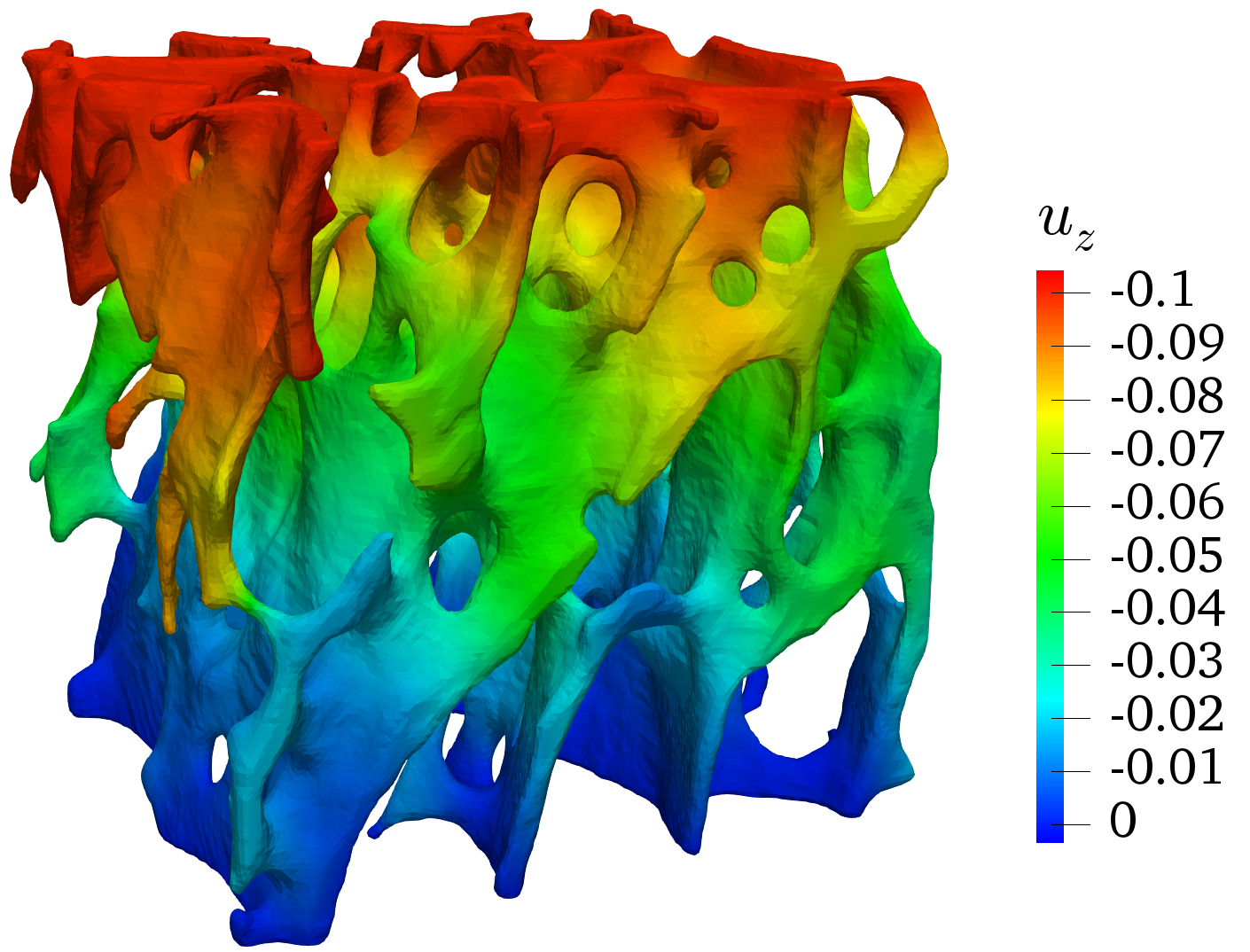}
     }
    \hfill
   \centering
    \subfloat[ \label{FEMc2}]{%
     \includegraphics[scale=0.23]{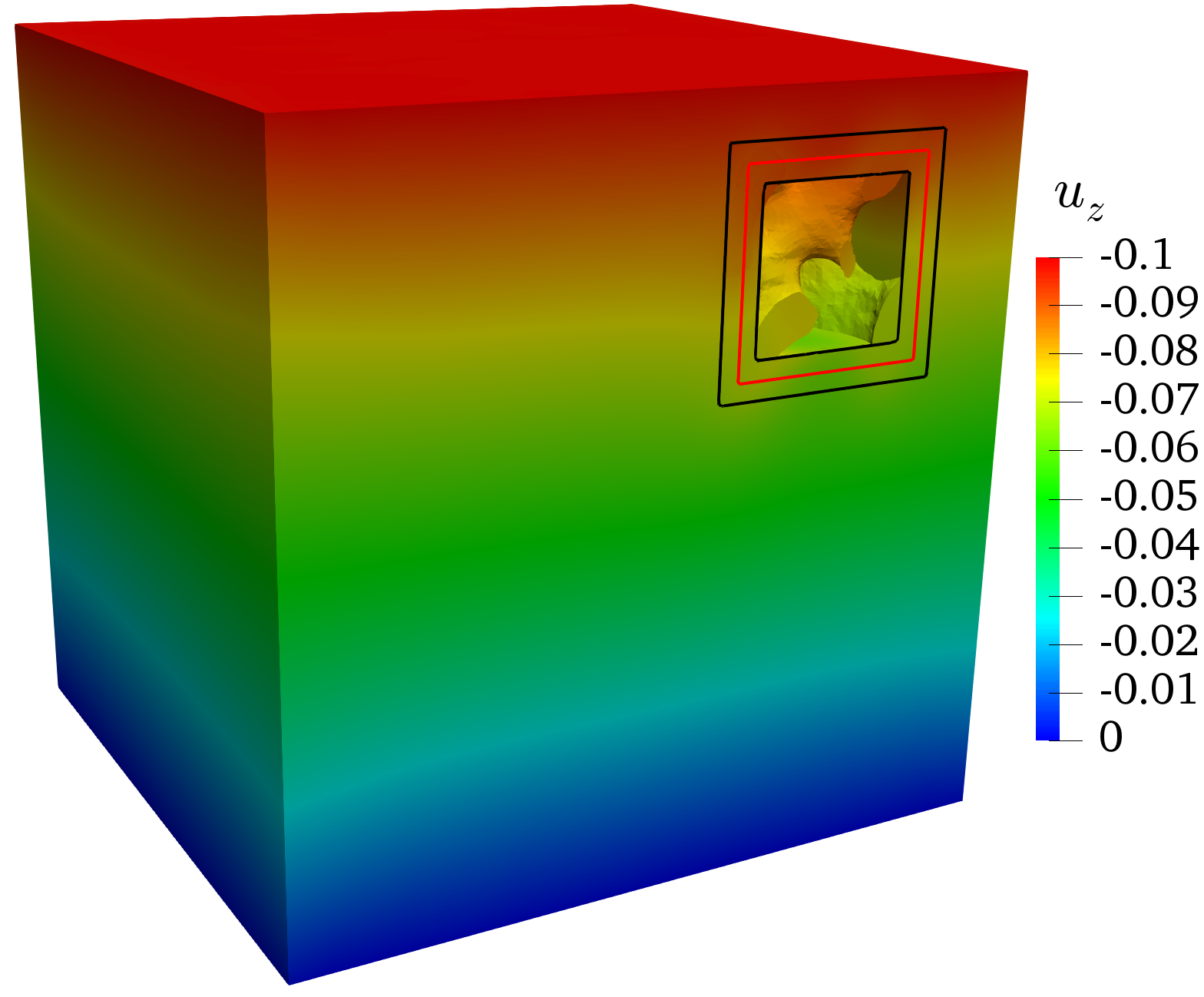}
     }
    \hfill
     \subfloat[ \label{CutFEMc3}]{%
      \includegraphics[scale=0.24]{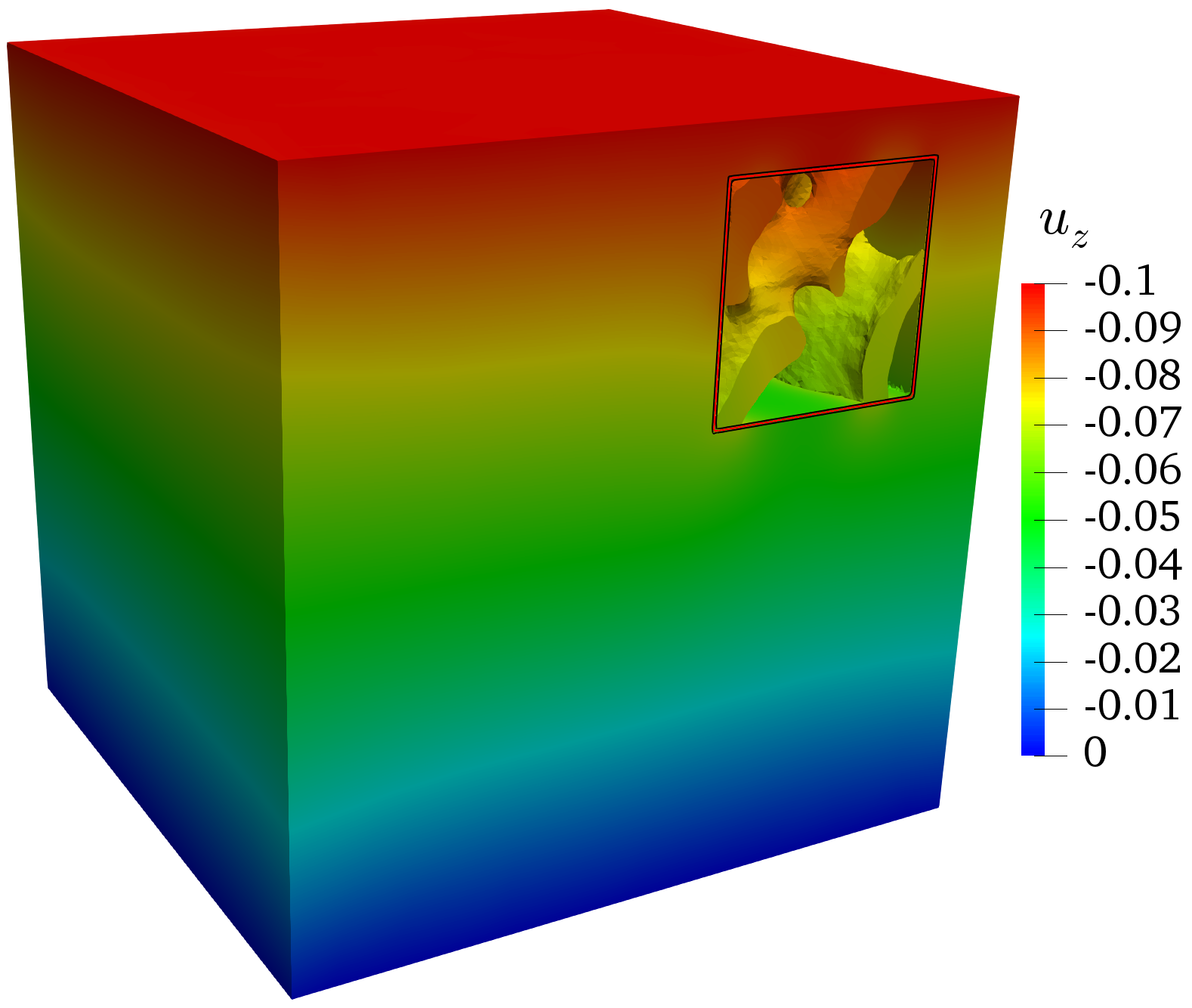}
     }
    \hfill
         \subfloat[ \label{CutFEMc4}]{%
       \includegraphics[scale=0.24]{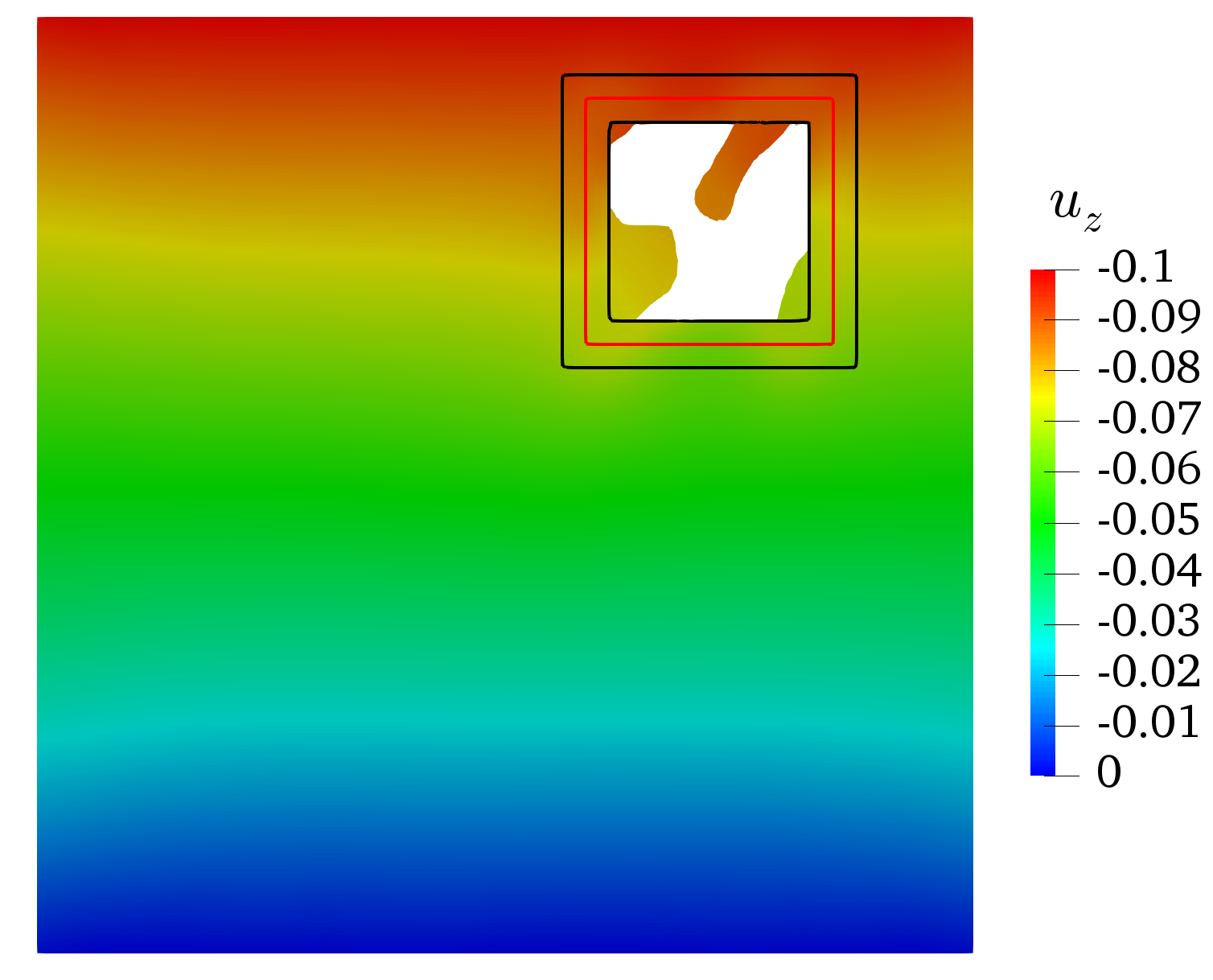}
     }
    \hfill
    \subfloat[ \label{CutFEMc5}]{%
     \includegraphics[scale=0.24]{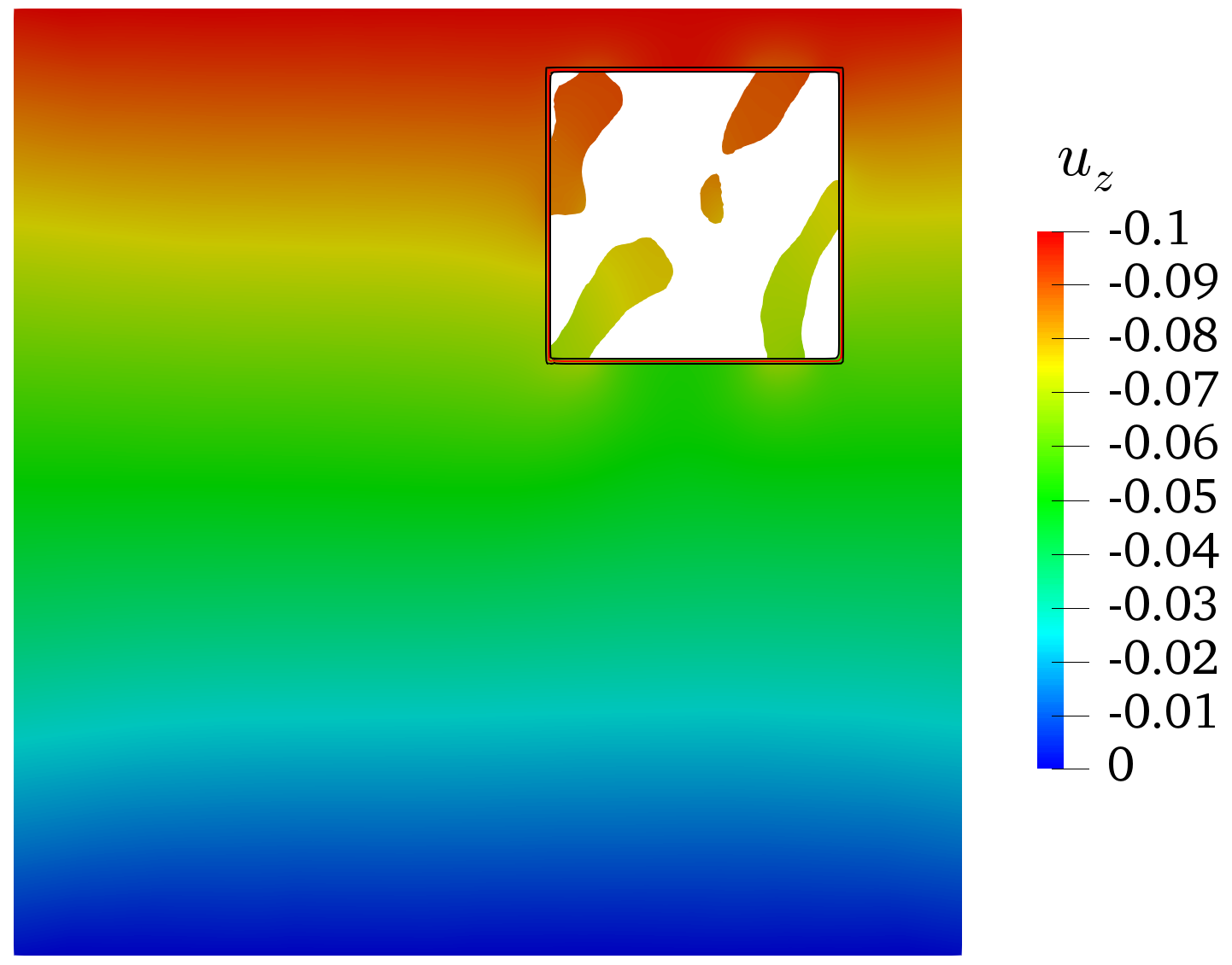}
     }
    \hfill
     \caption{3D and 2D representations of FEM and mixed multiscale displacement field component $u_y$, a) 3D FEM reference model, b) $2\epsilon=0.1$, c) $2\epsilon=0.01$, d) $2\epsilon=0.1$ and e) $2\epsilon=0.01$.}
     \label{fig:ModelCDisps}
\end{figure}

We perform a compression test for a full microscale FEM (as reference) and the mixed multiscale method with one zooming region in an arbitrary location. The displacement field component $u_y$ of these computations is shown in Figure \ref{fig:ModelCDisps}. For the mixed multiscale approach, the 3D simulations are carried out for two different smoothing lengths ($2\epsilon = 0.01, 0.1$) to study the mixing technique's stability for both sharp and wide transition regions. The comparison between full microscale and multiscale results shows that our level-set-based multiscale method can be successfully applied for 3D complex problems, in a mesh independent manner, and the mixing technique is stable for both types of transition regions.

\begin{figure}[!ht]
   \centering
    \subfloat[ \label{Indic001}]{%
     \includegraphics[scale=0.2]{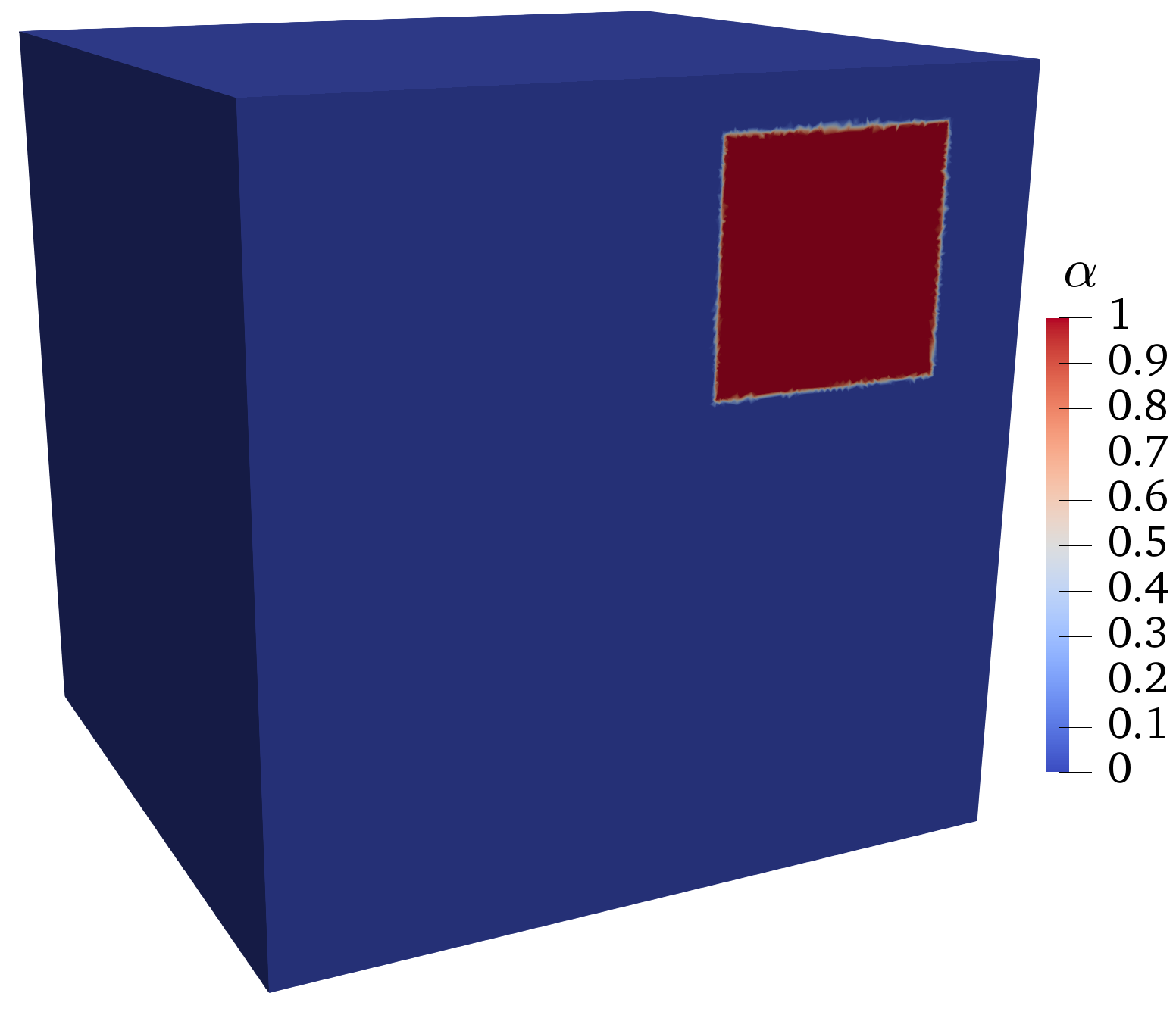}
     }
    \hfill
     \subfloat[ \label{indic01}]{%
       \includegraphics[scale=0.2]{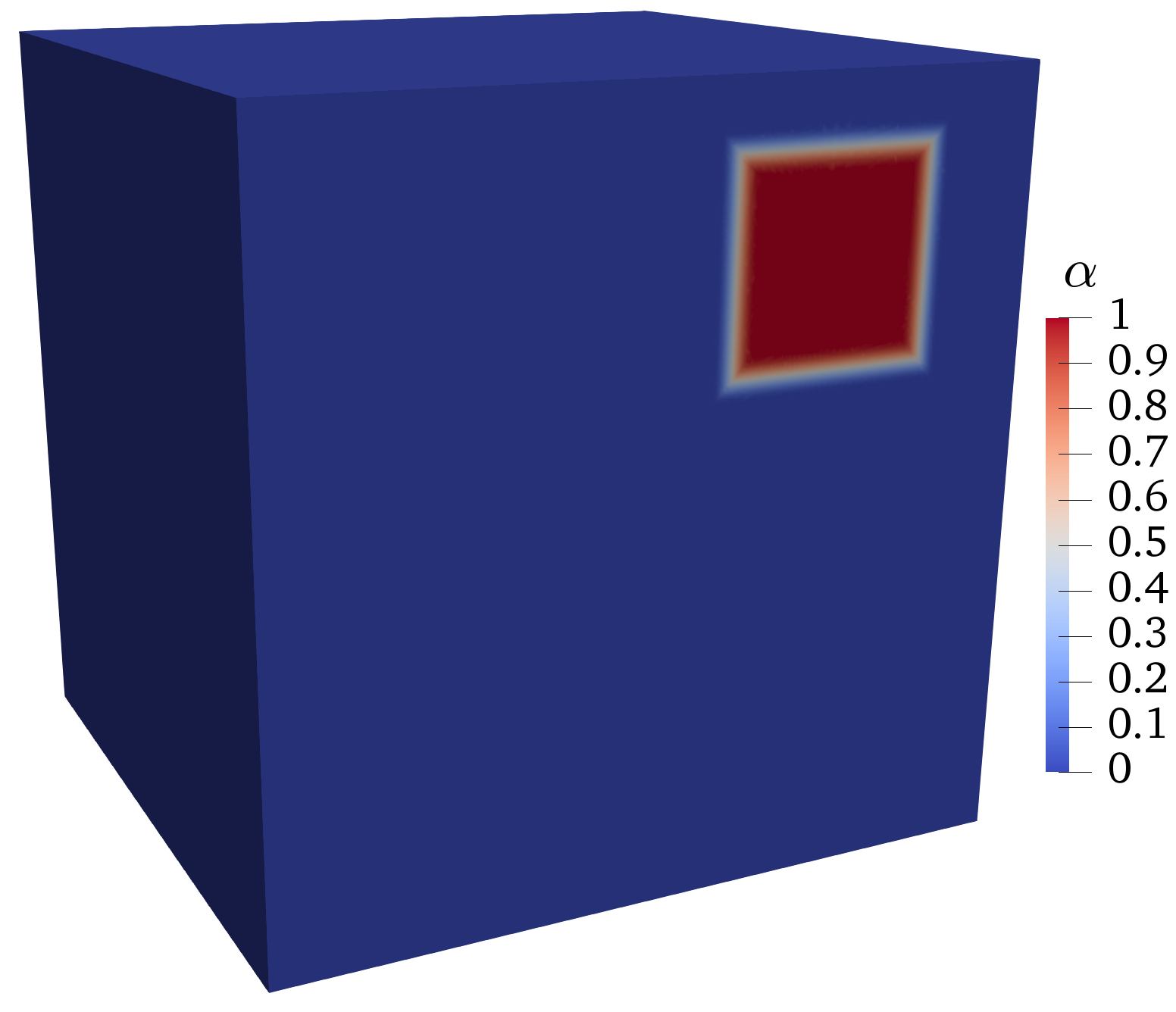}
     }
    \hfill
     \caption{Smoothing weight function $\alpha_h$ used for the 3D bone example with a) $2\epsilon=0.01$ and b) $2\epsilon=0.1$.}
     \label{fig:Meshess}
\end{figure}

\begin{figure}[!ht]
   \centering
    \subfloat[ \label{FEMcs}]{%
     \includegraphics[scale=0.25]{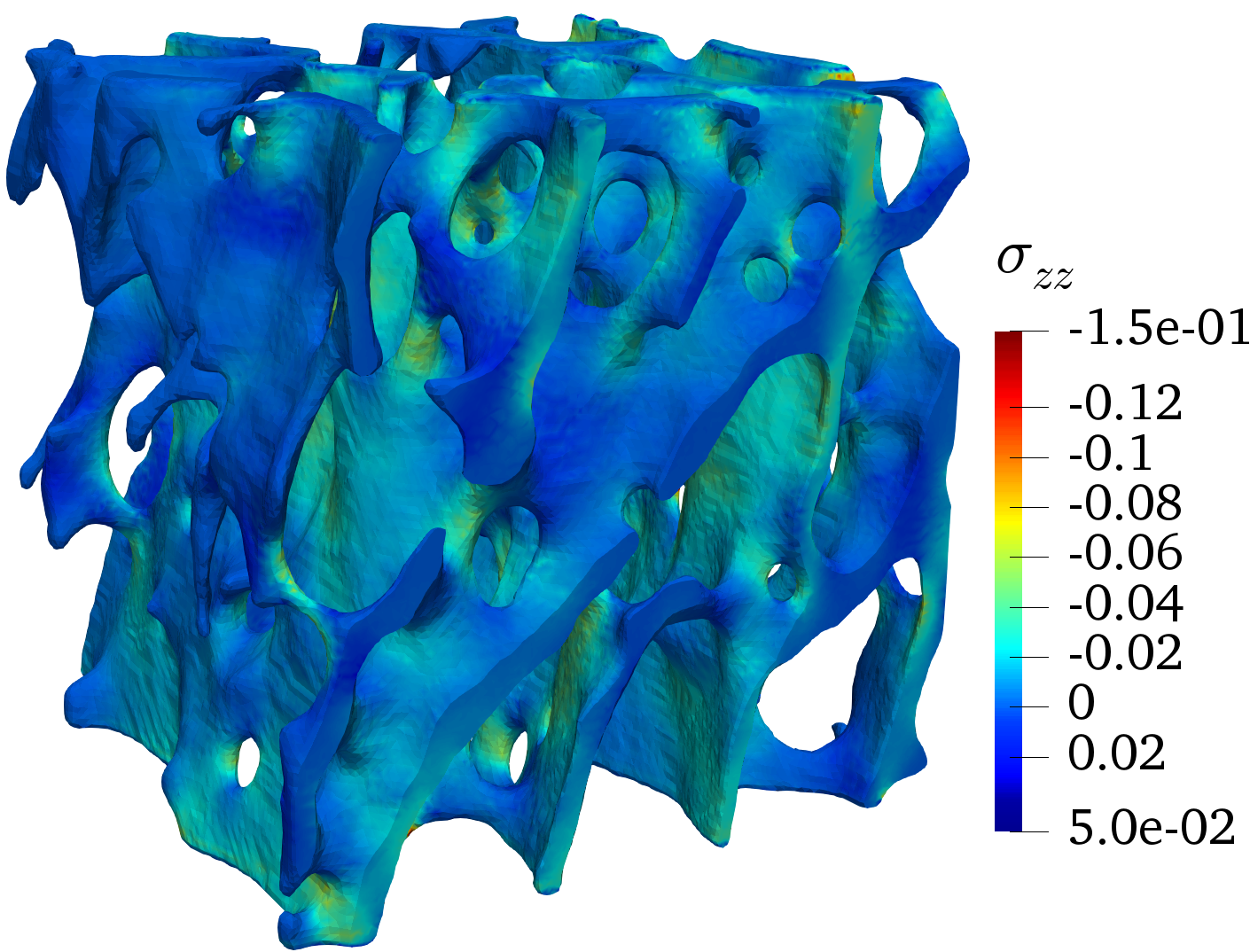}
     }
    \hfill
     \subfloat[ \label{Mixed001cs}]{%
       \includegraphics[scale=0.25]{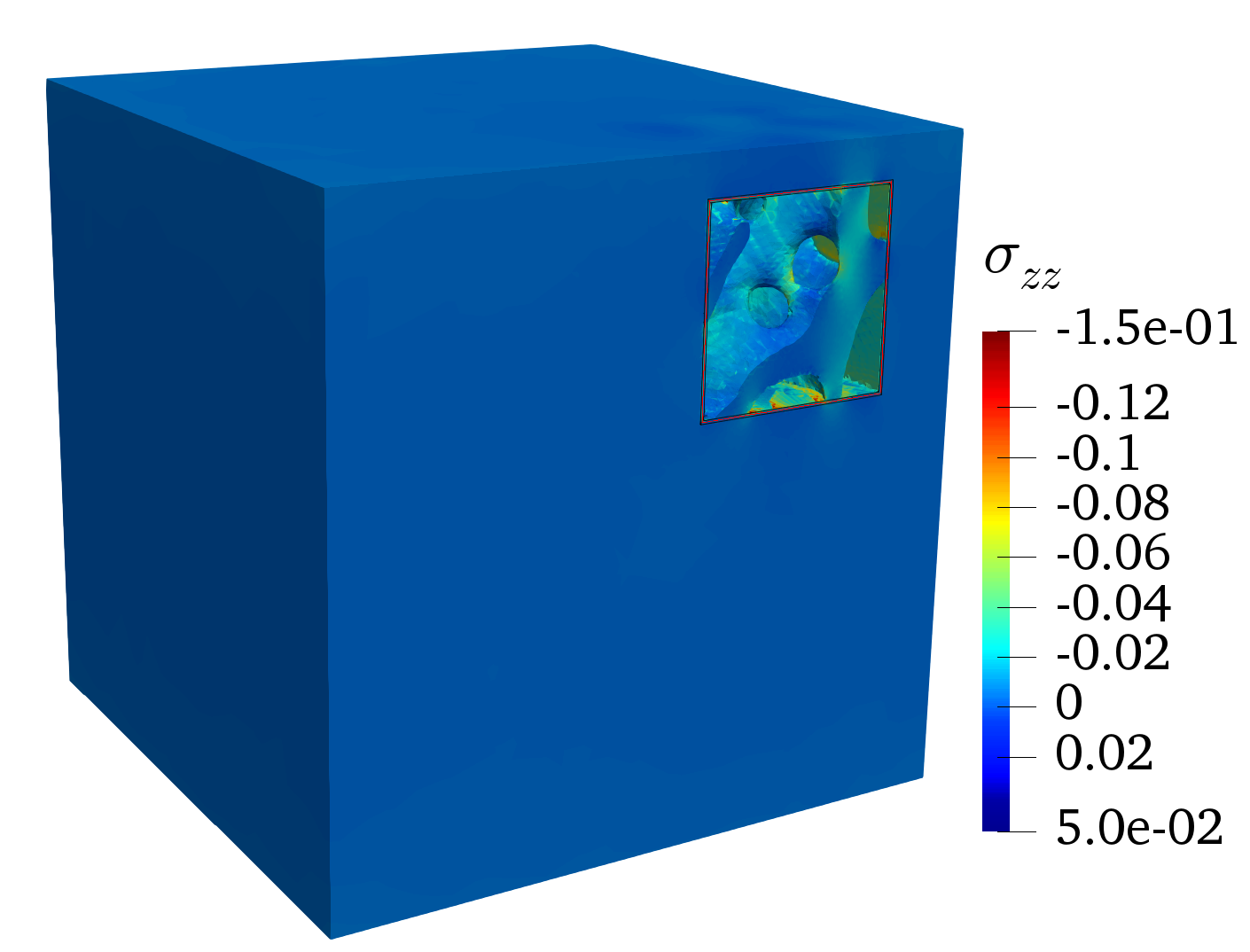}
     }
    \hfill
     \subfloat[ \label{Mixedcs}]{%
      \includegraphics[scale=0.25]{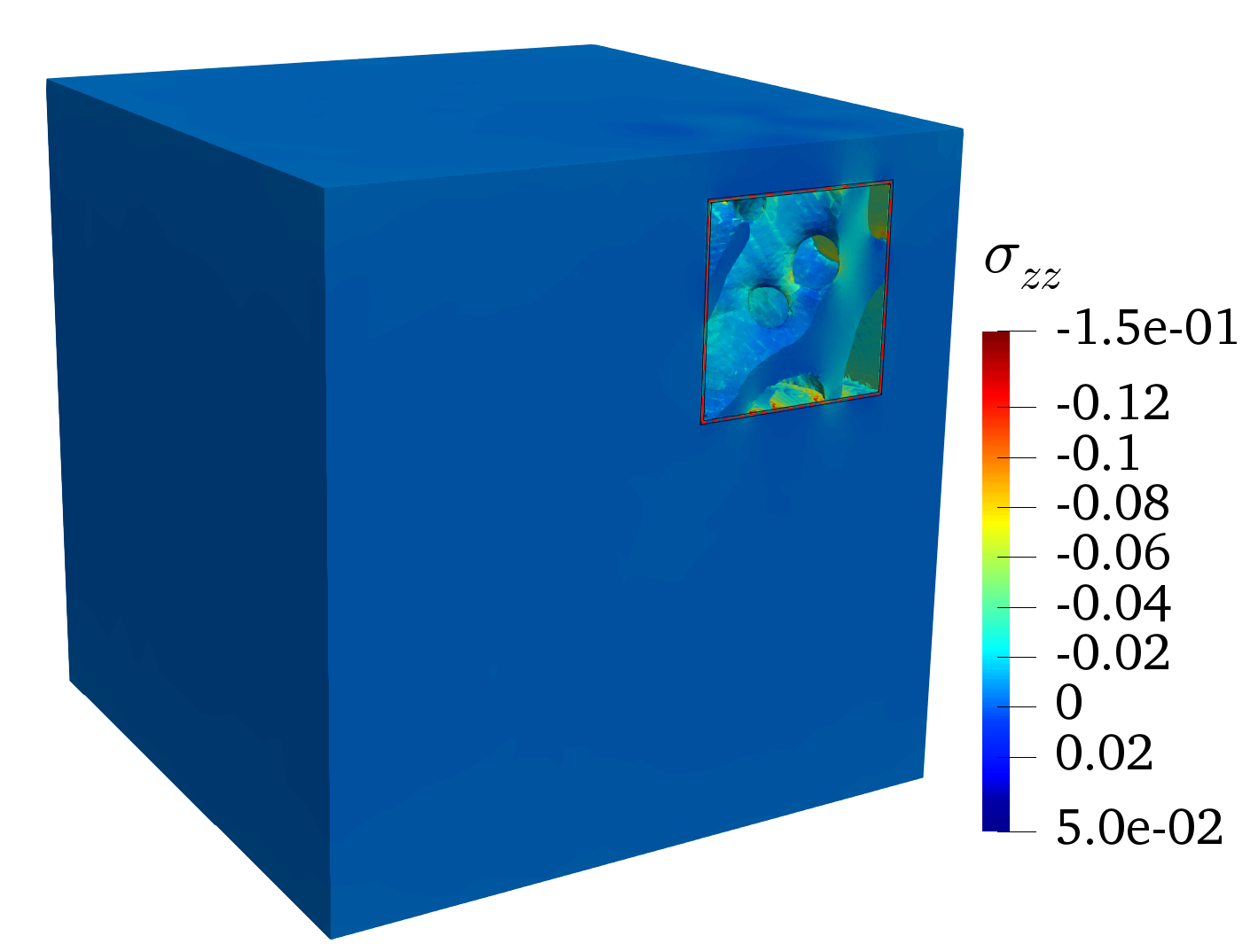}
     }
    \hfill
     \caption{Stress component $\sigma_{yy}$, a) FEM with $h=0.036$, b) mixed multiscale model-A with $h_{min}=0.036$, $\epsilon=0.01$ and only cut elements regularized, and c) mixed multiscale model-B with $h_{min}=0.036$, $\epsilon=0.01$, cut and porous elements are regularized.}
     \label{fig:ModelCStress}
\end{figure}

To further investigate the accuracy of numerical results, we show the variation of stress component $\sigma_{yy}$ for two smoothing lengths in Figure \ref{fig:ModelCStress}. The results show that the response inside the zoom is consistent with the corresponding FEM reference model.

\begin{figure}
\centering
  \includegraphics[scale=.25]{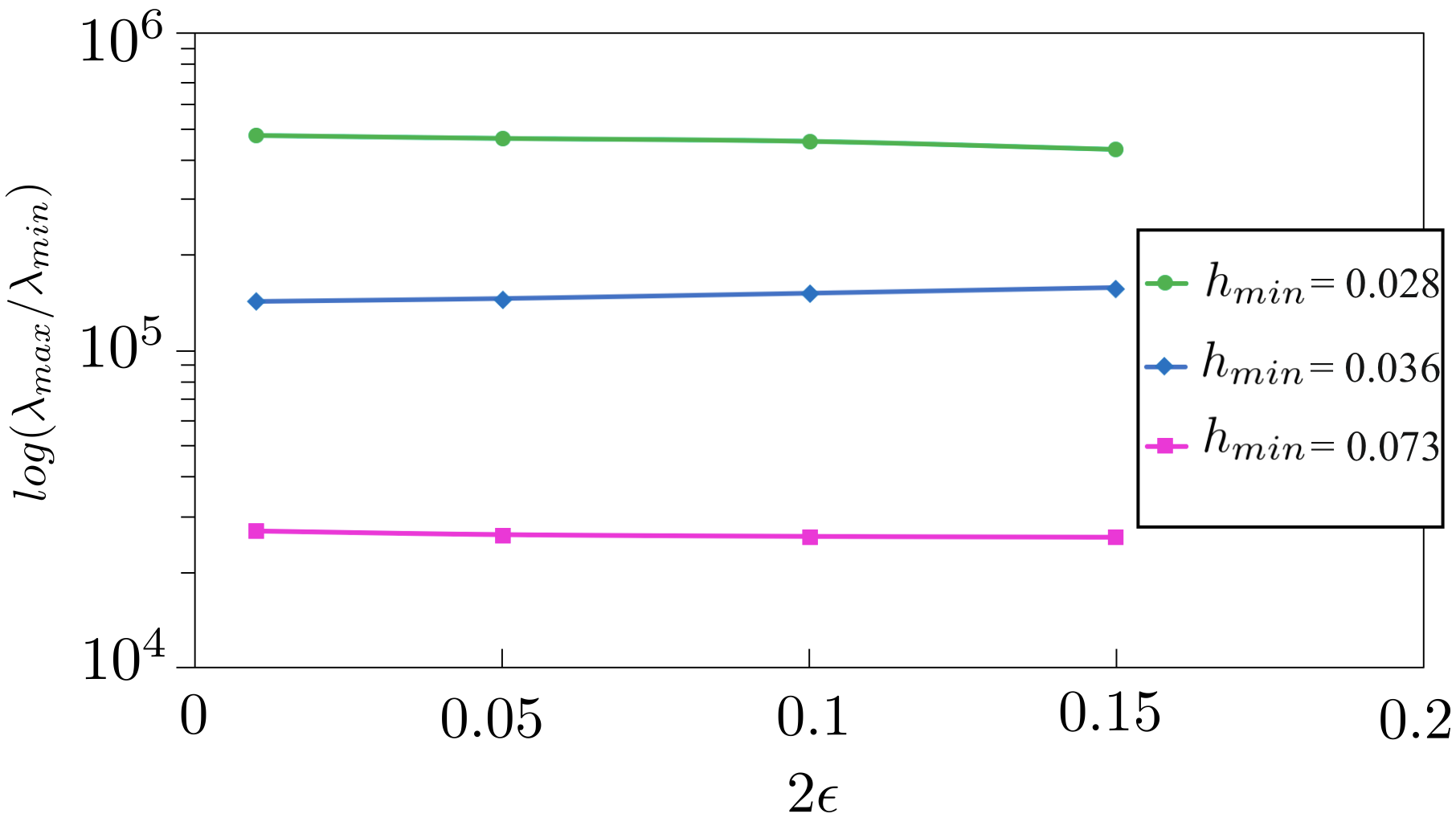}
  \caption{Condition numbers for mixed multiscale models with various mesh sizes and smoothing lengths.}
  \label{fig:ModelCCondition}
\end{figure}

We also study the condition number of the 3D mixed multiscale system matrix for different smoothing lengths and element sizes. The results in Figure \ref{fig:ModelCCondition} show that the condition numbers stay stable for various smoothing lengths and mesh sizes.

\clearpage

\section{Conclusions}

A framework was proposed to construct an unfitted concurrent multiscale model for heterogeneous structures. In our multiscale framework, we developed a mixing approach over a single background mesh to couple micro and macroscale models. Therefore, unlike domain decomposition methods where an interface condition is required between macro and microscale models, the interface constraint is not needed in our mixing approach.

We demonstrated the validity of our mixed multiscale framework for linear elasticity in 2D and 3D. We first tested the idea of a functional description of the whole heterogeneous structure by projecting it onto a background mesh which is fine in areas of interest and coarse outside. This projection of the functional description onto an adapted background mesh was done successfully by CutFEM, in which the geometry was approximated by a piecewise linear signed distance function in each background mesh element. We showed that the accuracy of results in the fine regions is good; however, the very coarse mesh cells outside of the regions of interest give rise to the random appearance of geometrical artifacts in the coarse region, yielding stress singularities. Next, we tested the same problem within the mixed multiscale framework where an equivalent homogenized domain was adopted in the coarse region. The results showed that employing the multiscale approach improves the results in the coarse domain. Next, we extended the application of our multiscale framework to 3D elasticity problems. Here, we employed a given surface mesh of trabecular bone to define the microscale geometry. The obtained results show a good agreement with the corresponding reference model in terms of global and local responses, where the corresponding multiscale system matrix remains well-conditioned under different mixing lengths. The numerical results in 2D and 3D simulations demonstrated the accuracy and robustness of our unfitted multiscale framework in modeling highly heterogeneous structures where the geometry and zoom location can be defined arbitrarily.

We intend to extend the current mixed multiscale framework for damage mechanics problems in the future. While in the current study, we define the zooming size and location arbitrarily, showing that the bridging scale works efficiently, the zoom region for the damage problems will be adaptively updated, keeping the damage/crack inside the zoom.

%\begin{center}
\section*{Acknowledgments} 
%\end{center}

The authors acknowledges the support of Cardiff University, funded by the European Union’s Horizon 2020 research and innovation program under the Marie Sklodowska-Curie grant agreement No. 764644. The authors also gratefully acknowledges Prof. Perilli from Flinders University and Dr Baruffaldi from Laboratorio di Tecnologia medica for providing the medical image that has been used in this work.

%\bibliographystyle{unsrt}
%\bibliographystyle{elsarticle-num}

%\begin{center}
\bibliography{ReferenceArlequin}  
%\end{center}

\end{document}